# Real Space Investigation of order-disorder transition of vortex lattice in Co-intercalated 2H-NbSe$_2$

*A thesis*

*Submitted to the*

*Tata Institute of Fundamental Research, Mumbai*

*for the degree of Doctor of Philosophy in Physics*

**By**

**Somesh Chandra Ganguli**

Department of Condensed Matter Physics and Materials Science

Tata Institute of Fundamental Research

Mumbai, India

September, 2016

*To my family (ma, baba, kaku, didi)*

# DECLARATION

This thesis is a presentation of my original research work. Wherever contributions of others are involved, every effort is made to indicate this clearly, with due reference to the literature, and acknowledgement of collaborative research and discussions.

The work was done under the guidance of Professor Pratap Raychaudhuri, at the Tata Institute of Fundamental Research, Mumbai.

**Somesh Chandra Ganguli**

In my capacity as supervisor of the candidate's thesis, I certify that the above statements are true to the best of my knowledge.

**Prof. Pratap Raychaudhuri**

Date:

# PREAMBLE

The work presented in my doctoral thesis is an experimental investigation of the nature of order to disorder transition of vortex lattice in a Type-II superconductor namely Co-intercalated 2H-NbSe$_2$ using scanning tunnelling spectroscopy and complimentary ac susceptibility measurements.

The thesis is organized as follows:

In Chapter I, I provide an introduction to various aspects of order to disorder transition and the vortex lattice in a Type-II superconductor as an ideal system to study the order to disorder transition in presence of random impurity.

Chapter II deals with the instrumentation required for real space imaging of vortex lattice namely scanning tunnelling spectroscopy (STS). Here I describe the basics of STS and also the technique of acquisition of real space image of vortex lattice and basic image analysis by Delaunay triangulation. It also contains the details of preparation and characterisation of Co-intercalated 2H-NbSe$_2$ single crystals.

In Chapter III, field driven disordering of vortex lattice at 350 mK is described. We also describe the presence of metastability in field cooled states.

In Chapter IV, we study the nature of field driven transformations by thermal hysteresis study namely superheating and supercooling measurements.

Chapter V contains the study of orientaional coupling between crystalline lattice and vortex lattice and its influence on the order to disorder transition.

Finally, in Chapter VI we conclude about our findings and possible future goals from both experimental and theoretical aspects.

# STATEMENT OF JOINT WORK

The experiments reported in this thesis have been carried out in the Department of Condensed Matter Physics and Materials Science at the Tata Institute of Fundamental Research under the supervision of Prof. Pratap Raychaudhuri. The results of the major portions of the work presented in this thesis have already been published in refereed journals.

All the STM/STS experiments and analysis discussed in this thesis were performed by me. The details of collaborative work are as follows:

Single crystals of Co-doped 2H-NbSe$_2$ were grown by Vivas Bagwe and Dr. Parasharam Shirage in the laboratory of Prof. Arumugam Thamizhavel. AC susceptibility measurement using 2-coil mutual inductance technique was done in collaboration with Harkirat Singh, Rini Ganguly and Indranil Roy. Some of the programmes for image analysis were coded by Garima Saraswat.

x

x

# Acknowledgements

I would like to express my sincere gratitude to my thesis advisor *Prof. Pratap Raychaudhuri* for providing continuous support and motivation during my Ph.D.

I thank my past and present lab mates *Dr. Anand Kamlapure, Dr. Garima Saraswat, Dr. Sanjeev Kumar, Dr. Harkirat Singh, Rini Ganguly, Indranil Roy, Dibyendu Bala, Vivas Bagwe, John Jesudasan, Dr. Parashram Shirage* and project students *Harsh Bhatt, Siddartha Vuppala, Atreyie Ghosh* for their collaboration, support and company.

I thank *Subash Pai* from Excel Instruments for all the prompt technical support. I also thank *Ganesh Jangam* and *Prof. P L Paulose* for help in SQUID magnetisation measurements, *Bagyshri Chalke, Rudhir Bapat* for characterizing the samples using EDX and *Atul Raut* for technical help. Most importantly I thank *Low Temperature Facility* team of TIFR for continuous supply of liquid He and Nitrogen.

I would like to thank *Prof. Shobo Bhattacharya, Prof. Srinivasan Ramakrishnan, Prof. Valerii Vinokur, Prof. Gautam Menon, Prof. Deepak Dhar, Prof. Rajdeep Sensarma, Prof. Victoria Bekeris, Prof. Hermann Suderow, Prof. Pradeep Kumar*, and *Prof. Gabriela Pasquini* for discussing and critiquing our work.

I thank all my teachers during my school, college and university days, especially *Dr. Zakir Hossain, Dr. Dipankar Chakrabarti, Dr. Kaushik Bhattacharya, Dr. Satyajit Banerjee, Dr Dipak Ghose, Dr. Basudev Ghosh, Dr. Kartick Chandra Pal, Dr. Subhankar Roy, Rathin babu, Sukumar babu, Chinmay Babu, Jayashree ma'am, Ujjal da, Swarup da, Subhash da, Mousumi di* and many more for their support, guidance and love during my formative years.

I thank all my friends that I had the good fortune of having: *Baibhab, Tamonash, Arpan, Susanta, Arnab, Debanjan, Subhajit, Sovon, Pramit, Swagatam, Chandan, Gouranga, Abhrajit, Abhijit, Kaushik, Kunal, Rupak, Kartick, Buddha, Rajat, Sujit, Debarchan, Uday, Tanay, Anshuman, Krishanu, Abhradip, Anondo, John, Arup, Sayandip, Arka, Sayanti, Gorky, Dibyendu, Kajal, Arindam, Sun, Ankita, Mrinmoyee, Buro, Soumo, Santanu, Malancha, Sourav, Kalyan, Supriyo, Nabarun, Bhanu, Nihit, Om Prakash, Subhrangshu, Soham, Varghese, Jayasuriya, Ajith, Deep, Nirupam, Swagata, Bodhayan, Saikat, Amlan, Sanat, Atul,*

*Soumyadip, Randhir, Sunil, Sachin, Gajendra, Aditya* and many more for all the moments we shared together.

I had great time playing Table Tennis, Cricket, Badminton, Football, Frisbee, Volleyball and Tennis in TIFR. I enjoyed the monsoon trekking in the Sahyadri. I thank all the players and trekkers in TIFR.

And last but not the least, I would like to express my deepest gratitude towards my family, *ma, baba, kaku* and *didi* who stood by my side through thick and thin. I dedicate this thesis to them.

# List of Publications in refereed Journal

## Related to the thesis

1. Disorder-induced two-step melting of vortex matter in Co-intercalated NbSe$_2$ single crystals

**Somesh Chandra Ganguli**, Harkirat Singh, Indranil Roy, Vivas Bagwe, Dibyendu Bala, Arumugam Thamizhavel, and Pratap Raychaudhuri

**Phys. Rev. B 93, 144503 (2016)**

**2.** Orientational coupling between the vortex lattice and the crystalline lattice in a weakly pinned Co$_{0.0075}$NbSe$_2$ single crystal

**Somesh Chandra Ganguli**, Harkirat Singh, Rini Ganguly, Vivas Bagwe, Arumugam Thamizhavel and Pratap Raychaudhuri

**J. Phys.: Condens. Matter 28, 165701 (2016)**

3. Disordering of the vortex lattice through successive destruction of positional and orientational order in a weakly pinned Co$_{0.0075}$NbSe$_2$ single crystal

**Somesh Chandra Ganguli**, Harkirat Singh, Garima Saraswat, Rini Ganguly, Vivas Bagwe, Parasharam Shirage, Arumugam Thamizhavel & Pratap Raychaudhuri

**Scientific Reports 5, 10613 (2015)**

4. A 350 mK, 9 T scanning tunneling microscope for the study of superconducting thin films on insulating substrates and single crystals

Anand Kamlapure, Garima Saraswat, **Somesh Chandra Ganguli**, Vivas Bagwe, Pratap Raychaudhuri, and Subash P. Pai

**Review of Scientific Instruments 84, 123905 (2013)**

# Not related to the thesis

1. Emergence of nanoscale inhomogeneity in the superconducting state of a homogeneously disordered conventional superconductor

Anand Kamlapure, Tanmay Das, **Somesh Chandra Ganguli**, Jayesh B. Parmar, Somnath Bhattacharyya and Pratap Raychaudhuri

**Scientific Reports 3, 2979 (2013)**

2. Universal scaling of the order-parameter distribution in strongly disordered superconductors

G. Lemarie, A. Kamlapure, D. Bucheli, L. Benfatto, J. Lorenzana, G. Seibold, **S. C. Ganguli**, P. Raychaudhuri, and C. Castellani

**Phys. Rev. B 87, 184509 (2013)**

3. Enhancement of the finite-frequency superfluid response in the pseudogap regime of strongly disordered superconducting films

Mintu Mondal, Anand Kamlapure, **Somesh Chandra Ganguli**, John Jesudasan, Vivas Bagwe, Lara Benfatto and Pratap Raychaudhuri

**Scientific Reports 3, 1357 (2013)**

4. Andreev bound state and multiple energy gaps in the noncentrosymmetric superconductor BiPd

Mintu Mondal, Bhanu Joshi, Sanjeev Kumar, Anand Kamlapure, **Somesh Chandra Ganguli**, Arumugam Thamizhavel, Sudhanshu S Mandal, Srinivasan Ramakrishnan and Pratap Raychaudhuri

**Phys. Rev. B 86, 094520 (2012)**

# List of symbols and abbreviations

## Symbols

| | |
|---|---|
| $\Phi_0$ | Magnetic flux quantum |
| $e$ | electronic charge |
| $h$ | Planck's constant |
| $a_0$ | Vortex lattice constant |
| $E_c$ | Condensation energy |
| $E_F$ | Fermi energy |
| $\omega_D$ | Debye cut-off frequency |
| N(0) | Density of states at Fermi energy |
| $\Delta$ | Superconducting energy gap |
| $H_{c1}$ | Lower critical field |
| $H_{c2}$ | Upper critical field |
| $J_s$ | Supercurrent density |
| $\lambda_L$ | London penetration depth |
| $\xi$ | Pippard coherence length |
| $\xi_{GL}$ | Ginzburg Landau coherence length |
| $H_P$ | Peak field |
| $H_P^{on}$ | Onset field of peak effect |
| $J_c$ | Critical current density |
| $V_c$ | Correlation volume |
| $T_c$ | Superconducting transition temperature |
| $G_{\vec{k}}(\vec{r})$ | Positional correlation function |
| $G_6(\vec{r})$ | Orientational correlation function |

# Abbreviations

| | |
|---|---|
| ODT | Order disorder transition |
| 2D | 2 dimensional |
| 3D | 3 dimensional |
| BKT | Berezenskii Kosterlitz Thouless |
| BKTHNY | Berezenskii Kosterlitz Thouless Halperin Nelson Young |
| VL | Vortex lattice |
| CL | Crystalline lattice |
| STM | Scanning tunnelling microscope/ microscopy |
| STS | Scanning tunnelling spectroscopy |
| LT-STM | Low temperature scanning tunnelling microscope |
| FC | Field cooled |
| ZFC | Zero field cooled |
| FT | Fourier transform |
| OS | Ordered state |
| OG | Orientational glass |
| VG | Vortex glass |
| MG | Multidomain glass |
| QLRPO | Quasi-long range positional order |
| SD | Spectral density |
| EDX | Energy dispersive x-ray spectroscopy |
| SQUID | Superconducting quantum interference device |

# Table of Contents

# Synopsis

## I. Introduction

Identical interacting particles form periodic structure below a certain temperature. This phenomenon has been observed in a variety of systems having different kinds of interactions; e.g. formation of crystalline solids below melting point, self-arrangement of various molecules over substrates etc. Though this formation of ordered structures is a universal phenomenon, till now no analytic derivation has found out the reason behind it and it still remains one of the outstanding problems in condensed matter physics. All of these periodic structures undergo an order-disorder transition (ODT). For example crystalline solids undergo melting to liquid state where the periodicity in the system is completely lost. Crystalline solid to liquid melting transition is a well-known 1st order phase transition having latent heat of transition. Berezenskii, Kosterlitz, Thouless and later Halperin, Nelson and Young predicted that a 2 dimensional (2-D) system having logarithmic interactions can undergo a 2-step continuous phase transition[1]. This is known as BKT (or BKTHNY) transition and has been observed in various 2-D systems[2]. But nature of ODT in most periodic system has remained controversial.

## I. *a*. Our model system: Vortex lattice in Type-II superconductor:

Type-II superconductors can withstand magnetic flux lines each having integer flux quanta $\Phi_0 = \frac{h}{2e}$ threading through it above a characteristic field value called lower critical field ($H_{c1}$). Due to each flux line having shielding current circling in the same direction (depends on the applied magnetic field), they experience mutual repulsion among them resulting in periodic arrangement of the flux lines to minimise the energy. It can be shown that for a conventional S-wave superconductor, the flux lines arrange themselves in a hexagonal (triangular) lattice with lattice constant $a_0 = (\frac{4}{3})^{\frac{1}{4}}\sqrt{\frac{\Phi_0}{B}}$ ($B$ is applied magnetic field). These flux lines are called vortices and the lattice formed by them is called Abrikosov vortex lattice (VL). For an ideal Type-II superconductor VL should be perfectly periodic. But in a real system, there are various types of defects and impurities present creating local non superconducting (normal) region. To avoid the cost of condensation energy $E_c = \frac{1}{2}N(0)\Delta^2$, it is energetically favourable for vortices to pass through these normal regions, thus breaking the constraint of positional order and creating defects in otherwise periodic VL. It is called pinning of vortex lattice and the local impurities are called pinning centres. Creation of such defects in an ordered VL gradually

# Synopsis

drives it into a disordered state. VL thus provides a versatile model system to study ODT in a periodic medium in the presence of quenched random disorder.

At a characteristic magnetic field called onset peak field ($H_p^{on}$), the VL starts to get disordered. Phenomenological argument for this VL disordering was given by Pippard[3]. He argued that the vortex-vortex interaction which is the elastic deformation energy varies with applied magnetic field H as $(H-H_{c2})^2$ ($H_{c2}$: upper critical field); whereas the vortex impurity interaction goes as $(H-H_{c2})$. So as H→$H_{c2}$, vortex-vortex interaction energy falls faster to zero and at a particular field $H_p^{on}$, the vortex-impurity energy wins over so the vortices prefer to pass through pinning centres. As a result, the VL starts to disorder.

There are various macroscopic signatures of this ODT in VL. The most well-known signature is 'Peak effect' in bulk magnetization/critical current density[4] manifested as a non-monotonic increase in the bulk pinning and consequently of critical current density and diamagnetic response in ac magnetic susceptibility with field. Now, critical current density ($J_c$) is defined as the current at which vortices start moving. In an ideal defect-free Type-II superconductor, upon application of infinitesimal current the vortices will start moving due to no opposing force. Due to Lorentz interaction between moving vortices and applied magnetic field, there will be an induced emf along the direction of applied current. This will lead to dissipation and hence there will be no zero resistance state. However in a real system, there is always finite pinning which leads to non-zero $J_c$. This can be quantitatively understood using collective pinning model proposed by Larkin and Ovchinnikov[5]. In this description, critical current density $J_c = \frac{1}{B}(\frac{n_p\langle f^2\rangle}{V_c})^{\frac{1}{2}}$. Where $n_p$ is density of pinning centres, f is the force exerted by a single pinning centre and $V_c$ is the correlation volume, which is the maximum volume within which order is maintained. As for an ordered vortex lattice $V_c$→∞, so $J_c$→0. As the VL becomes more disordered with increasing field and temperature $V_c$ decreases and hence $J_c$ increases. As a consequence, near $H_p^{onset}$ where the VL starts to disorder, there is a non-monotonic increase in $J_c$.

Upon increase in field/temperature the VL is proliferated by defects due to individual vortices getting randomly pinned and at a characteristic field and temperature it becomes an isotropic liquid. Therefore VL has been widely studied as a model system to understand the order to disorder transition (ODT) in the presence of random pinning[6]. Direct experimental evidence of VL melting has been observed in layered high-$T_c$ cuprates[7]. However, in conventional superconductors, the nature of VL melting has remained controversial. Thermodynamic signatures of a first-order ODT were found by many studies[8] in the presence of weak or moderate pinning. However, no evidence of VL melting below $H_{c2}$ was found on extremely pure Nb single crystals[9]. Also, since signatures of the ODT in conventional superconductors get considerably broadened in the presence of random pinning, it has been suggested by some authors that the ODT could be a continuous crossover rather than a phase transition[10]. Presence of random pinning potential also prevents establishing perfect crystalline or liquid phase by making the states glassy in nature with very slow kinetics. We address all these issues by imaging the VL across peak effect in real space using scanning tunnelling microscope (STM).

## II. a. Sample preparation and characterisation:

The samples used in this study consist of pure and Co-intercalated 2H-NbSe$_2$ single crystals. NbSe$_2$ is a conventional Type-II superconductor. The random intercalation of Co provides us a handle to control the degree of pinning. Earlier work[11] showed intercalated Co atoms apart from generating random pinning centres for the vortices, reduce anisotropy in the upper critical field compared to undoped NbSe$_2$, thereby making the vortex lines stiffer and hence less susceptible to bending. The Co$_{0.0075}$NbSe$_2$ single crystal was grown by iodine vapor transport method. Stoichiometric amounts of pure Nb, Se and Co together with iodine as the transport agent were mixed and placed in one end of a quartz tube, which was then evacuated and sealed. The sealed quartz tube was heated up in a two zone furnace between 5 to 10 days, with the charge-zone and growth-zone temperatures kept at, 800 °C and 720 °C respectively. We obtained single crystals with lateral size (in the a-b plane) of 2-5 mm and typical thickness varying between 60-150 µm. For the first set of crystals, we started with a nominal composition Co$_{0.0075}$NbSe$_2$ and the growth was continued for 5 days. We obtained single crystals with narrow distribution of $T_c$, in the range 5.3 – 5.93 K. The second set of crystals were also grown with the same nominal composition, but the growth was continued for 10 days. Here we obtained crystals with $T_c$ varying in the range 5.3 – 6.2 K. We conjecture that this larger variation of $T_c$ results from Co gradually depleting from the source such that crystal grown in later periods of the growth run have a slightly lower Co concentration. The third set of samples were pure 2H-NbSe$_2$ single crystal with $T_c$ ranging from 6.9 – 7.2 K.

Compositional analysis of few representative crystals from each batch was performed using energy dispersive x-ray analysis (EDX). We obtained a Co concentration of 0.45 atomic % for sample having $T_c$=5.88 K from the first set and 0.31 atomic % for sample having $T_c$=6.18 K from the second set. These compositions are marginally higher than the ones reported in ref.[12].

Synopsis

Since these measurements are close to the resolution limit of our EDX machine (below 1 atomic %), precise determination of the absolute value is difficult. However, measurements at various points on the crystals revealed uniform composition. This is also manifested by the sharp superconducting transitions observed from a.c. susceptibility in these crystals (Fig. 1(a)). DC magnetization measurements (Fig. 1(b)) using a Quantum Design superconducting quantum interference device (SQUID) magnetometer showed that with increasing Co concentration, peak effect gets more pronounced, i.e. critical current density increases (Fig. 1(c)) indicating stronger pinning. So, we have a control over the pinning strength by changing the Co-concentration.

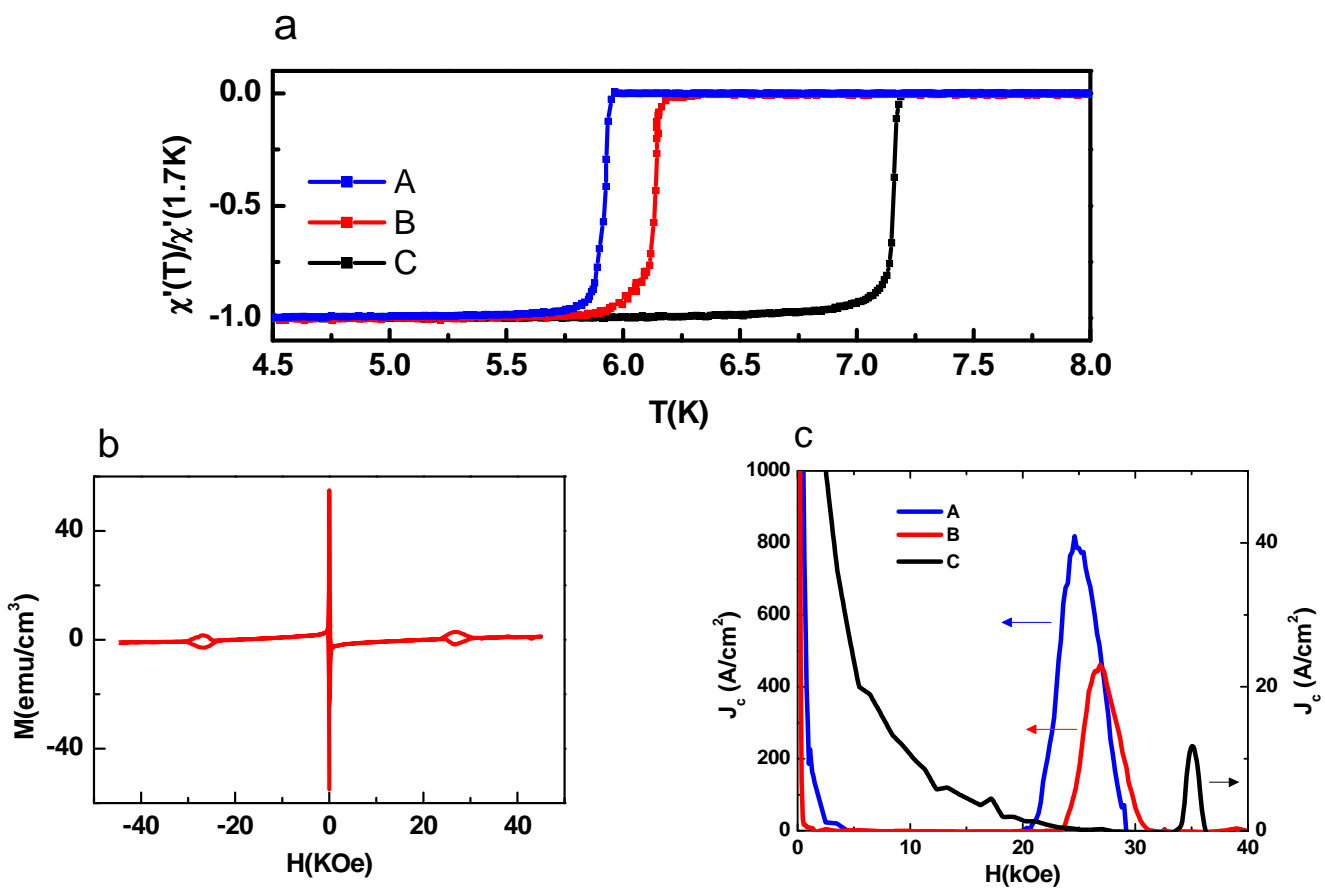

**Figure 1**. (a) Temperature variation of $\chi'$ in a zero applied dc magnetic field for 3 samples having $T_c$=5.9 K (A), 6.18 K (B) and 7.2 K (C). $\chi'$ is normalised to -1 in superconducting state and 0 in normal state. (b) Five-quadrant M-H loop for sample B at 1.8 K. (c) Variation of $J_c$ with magnetic field at 1.8 K for crystals A, B, C.

## II. b. Instrumentation: The Scanning Tunnelling Microscope

Scanning tunnelling microscope (STM) is an extremely versatile tool to probe the electronic structure of the material at the atomic scale. It works on the principle of quantum mechanical tunnelling between two electrodes namely a sharp metallic tip and the sample through vacuum as barrier. The tip is brought near the sample (within a few nm) using positioning units consisting of piezo-electric material. Tunnelling current flowing between tip and sample upon application of bias is amplified and recorded. Tunnelling current depends exponentially on the distance between tip and sample. By keeping the current constant using feedback loop and by scanning over the sample, the topographic image of the sample is generated.

The tunnelling conductance (*G*)) between the normal metal tip and the superconductor is given by,

$$\frac{dI}{dV}\bigg|_V = G(eV) \propto \int_{-\infty}^{\infty} N_S(E) \frac{\partial f(E-eV)}{\partial E} dE$$

At sufficiently low temperatures, Fermi function becomes step function; hence $G(eV) \propto N_s(eV)$. So, the tunnelling conductance is proportional to the local density of states of the sample at energy E = eV. Thus STM is able to measure local density of states through tunnelling conductance measurements. This method is called scanning tunnelling spectroscopy (STS). To measure the tunnelling conductance, tip sample distance is fixed by switching off the feedback loop and a small alternating voltage is modulated on the bias. The resultant amplitude of the current modulation as read by the lock-in amplifier is proportional to the *dI/dV* as can be seen by Taylor expansion of the current,

$$I(V + dV \sin(\omega t)) \approx I(V) + \frac{dI}{dV}\bigg|_V \cdot dV \sin(\omega t)$$

The modulation voltage used in the measurement is 150 $\mu V$ and the frequency used is 2.67 KHz.

Our home-built scanning tunnelling microscope (STM) operates down to 350mK and fitted with an axial 90 kOe superconducting solenoid[13]. Prior to STM measurements, the crystal is cleaved using a double-sided tape in-situ in vacuum of ~ $10^{-7}$ mbar, giving atomically smooth facets larger than 1.5μm × 1.5μm. The magnetic field is applied along the six-fold symmetric *c*-axis of the hexagonal 2H-NbSe$_2$ crystal. Well resolved images of the VL are obtained by measuring the tunnelling conductance (G(V) = dI/dV) over the surface at a fixed bias voltage (V~1.2mV) close to the superconducting energy gap, such that each vortex core manifests as a local minimum in G(V). Each image was acquired after stabilizing to the magnetic field. The precise position of the vortices are obtained from the images after digitally removing scan lines and finding the local minima in G(V) using WSxM software[14]. To identify topological defects, we Delaunay triangulated the VL and determined the nearest neighbor coordination for each flux lines. Topological defects in the hexagonal lattice manifest as points with 5-fold or 7-fold coordination number. Since, the Delaunay triangulation procedure gives some spurious bonds at the edge of the image we ignore the edge bonds while calculating the average lattice constants and identifying the topological defects. We perform the complimentary ac susceptibility measurements by 2-coil mutual inductance technique using our home-built ac susceptometer.

## III. Two step disordering of vortex lattice:

Here, we study the sequence of field induced disordering of VL across the peak effect in real space using STS. The measurements are performed at the lowest temperature, i.e. 350 mK. At first, the peak effect regime is established from isothermal ac susceptibility measurements. Then we track the VL is real space at various points on this curve. In addition to the order-disorder transition, we also investigate the existence of different metastable state which differ from their corresponding equilibrium states in the degree of positional and orientational order.

## III. *a*. Bulk pinning properties

The bulk pinning response of the VL was measured using 2-coil mutual inductance technique. The sample (having $T_c$ = 5.3K and $\Delta T_c$ = 200mK) is sandwiched between a quadrupolar primary coil and a dipolar secondary coil. The ac excitation amplitude in the primary coil is 10 mOe at 60 KHz. The response is measured using lock-in technique. Figure 2(a) shows the real part of the linear ac susceptibility ($\chi'$) at 350 mK when the sample is cycled through different thermomagnetic histories. The $\chi'$-$H$ for the zero field cooled (ZFC) state (red line) is obtained after cooling the sample to 350 mK in zero magnetic field and then ramping up the magnetic field. We observe "peak effect" as the diamagnetic response suddenly increases between 16 kOe ($H_p^{on}$) to 25 kOe ($H_p$) after which $\chi'$ monotonically increases up to $H_{c2}$ ~ 38 kOe. The ramp down branch is obtained by ramping down the magnetic field after it reaches a value $H > H_{c2}$ (black line). Between ZFC and ramp down branch, we observe a hysteresis starting below $H_p$ and extending well below $H_p^{on}$. The sample is said to be field cooled (FC) when after applying a field at 7 K (>$T_c$), it is cooled to 350 mK in the presence of the field. This state (solid squares) is much more disordered with a stronger diamagnetic response. But this is a non-equilibrium state. When the magnetic field is ramped up or ramped down from the pristine FC state, $\chi'$ merges with the ZFC branch or the ramp down branch respectively. Fig. 2(b) shows the phase diagram with $H_p^{on}$, $H_p$ and $H_{c2}$, obtained from isothermal χ'-H scans at different temperatures.

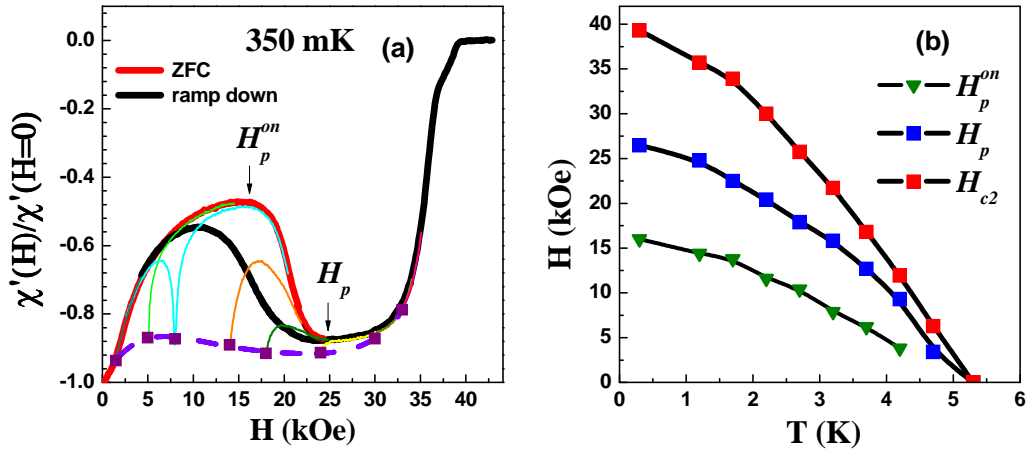

**Figure 2** (a) Magnetic field *(H)* dependence of the real part of linear ac susceptibility (χ') (normalised to its value in zero field) at 350 mK for the VL prepared using different thermomagnetic cycling. The red line is χ'-H when the magnetic field is slowly ramped up after cooling the sample in zero field (ZFC state). The black line is χ'-H when the magnetic field is ramped down from a value higher than $H_{c2}$. The square symbols stand for the χ' for the FC states obtained by cooling the sample from $T > T_c$ in the corresponding field; the dashed line shows the locus of these FC states created at different *H*. The thin lines starting from the square symbols show the evolution of χ' when the magnetic field is ramped up or ramped down (ramped down segment shown only for 0.8 T), after preparing the VL in the FC state. (b) Phase diagram showing the temperature evolution of $H_p^{on}$, $H_p$ and $H_{c2}$ as a function of temperature.

## III. *b*. Real space imaging of the VL

The VL was imaged in real space across the peak effect. All the images (Fig (3)) were taken at 350 mK which is the lowest temperature achievable in our system. So we can assume that the VL is at its ground state. The images were acquired over 1 µm × 1 µm area. Except at 32 and 34 KOe which were acquired over 400 nm × 400 nm area. Three distinct regions were identified: a) for H<$H_p^{onset}$ i.e. before the onset of peak effect where the shielding response monotonically increases, i.e. the sample becomes progressively less diamagnetic. In this region we have two representative points at H=10 KOe (not shown in the figure) and H=15 KOe (Fig 3(a)). At both these points no defect was identified in our field of view, b) $H_p^{onset}$<H<$H_p$, i.e. at the region where the shielding response decreases non-monotonically. Here we have real space VL image at 20, 24 and 25 KOe (Fig 3(b),(c),(d)). At 20 KOe we observe a single vortex with 5-fold coordination and its nearest neighbour vortex having 7-fold coordination within our field of view. This defect-pair is called a dislocation. We have few more dislocations at 24 KOe. The number of dislocations increases with increasing magnetic field. This is reflected in the Fourier transform of the VL image also. As the field is increased the individual Fourier spots get broader. But the pattern still retains the six-fold symmetry. c) H>$H_p$, at 26 KOe (Fig 3(e))

## Synopsis

we observe that apart from 5-fold and 7-fold coordinated vortices appearing as nearest neighbours, there are individual 5-fold defects which have no corresponding 7-fold defect at nearest neighbour position and vice-versa. These individual defects are called disclination. Appearance of disclination reflect on the Fourier transform as it becomes isotropic with no clear maxima indicating broken orientational order. The number of disclinations also increase with increasing field as seen in 30 KOe data (Fig 3(f)). At fields above 30 KOe, we see (Fig 3(g),(h)) that vortices are no longer appearing as sharp minima; instead they appear as long streak-like pattern. This indicates the melting of VL because the individual vortices are moving within our field of view.

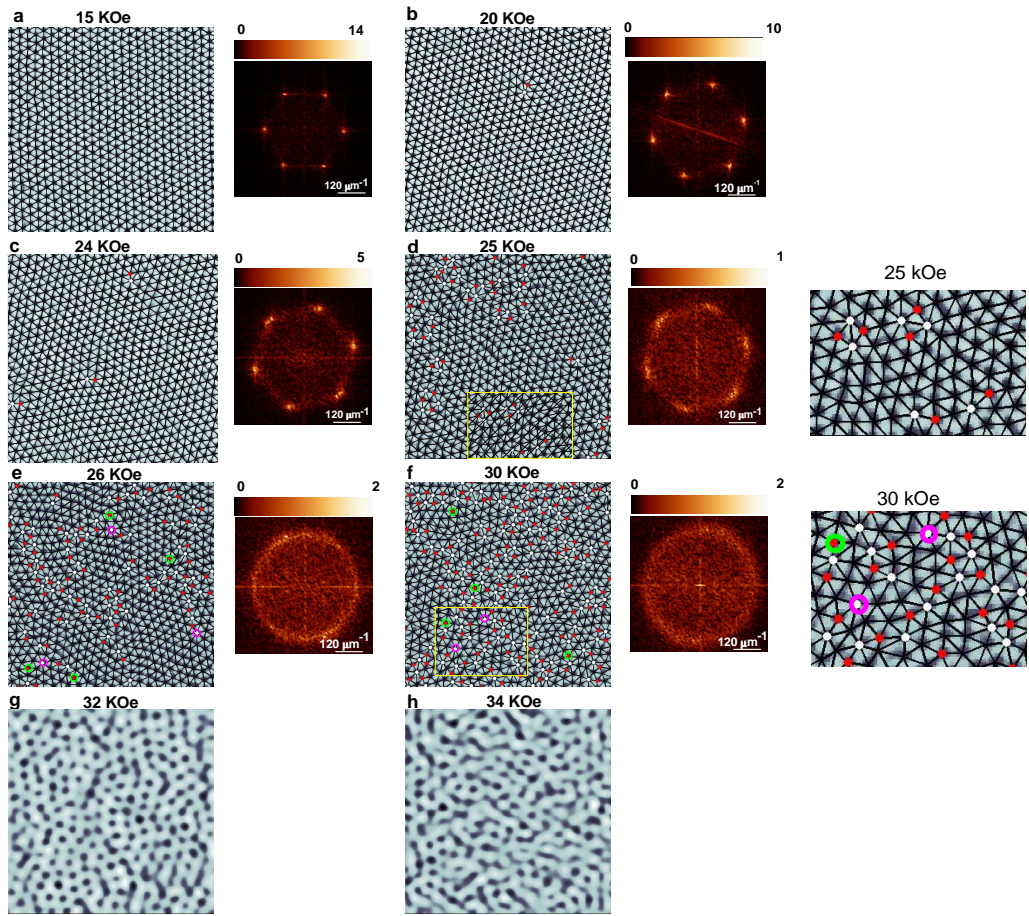

**Figure 3**. (a)-(f) STS conductance maps showing real space ZFC vortex lattice image at 350mK along with their Fourier transforms. Delaunay triangulation of the VL are shown as solid lines joining the vortices and sites with 5-fold and 7-fold coordination are shown as red and white dots respectively. The disclinations (unpaired 5-fold or 7-fold coordination sites) observed at 26 and 30 kOe are highlighted with green and purple circles. Images shown here have been zoomed to show around 600 vortices for clarity. The Fourier transforms correspond to the unfiltered images; the color scales are in arbitrary units. The expanded view of the defect structure inside the region bounded by the yellow boxes are shown for the VL at 25 and 30 kOe next to panels (d) and (f) respectively. (g)-(h) VL images (400nm× 400nm) at 32 kOe and 34 kOe.

Quantitative information on this sequence of disordering is obtained from the orientational ($G_6(\bar{r})$) and positional correlation ($G_{\bar{K}}(\bar{r})$) functions. $G_6(\bar{r})$ measures the degree of misalignment of the lattice vectors separated by distance r, with respect to the lattice vectors of an ideal hexagonal lattice. It is defined as, $G_6(r) = (1/n(r,\Delta r))\left(\sum_{i,j}\Theta\left(\frac{\Delta r}{2} - |r - |\bar{r}_i - \bar{r}_j||\right)\cos 6(\theta(\bar{r}_i) - \theta(\bar{r}_j))\right)$, where $\Theta(r)$ is the Heaviside step function, $\theta(\bar{r}_i) - \theta(\bar{r}_j)$ is the angle between the bonds located at $\bar{r}_i$ and the bond located at $\bar{r}_j$, $n(r,\Delta r) = \sum_{i,j}\Theta\left(\frac{\Delta r}{2} - |r - |\bar{r}_i - \bar{r}_j||\right)$, $\Delta r$ defines a small window of the size of the pixel around $r$ and the sums run over all the bonds. We define the position of each bond as the coordinate of the mid-point of the bond. $G_{\bar{K}}(\bar{r})$ measures the relative displacement between two vortices separated by distance $r$, with respect to the lattice vectors of an ideal hexagonal lattice. It is defined as, $G_{\bar{K}}(r) = (1/N(r,\Delta r))\left(\sum_{i,j}\Theta\left(\frac{\Delta r}{2} - |r - |\bar{R}_i - \bar{R}_j||\right)\cos \bar{K}\cdot(\bar{R}_i - \bar{R}_j)\right)$, where $\bm{K}$ is the reciprocal lattice vector obtained from the Fourier transform, $R_i$ is the position of the $i$-th vortex, $N(r,\Delta r) = \sum_{i,j}\Theta\left(\frac{\Delta r}{2} - |r - |\bar{R}_i - \bar{R}_j||\right)$ and the sum runs over all lattice points. The range of $r$ is restricted to half the lateral size (1 µm) of each image, which corresponds to $11a_0$ (where $a_0$ is the average lattice constant) at 10 kOe and $17a_0$ for 30 kOe. For an ideal hexagonal lattice, $G_6(r)$ and $G_{\bar{K}}(r)$ shows sharp peaks with unity amplitude around 1st, 2nd, 3rd etc… nearest neighbour distance for the bonds and the lattice points respectively. As the lattice disorder increases, the amplitude of the peaks decay with distance and neighbouring peaks at large $r$ merge with each other.

At 10 kOe and 15 kOe, $G_6(r)$ saturates to a constant value of ~0.93 and ~0.86 respectively after 2-3 lattice constants, indicating long-range orientational order. The envelope of $G_{\bar{K}}(\bar{r})$ decays slowly but almost linearly with r. But this linear decay cannot continue for large r. It is due to our limited field of view, we are not being able to capture the asymptotic behaviour at large r at low fields. Though we cannot say for sure that $G_{\bar{K}}(\bar{r})$ decays as a power-law for large r as predicted for a Bragg glass (BG), the slow decay of $G_{\bar{K}}(\bar{r})$ along with the long-range orientational order indicates it is a state of quasi long-range positional order (QLRPO). This state is the ordered state (OS) of the VL. For (20-25) kOe, $G_6(r)$ decays slowly with increasing r, consistent with a power-law ($G_6(r) \propto 1/r^\eta$), characteristic of quasi-long-range orientational order. On the other hand, $G_{\bar{K}}(\bar{r})$ displays a more complex behaviour. At 20 kOe, within our field of view the $G_{\bar{K}}(\bar{r})$ envelope decays exponentially with positional decay length, $\xi_p$~6.7. However for 24 and 25 kOe the initial decay is faster, but the exponential decay is only up to small values of r/a$_0$. At higher values $G_{\bar{K}}(\bar{r})$ decays as a power-law. So this state having short range positional order and quasi-long range orientational order is a unique state (which is very similar to 'Hexatic phase' in 2-D systems) is called orientational glass (OG). The OG state thus differs from the QLRPO state in that it does not have a true long-range orientational order. It also differs from the hexatic state in 2-D systems, where $G_{\bar{K}}(\bar{r})$ is expected to decay exponentially at large distance. Finally, above 26kOe, $G_6(r) \propto e^{-r/\xi_{or}}$ ($\xi_{or}$ is the decay length of orientational order). This is amorphous vortex glass (VG) state with short-range positional and orientational order.

The VL structures for the ramp down branch are similar to ZFC: At 25 and 20 kOe the VL shows the presence of dislocations and at 15 kOe it is topologically ordered. At 25 kOe ≈ $H_p$, we observe that $G_{\bar{K}}(r)$ for ZFC and ramp down branch are similar whereas $G_6(r)$ decays faster for the ramp down branch. However, in both cases $G_6(r)$ decays as a power-law characteristic of the OG state. At 15 kOe, which is just below $H_p^{on}$, both ZFC and ramp down branch show long-range orientational order, while $G_{\bar{K}}(r)$ decays marginally faster for the ramp down branch. Thus, while the VL in the ramp down branch is more disordered, our data do not provide any evidence of supercooling with isothermal field ramping across either OS→OG or OG→VG transitions as expected for a first order phase transition.

The FC state show an OG at 10 kOe and 15 kOe (free dislocations), and a VG above 20 kOe (free disclinations). The FC OG state is however extremely unstable. This is readily seen by applying a small magnetic pulse (by ramping up the field by a small amount and ramping back), which annihilates the dislocations in the FC OG eventually causing a dynamic transition to the QLRPO state. The metastability of the VL persists even above $H_p$ where the ZFC state is a VG. The FC state is more disordered with a faster decay in $G_6(r)$.

## IV. Superheating and supercooling across the phase transition lines:

Having established that the VL disorders in two steps we now investigate the nature of these transformations. The 2-step disordering of VL is somehow reminiscent of 2-step BKT transition in 2-D system. However, for our sample thickness (~ 60-100 µm) the vortices can bend significantly along the length of the vortex. Thus the VL is in the 3-dimensional (3-D) limit. However, since we do not expect BKT transition for a 3-D system we explore here whether the two transformations from OS→OG and OG→VG correspond to two separate first order phase transitions.

One of the distinctive feature of a first order phase transition is the presence of superheating and supercooling. In this section, we perform thermal hysteresis measurements across the two transition to investigate if we can identify the corresponding superheated and supercooled states. The experiments were performed on crystal having $T_c$ = 5.88 K. At first, we concentrate on the OS-OG phase boundary. At first, ZFC state was prepared at 8 KOe, 420 mK (Fig. 4(a)). As the state lies well within OS, the VL is ordered with each vortex having six-fold coordination. The VL is then heated up to 4.14 K thus crossing the OS-OG phase boundary. This warmed-up ZFC state also continues to remain ordered (Fig. 4(b)). But, if we apply a dc magnetic field pulse of amplitude 300 Oe and image the resulting state, we observe that topological defects have proliferated the VL (Fig.4(c)). Corresponding autocorrelation functions defined as $G(\bar{r}) = \sum_{r'} f(\bar{r}+\bar{r}')f(\bar{r})$ (where $f(\bar{r})$ is the image matrix) are calculated. A faster radial decay of the autocorrelation function implies a more disordered state. We therefore have a superheated OS. Warming up the sample up to 4.35 K and then cooling it down to 1.6 K, we see that the VL continues to remain topologically disordered (Fig. 4(d)) indicating supercooled OG. But, again upon application of dc magnetic field pulse of 300 Oe, the VL becomes topologically ordered (Fig. 4(e)) thus going to equilibrium OS.

Synopsis

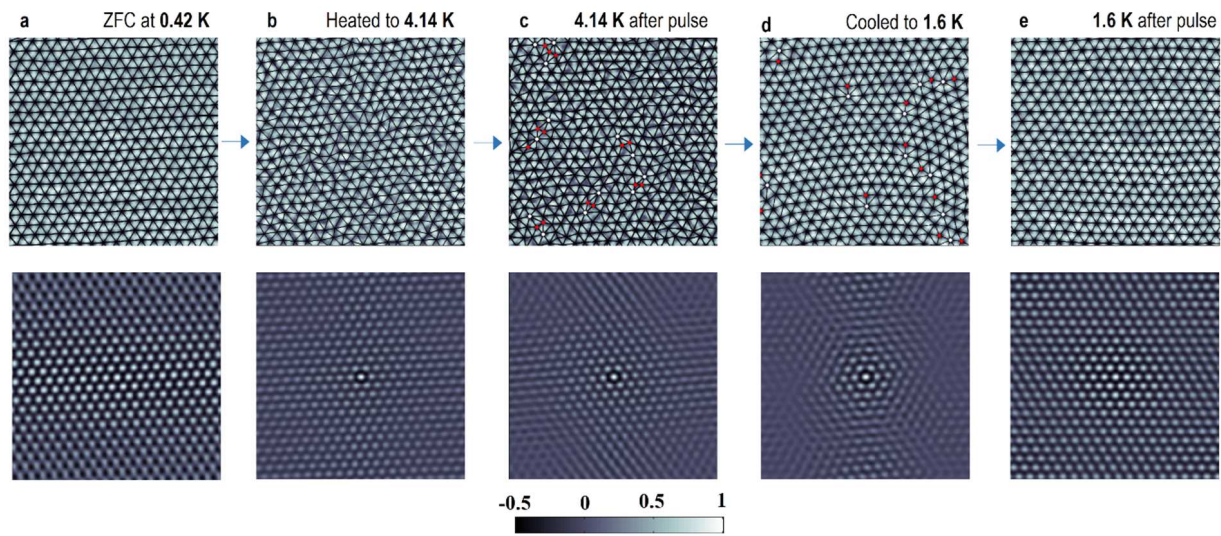

**Figure. 4.** Hysteresis of the VL across the OS-OG boundary. Conductance maps (upper panel) and the corresponding autocorrelation function (lower panel) showing (a) the ZFC VL created at 0.42 K in a field of 8 kOe, (b) the VL after heating the crystal to 4.14 K keeping the field constant, (c) after applying a magnetic-field pulse of 300 Oe at the same temperature, (d) the VL at 1.6 K after the crystal is heated to 4.35 K and cooled to 1.6 K, and (e) VL at 1.6 K after applying a magnetic-field pulse of 300 Oe. In the upper panels, Delaunay triangulation of the VL is shown with black lines and sites with fivefold and sevenfold coordination are shown with red and white dots, respectively. The color scale of the autocorrelation functions is shown in the bottom.

We then come to OG-VG phase boundary. We prepare the ZFC state at 24 KOe, 420 mK. As the system is in OG phase, it contains dislocations (Fig. 5(a)). We then warm the state up to 2.2 K thus crossing the phase boundary. Here, the number of dislocations greatly increases (Fig. 5(b)). In addition to dislocations composed of nearest-neighbor pairs of fivefold and sevenfold coordinated vortices, we also observe dislocations composed of nearest-neighbor pairs with fourfold and eightfold coordinations, an eightfold coordinated site with two adjacent fivefold coordinated sites, and a fourfold with two adjacent sevenfold coordinated sites. In addition, we also observe a small number of disclinations in the field of view. To determine the nature of this state, we examine the 2D Fourier transform (FT) of the VL image. The FT of the image shows six diffuse spots showing that the orientational order is present in the VL. This is not unexpected since a small number of disclinations does not necessarily destroy the long-range orientational order. This state is thus a superheated OG state. But upon application of 300 Oe field pulse, large number of disclinations proliferate the system (Fig.5(c)). The FT becomes an isotropic ring indicating the state is amorphous VG. The system is then cooled to 1.5 K. The FT continues to remain isotropic ring corresponding to a supercooled VG (Fig. 5(d)). Upon application of 300 Oe field pulse, all the disclinations disappear (Fig. 5(e)). FT recovers the clear sixfold pattern indicating equilibrium OG state. For the superheated OG state at 2.2 K and the equilibrium OG state at 1.5 K, the orientational correlation function $G_6(r)$ tends towards a constant value for large $r$ (Fig.5(f)) showing long range orientational order. On the other hand for the supercooled VG state at 1.5 K and the equilibrium VG state at 2.2 K, $G_6(r)$ tends towards zero for large $r$, characteristic of an isotropic amorphous state.

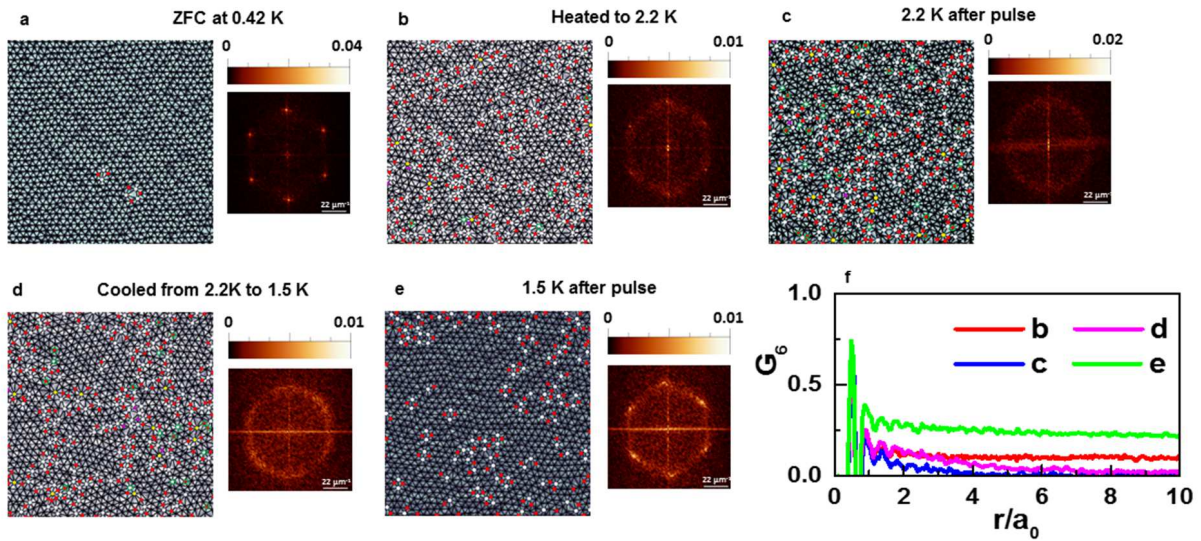

**Figure 5**. Hysteresis of the VL across the OG-VG boundary. Conductance map showing (a) the ZFC VL created at 0.42 K in a field of 24 kOe, (b) the VL after heating the crystal to 2.2 K keeping the field constant, (c) after applying a magnetic-field pulse of 300 Oe at the same temperature, (d) the VL after the crystal is subsequently cooled to 1.5 K, and (e) the VL at 1.5 K after applying a magnetic field pulse of 300 Oe. The right-hand panels next to each VL image show the 2D Fourier transform of the image; the color bars are in arbitrary units. Delaunay triangulation of the VL is shown with black lines, sites with fivefold and sevenfold coordination are shown with red and white dots, respectively, and sites with fourfold and eightfold coordination are shown with purple and yellow dots, respectively. The disclinations are circled in green. (f) Variation of $G_6$ as a function of $r/a_0$ for the VL shown in panels (b)–(e).

In this context it is important to note that for a glassy system, due to random pinning the VL might not be able to relax to its equilibrium configuration with change in temperature even if we do not cross any phase boundary. We have also observed this kind of metastable states in our experiments. So, the presence of thermal hysteresis alone does not necessarily imply a phase transition. However, these kind of metastable states vary from the corresponding equilibrium state only in the number of dislocations (and in the asymptotic value of $G_6(r)$). On the other hand, the superheated/supercooled states are distinct from the corresponding equilibrium states both in the nature of topological defects and consequently in their symmetry properties.

When the OS-OG and OG-VG phase boundaries were crossed by isothermal field ramping at 350 mK, there was no evidence of superheating/supercooling though a significant hysteresis was observed between the field ramp up and ramp down branch. Probably this is due to field

# Synopsis

ramping changing the density of vortices which involves large scale movement of vortices and thereby providing the activation energy to drive the VL into its equilibrium state. In contrast temperature sweeping does not significantly perturb the vortex lattice owing to the low operating temperatures making these metastable states observable.

## V. Orientational coupling between vortex lattice and atomic lattice:

While establishing vortex phase diagram of a homogeneous Type-II superconductor the interactions considered are vortex-vortex interaction, which stabilizes the VL by making it ordered and interaction of vortices with random impurity potential which tend to destroy the order. In single crystals of conventional superconductors, apart from one case[15], most theories regarding vortex phase diagram[16] consider the VL to be decoupled from the crystal lattice (CL) except for the random pinning potential created by defects. The VL can also get coupled to the underlying substrate. There have been experiments done[17] where it was shown that the VL orients itself in particular with respect to pinning potential in artificially engineered periodic pinning when the lattice constant is commensurate with the pinning potential. This gives rise to matching effects. In cubic and tetragonal systems, it has been theoretically[18] and experimentally[19] shown that non-local corrections to the vortex–vortex interaction can carry the imprint of crystal symmetry. Recent neutron diffraction experiments[20] on Nb single crystal also show that the structure of the VL varies on altering the symmetry of the crystalline axis along which the magnetic field is applied. So the influence of the symmetry of the CL on the VL and its subsequent effect on the ODT needs to be carefully looked into.

Here, we investigate the coupling between the VL and CL by simultaneously imaging them using STS. The experiments were done on the sample having $T_c$ = 5.88 K. Before STM measurements, the sample was cleaved in-situ in a vacuum of $10^{-7}$ mbar. 2H-NbSe$_2$ has layered hexagonal structure. In its unit cell, there are 2 sandwiches of hexagonal Se-Nb-Se layers. So during cleaving, it cleaves between the weakly coupled neighbouring Se-layers exposing Se-terminated surface. Near the centre of the crystal, we obtained an area of 1.5 μm × 1.5 μm with surface height variation < 2Å. Atomic resolution images were taken at different points within the area to determine the orientation of the crystalline lattice vectors. The Fourier transform of the CL revealed 6 clear spots indicating the hexagonal symmetry of the CL. It also features 6 diffuse spots appearing at $1/3^{rd}$ of the K-value of crystalline lattice. They indicate the 3×3 charge density wave (CDW) structure present in the system which gets diffuse due to the presence of impurity Co-atoms.

We now focus on the VL created at 2.5 KOe in ZFC protocol. In figure 6 such representative images of 2.5 KOe ZFC states created at 350 mK are shown. The figures are obtained over 3 different areas of size 1.5 µm × 1.5 µm. We observe that orientation of these ZFC images are different from each other and to that of the CL. Moreover the ZFC state in figure 6(c) shows domain structure with two domains being separated by a line of dislocations, i.e. nearest neighbour pair of 5-fold and 7-fold coordinated vortices. Applying a dc magnetic field pulse at 350 mK does not alter the orientation. But if we go to a temperature higher than 1.5 K, ramp the field up to 2.8 KOe and come back, we observe in the image that in all the 3 cases VL has reoriented itself in the direction of the CL (figure 6(d),(e),(f)) and the domain structure has broken. To observe this rotation of VL at 350 mK, we need to ramp the field up to 7.5 KOe. We observe that the VL orientation gradually rotates with increasing field and at 7.5 kOe the VL is completely oriented along the CL. Upon decreasing the field, the VL maintains its orientation and remains oriented along the CL. We also observed that the orientation of the VL does not change when the crystal is heated up to 4.5 K without applying any magnetic field perturbation. Therefore, the domain structure in the ZFC VL corresponds to a metastable state, where different parts of the VL get locked in different orientations.

To see the signature of this orientational ordering on the bulk pinning properties of the VL we performed ac susceptibility measurements on the same crystal (figure 6(g)) using three protocols: In the first two protocols, the VL is prepared in the FC and ZFC state (at 2.5 kOe) and the real part of susceptibility ($\chi'$) is measured while increasing the temperature; in the third protocol ZFC state of the vortex lattice is prepared at 1.6 K and the $\chi'$-T is measured while a magnetic field pulse of 300 Oe is applied at regular intervals of 100 mK while warming up. As expected the disordered FC state has a stronger diamagnetic shielding response than the ZFC state indicating stronger bulk pinning. For the pulsed-ZFC state, $\chi'$-T gradually diverges from the ZFC warmed up state and shows a weaker diamagnetic shielding response and exhibits a more pronounced dip at the peak effect. Both these show that the pulsed-ZFC state is more ordered than the ZFC warmed up state, consistent with the annihilation of domain walls with magnetic field pulse.

Synopsis

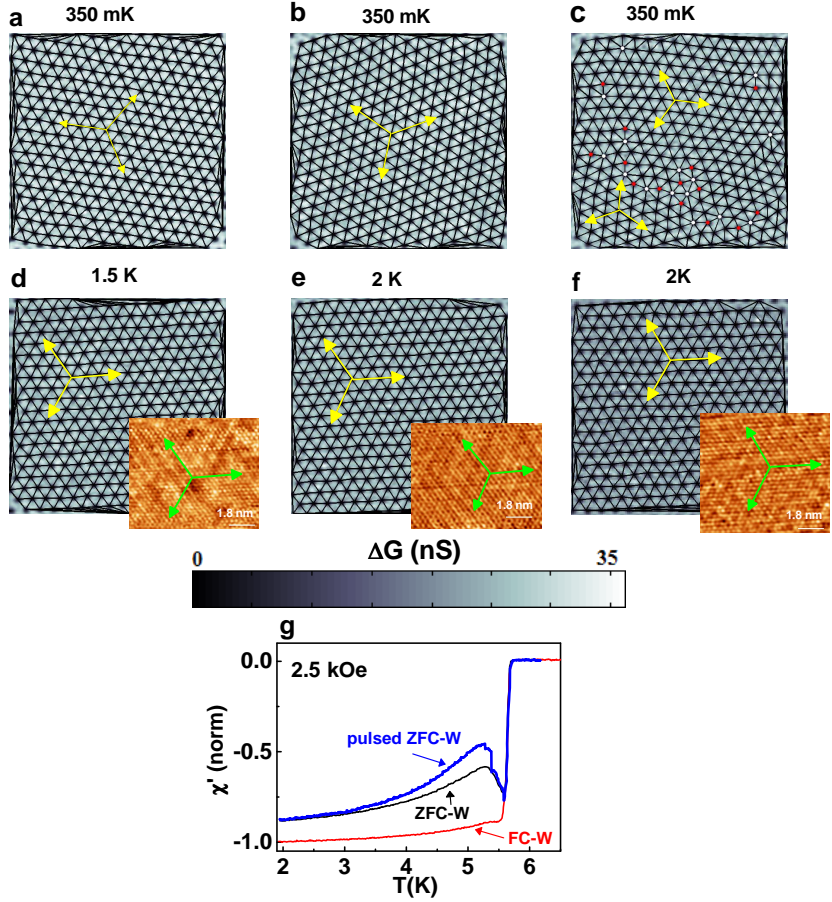

**Figure 6.** (a)-(c) Differential conductance maps showing the ZFC VL images (1.5 µm × 1.5 µm) recorded at 350 mK, 2.5 kOe at three different places on the crystal surface. (d)-(f) VL images at the same places as (a)-(c) respectively after heating the crystal to 1.5 K (for (d)) or 2 K ((e) and (f)) and applying a magnetic pulse of 300 Oe. Solid lines joining the vortices show the Delaunay triangulation of the VL and sites with 5-fold and 7-fold coordination are shown as red and white dots respectively. The direction of the basis vectors of the VL are shown by yellow arrows. In figure (c) a line of dislocations separate the VL into two domains with different orientations. The right inset in (d)-(f) show the orientation of the lattice, imaged within the area where the VL is imaged. (g) Susceptibility ($\chi'$) as a function of temperature (T) measured at 2.5 kOe while warming up the sample from the lowest temperature. The three curves correspond to $\chi'$-T measured after field cooling the sample (FC-W), after zero field cooling the sample (ZFC-W) and zero field cooling the sample and then applying a magnetic pulse of 300 Oe at temperature intervals of 0.1 K while warming up (pulsed ZFC-W). The y-axis is normalized to the FC-W $\chi'$ at 1.9 K.

We next explore the effect of this orientational coupling on the ODT of VL. As the field is increased further the VL remains topologically ordered up to 24 kOe. At 25 kOe dislocations proliferate in the VL, in the form of neighboring sites with five-fold and seven-fold coordination. At 27 kOe, we observe that the disclinations proliferate into the lattice. However, the corresponding FT show a six-fold symmetry all through the sequence of disordering of the VL. Comparing the orientation of the principal reciprocal lattice vectors of the VL to that of the FT of CL, we observe that the VL is always oriented along the CL even at 28 KOe where there are a lot of disclinations within our field of view. So, in this case the amorphous VG state is not realised even though disclinations have proliferated the VL. This is in contrast to our earlier 2-step melting observation (Fig. 3). The difference could be attributed to weaker defect pinning in the present crystal, which enhances the effect of orientational coupling in maintaining the orientation of the VL along the crystal lattice.

This orientational coupling cannot be explained by conventional pinning which requires modulation of superconducting order parameter over a length scale of the size of the vortex core, which is an order of magnitude larger than interatomic separation and the CDW modulation. One likely origin can be due to anisotropic vortex cores whose orientation is locked along a specific direction of the CL. In unconventional superconductors (e.g. $(La,Sr)CuO_4$, $CeCoIn_5$) such anisotropic cores arises from the symmetry of the gap function, which has nodes along specific directions[21]. However even in an s-wave superconductor such as $YNi_2B_2C$ vortex cores with four-fold anisotropy has been observed[22], and has been attributed to the anisotropic superconducting energy gap resulting from Fermi surface anisotropy.

To explore this, we performed high resolution spectroscopic imaging of a single vortex core at 350 mK. We first created a ZFC vortex lattice at 350 mK (figure 7(a)) in a field of 700 Oe, for which the inter vortex separation (177 nm) is much larger than the coherence length. It minimizes the influence of neighboring vortices. As expected at this low field the VL is not aligned with the CL. We then chose a square area enclosing a single vortex and measured the full G(V)-V curve from 3 mV to −3 mV at every point on a 64 × 64 grid. We then plot the normalized conductance G(V)/G(3 mV) at 3 bias voltages (figure 7(b),(c),(d)). The normalized conductance images reveal a hexagonal star shape pattern consistent with previous measurements in undoped $2H-NbSe_2$ single crystals[23]. Atomic resolution images captured within the same area (figure 7(e)) shows that the arms of the star shape in oriented along the principal directions of the CL, but not along any of the principal directions of the VL. It rules out the possibility that the shape arises from the interaction of supercurrents surrounding each vortex core. The reduced contrast in our images compared to undoped $2H-NbSe_2$ is possibly due to the presence of Co impurities which smears the gap anisotropy through intra and inter-band scattering.

Synopsis

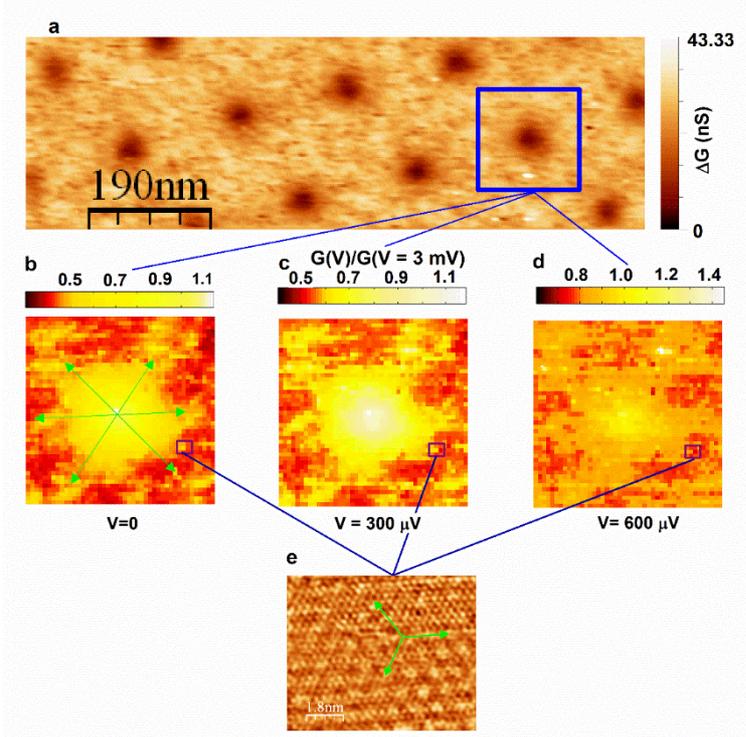

Figure 7. (a) Differential conductance map showing the ZFC VL at 700 Oe and 350 mK. (b) High resolution image of the single vortex (114 nm ×114 nm) highlighted in the blue box in panel (a) obtained from the normalized conductance maps ( G(V)/G(V = 3mV) ) at 3 different bias voltages. The vortex core shows a diffuse star shaped patters; the green arrows point towards the arms of the star shape from the center of the vortex core. (c) Atomic resolution topographic image of the CL imaged within the box shown in (b); the green arrows show the principal directions of the crystal lattice.

As the magnetic field is increased the vortices come closer and feel the star shape of neighboring vortices. Since the star shape has specific orientation with respect to the CL, the VL also orients in a specific direction with respect to the CL. When the field is reduced, the vortices no longer feel the shape of neighboring vortices, but the lattice retains its orientation since there is no force to rotate it back.

## VI. Conclusion and future directions:

In our experiment, we provide structural evidence that across the peak effect, the VL in a weakly pinned 3D superconductor disorders through two first order thermodynamic phase transitions. The sequence of disordering of VL observed in our experiment is reminiscent of the two-step BKTHNY transition observed in 2-D systems, where a hexatic fluid exists as an intermediate state between the solid and the isotropic liquid. However, The BKTHNY mechanism of two-step melting is not applicable in this system since it requires logarithmic interaction between vortices which is not realized in a 3-D VL.

In our case, the two-step disordering is essentially induced by the presence of quenched random disorder in the crystalline lattice, which provides random pinning sites for the vortices. This alternative viewpoint to the thermal route to melting is the disorder-induced ODT originally proposed by Vinokur et al.[24]. Here it was speculated that in the presence of weak pinning the transition can be driven by point disorder generating positional entropy rather than temperature. In this scenario topological defects proliferate in the VL through the local tilt of vortices caused by point disorder, creating an "entangled solid" of vortex lines. Here, in contrast to conventional thermal melting, the positional entropy generates instability in the ordered VL driving it into a disordered state, even when thermal excitation alone is not sufficient to induce a phase transition. Further evidence for this is obtained by comparing χ' as a function of reduced magnetic field $h = H/H_{c2}$ using samples with different pinning strengths. We observe as the pinning gets weaker the difference $h = h_p - h_p^{on}$ decreases thereby shrinking the phase space over which the OG state is observed. We speculate that in the limit of infinitesimal small pinning $h \to 0$, thereby merging the two transitions into a single first-order transition possibly very close to $H_{c2}$.

In future, it could be worthwhile to look for signatures of these transitions using thermodynamic measurements (e.g. specific heat), although such signatures are likely to be very weak. Theoretically, it would be interesting to investigate the role of disclinations in the VL, which has not been explored in detail so far. It would also be interesting to explore to what extent these concepts can be extended to other systems, such as colloids, charge-density waves and magnetic arrays where a random pinning potential is almost always inevitably present. We also hope that future theoretical studies will quantitatively explore the magnitude CL-VL coupling energy scale with respect to vortex–vortex and pinning energies and its effect on the vortex phase diagram of Type-II superconductors.

Synopsis

# Chapter 1

# Introduction

Order to disorder transition (ODT) lies at the heart of condensed matter physics. It is a well-known phenomenon that identical particles having some sort of interaction among themselves are bound to form an ordered periodic structure below a certain temperature. Formation of crystalline solids below melting point is the most familiar example. Furthermore there are numerous examples of self-assembled structures e.g. molecular self-assembly over substrates[1,2,3], formation of colloidal crystal on solvent[4], anodisation of alumina template[5] etc. All these periodic structures are characterised by their long-range order. However above a certain temperature all these structures undergo a transformation upon which they lose the order and becomes isotropic liquid. This phenomenon is known as order-disorder transition (ODT)[6,7,8].

Presence of defects play an important role in condensed matter systems[9]. Dislocations determine the strength of crystalline materials, vacancies and interstitials affect diffusion, vortex motion controls resistance of a superconductor etc. Theoretically, dynamics of defect mediated phase transitions were treated by renormalization-group and dynamical scaling theories especially in two-dimensions[10]. These investigations showed the presence of a fourth 'hexatic' phase of matter apart from the basic solid, liquid and gaseous phases. We cannot distinguish between the liquid and gaseous phases above critical point. But, the solid and liquid phases are distinguishable by the orientational symmetry. Regular arrangement of atoms in solids reflect broken translational invariance implying long-range orientational order which is absent in liquids. Now, it is possible to imagine an intermediate state with short range translational order characteristic of a liquid coexisting with broken rotational symmetry. Such an intermediate phase was theoretically predicted for melting of 2 dimensional (2D) crystals independently by Berezensky, Koterlitz, Thouless, Halperin, Nelson and Young[11,12]. In their model, the 2D crystalline solid could melt into isotropic liquid in 2 steps with each step being a continuous phase transition going through intermediate hexatic phase. This 2 step continuous transition is known as BKT (BKTHNY) transition. The hexatic phase has been observed in various systems such as free standing liquid crystal films, colloidal crystals etc.

The hexatic phase is characterised by the presence of dislocations (Figure 1.1 a) which breaks the positional order but does not break the orientational order. For an underlying triangular

Chapter 1

lattice a dislocation is a nearest neighbour pair of five-fold or seven-fold coordinated particles. Whereas the liquid phase has disclinations, which for a triangular lattice is an isolated five-fold or seven-fold coordinated particle (Figure 1.1 (b), (c)). Presence of disclinations breaks the orientational order. So in BKT transition, first dislocations proliferate the system breaking the positional order and driving (quasi-) periodic solid → hexatic phase transition. Finally the dislocations dissociate into isolated disclinations driving hexatic → isotropic liquid phase transition.

For most periodic systems, the nature of ODT has remained controversial. The presence of disorder in a real system complicates the scenario. It can drive the system into disordered state even when thermal fluctuations are not enough to actuate the ODT. Such disorder driven transition has been observed in soft matter system like 2D colloidal crystal[13].

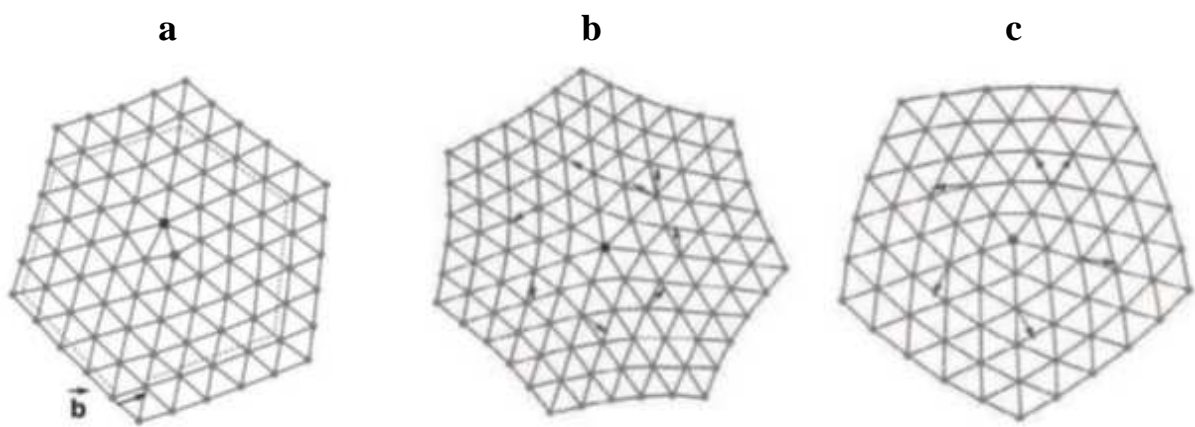

**Figure 1.1** (a) A dislocation in a triangular lattice having five- and a seven-fold coordinated particles one lattice spacing apart. The dashed line is a Burgers circuit which would close in a perfect lattice but fails to close by a Burger's vector **b** perpendicular to the bond joining five- and a seven-fold coordinated particles. (b), (c) ± $\pi$ /3 disclinations (isolated seven-fold, five-fold coordinated particles) in an underlying triangular lattice. Image courtesy Ref. [1].

## 1.1 Thermodynamics of order disorder transition

To understand ODT from thermodynamic point of view, let us consider a closed system of constant volume connected to a thermal bath having temperature $T$ with which it can exchange heat. The Helmholtz free energy of the system is given by

$$F = E - TS \quad (1.1)$$

Where $E$ is the internal energy of the system. $S$ is the entropy which quantifies the disorder. Now in equilibrium, the system will try to minimize $F$. At low $T$, the contribution of entropic term is negligible and the minima of $F$ is determined by the minina of $E$. Internal energy $E$ is minimum when the constituent units (molecules/atoms) of the system are in a periodic order so that the interaction energy among them is minimised. So, at low temperature the system will prefer to from an ordered periodic structure. Whereas at high $T$, the entropic term is more dominant. So, the system prefers to increase S, so it becomes more disordered. So, temperature of the system $T$ drives the ODT which occurs at some characteristic temperature $T_c$ called the critical temperature.

For a closed, expansible and diathermic system, the equilibrium is determined by the minima in Gibbs free energy

$$G = E + PV - TS \quad (1.2)$$

Across the phase transition, the free energy is continuous. If the first derivatives of the free energy, $V = (\frac{\partial G}{\partial P})_T$, $S = -(\frac{\partial G}{\partial T})_P$ has a discontinuity then the phase transition is said to be of 'First order'. So, the first order phase transition at constant ($T$, $P$) is characterised by change in molar volume and having latent heat of transition. If the first derivatives of free energy are continuous across the transition but the second derivatives have discontinuity then the phase transition is said to be continuous or 'Second order'.

## 1.2 Ginzburg-Landau theory of phase transition

The phenomenological theory for phase transition is given by Ginzberg and Landau[14]. It is based on the assumption that the free energy near the critical temperature can be expressed as a Taylor's series expansion in terms of 'order parameter' of the system.

$$F = F_0(T) + \frac{\alpha(T)}{2}\psi^2 + \frac{\beta(T)}{4}\psi^4 + \frac{\gamma(T)}{6}\psi^6 \quad (1.3)$$

Where $\alpha(T) = \alpha_0(T - T_c)$, $\alpha_0 > 0$ and $\beta > 0$ is independent of the temperature.

The stable equilibrium state is given by $(\frac{\partial F}{\partial \psi})_T = 0$ and $(\frac{\partial^2 F}{\partial \psi^2})_T > 0$

At $T \geq T_c$ the stable equilibrium occurs at $\psi = 0$.

At $T < T_c$ the stable equilibrium occurs at $\psi = \pm\sqrt{(\frac{\alpha_0(T-T_c)}{\beta})}$.

Chapter 1

So, at $T=T_c$, $\psi$ is continuous. As a consequence, the first order derivatives of the free energy are also continuous. So the phase transition described above is continuous or 2$^{nd}$ order. This case is pictorially described in Fig. (1.2(a), (b)).

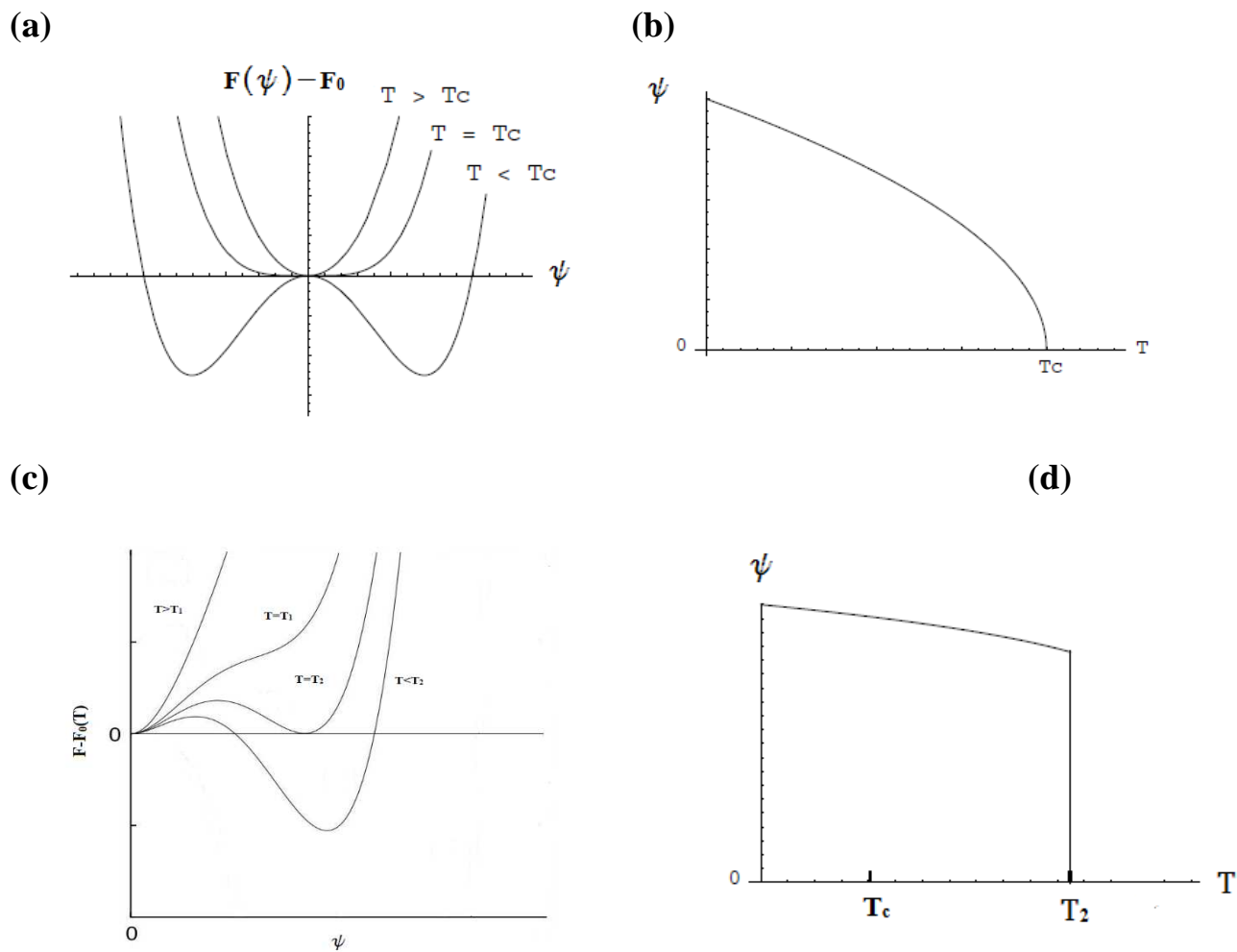

**Figure 1.2** (a) Free energy vs. order parameter at various temperatures. (b) Order parameter vs. temperature showing continuous transition at T$_c$ (c) Free energy vs. order parameter at various temperatures. (d) Order parameter vs. temperature showing a sudden jump at T$_2$ characteristic of a first order phase transition.

Now if the free energy is not symmetric with respect to the order parameter, then the Taylor's expansion can have odd power term.

$$F = F_0(T) + \frac{\alpha(T)}{2}\psi^2 + \frac{\beta(T)}{4}\psi^4 + \frac{\gamma(T)}{3}\psi^3 \quad (1.4)$$

For $T>T_c$, the secondary minima is only possible when $\gamma(T) < 0$

The non-trivial solutions $\psi = \frac{-\gamma \pm \sqrt{\gamma^2 - 4\alpha\beta}}{2\beta}$ are valid for $\gamma^2 \geq 4\alpha\beta \Rightarrow T \leq T_c + \frac{\gamma^2}{4\alpha_0\beta} = T_1$

For $T > T_1$ there exists only minima corresponding to $\psi = 0$. So, the disordered phase is stable. The ordered state corresponding to the non-trivial minima is metastable.

At $T = T_2$ where non-trivial minima touches $\psi$-axis is $T_2 = T_c + \frac{2\gamma^2}{9\alpha_0\beta}$. At this temperature both ordered and disordered phases coexist. Below $T_2$ disordered phase becomes unstable and the ordered state is stable. So, at $T = T_2$ we have a phase transition. The order parameter as well as the specific heat are discontinuous at the transition temperature. This transition is a first order transition. This case is pictorially described in Fig. (1.2(c), (d)).

One of the distinctive feature of a first order phase transition is the presence of superheating and supercooling. Due to the presence of an activation energy barrier between the ordered and disordered states, the system could remain at the ordered state even slightly above the phase transition temperature resulting in a metastable ordered superheated state. In the same way, the system could remain at the disordered state even slightly below the phase transition temperature resulting in a metastable disordered supercooled state. Upon external perturbation, these metastable states shows a dynamic transition to their corresponding stable state configurations.

## 1.3 Our model system: The Abrikosov vortex lattice

We have used vortex lattice in a Type-II superconductor as our model system to study the ODT.

### 1.3.1 Basics of superconductivity

Kemerlingh Onnes observed in 1911 that the electrical resistance of various metals like mercury, lead and tin suddenly become zero (for highest possible experimental resolution) within a certain temperature range of critical temperature $T_c$ characteristic of the material[15]. He named this new phase of matter as superconductor. Apart from being a perfect conductor it is also a perfect diamagnet as discovered by Meissner and Oschenfeld[16]. The complete flux repulsion by a superconducting material is reversible and is called Meissner effect.

### 1.3.2 London equations:

The phenomenological theory developed by F. London and H. London yields the perfect conductance and Meissner effect[17].

# Chapter 1

If $n_s$ is the number density of superconducting electrons and $J_s$ is the supercurrent density, then from Drude model for electrical conductivity we obtain

$$\frac{dJ_s}{dt} = \frac{n_s e^2}{m} E \quad (1.5)$$

Now from Maxwell's equation,

$$\nabla \times \mathbf{E} = -\frac{d\mathbf{B}}{dt} \Rightarrow \frac{d\mathbf{B}}{dt} = \frac{m}{\mu_0 n_s e^2} \nabla^2 \left(\frac{d\mathbf{B}}{dt}\right) \Rightarrow \nabla^2 \dot{\mathbf{B}} = \frac{\dot{\mathbf{B}}}{\lambda_L^2} \quad (1.6)$$

$$\text{Where } \lambda_L = \sqrt{\frac{m}{\mu_0 n_s e^2}} \quad (1.7)$$

So, we have $\dot{\mathbf{B}} \rightarrow 0$ inside the bulk of a perfect conductor. London argued that as the magnetic induction $\mathbf{B}$ itself is zero within a superconductor, so the equation must be recast as

$$\nabla^2 \mathbf{B} = \frac{\mathbf{B}}{\lambda_L^2} \quad (1.8)$$

So, we have the magnetic induction decaying within a superconductor having a characteristic decay length $\lambda_L$ called the London penetration depth.

Assuming a semi-infinite superconductor for x ≥ 0

If the applied magnetic field $\mathbf{B} = B_0 \hat{y}$ for x < 0

Then $\mathbf{B}(x) = B_0 e^{-\frac{x}{\lambda_L}} \hat{y}$ and $\mathbf{J}_s(x) = -\frac{B_0}{\lambda_L} \hat{z} e^{-\frac{x}{\lambda_L}}$ for x ≥ 0.

Thus the supercurrents flow in the direction parallel to the surface and perpendicular to $\mathbf{B}$ and decrease into the bulk over the same scale $\lambda_L$.

## 1.3.3 Non-local response and Pippard coherence length

Assuming the supercurrent density $\mathbf{J}_s$ and magnetic vector potential $\mathbf{A}$ is non-local, Pippard obtained a length scale over which the superfluid density changes[18]. He estimated this length scale by uncertainty principle argument. Electrons within the energy range $k_b T_c$ of Fermi energy take part in superconductivity and these electrons have a momentum range $\Delta p \approx \frac{k_b T_c}{v_f}$. This gives $\Delta x \gtrsim \frac{\hbar}{\Delta p} \approx \frac{\hbar v_f}{k_b T_c}$. The chararacteristic length scale is called coherence length and is given by

$$\xi_0 = a\frac{\hbar v_f}{k_b T_c}, \text{ where a is a numerical constant of order unity.}$$

Coherence length in presence of scattering is given by

$$\frac{1}{\xi} = \frac{1}{\xi_0} + \frac{1}{l} \quad (1.9) \text{ where l is the mean free path}$$

## 1.3.4 Ginzburg-Landau theory for superconductivity

For inhomogeneous superconductors where the superconducting properties vary spatially, microscopic theory is difficult to establish. It's easier to resort to the phenomenological mean field description by Ginzburg and Landau. It has a complex order parameter $\psi = |\psi_0|e^{i\theta}$ whose amplitude $|\psi|^2 = n_s$ (superconducting electron density).

In presence of magnetic field, free energy of ground state of the superconductor can be written as a power series in $\psi$ as

$$f = f_0 + \alpha|\psi|^2 + \frac{\beta}{2}|\psi|^4 + \frac{1}{2m^*}\left|\left(\frac{\hbar}{i}\nabla - e^*A\right)\psi\right|^2 + \frac{B^2}{2\mu_0} \quad (1.10)$$

$\alpha, \beta$ are GL parameters where $\beta > 0$ and $\alpha = \alpha_0(\frac{T}{T_c} - 1)$

Minimising f with respect to $\psi$ gives us GL equation

$$\alpha|\psi| + \beta|\psi|^2\psi + \frac{1}{2m^*}\left|\left(\frac{\hbar}{i}\nabla - e^*A\right)\right|^2 \psi = 0 \quad (1.11)$$

In the absence of external magnetic field it becomes

$$\alpha|\psi| + \beta|\psi|^2\psi - \frac{\hbar^2}{2m^*}\nabla^2\psi = 0 \quad (1.12)$$

It gives us a length scale over which the superconducting order parameter varies spatially,

$$\xi_{GL} = \sqrt{\frac{\hbar^2}{2m^*|\alpha(T)|}} \quad (1.13)$$

This GL coherence length is of the order of Pippard coherence length for a clean superconductor well below $T_c$.

London penetration depth $\lambda = \sqrt{\frac{m^*}{\mu_0 e^2 n_s}} = \sqrt{\frac{m^*}{\mu_0 e^2 (-\frac{\alpha}{\beta})}}$ (As for $T < T_c$, $n_s = |\psi|^2 = -\frac{\alpha}{\beta}$)

Chapter 1

The ratio $\kappa = \frac{\lambda}{\xi}$ is called the GL parameter and it is independent of temperature within GL theory.

## 1.3.5 Type I and Type II superconductors

Based on the ratio $\kappa = \frac{\lambda}{\xi}$, superconducting materials are classified into two broad categories.

For $\kappa < \frac{1}{\sqrt{2}}$ The superconductor shows complete flux exclusion up to some critical field $H_c$, above which it becomes normal. These are called Type I superconductors.

For $\kappa > \frac{1}{\sqrt{2}}$, up to certain field $H_{c1}$ called lower critical field the response is completely diamagnetic. Above $H_{c1}$ it is energetically favourable for the superconductor to form normal regions within which the external field penetrates. These normal regions are separated by superconducting regions between them where there is no flux penetration. The volume ratio between normal to superconducting region grows with increasing magnetic field. Above $H_{c2}$, called upper critical field, the material becomes completely normal.

To maintain the single-valued-ness of superconducting wave function, the phase change around a closed path surrounding the normal region must be integral multiple of $2\pi$. Therefore, we have,

$$2n\pi = \frac{4\pi m}{h} \oint \boldsymbol{J_s}. d\boldsymbol{l} + \frac{4\pi e}{h} \oint \boldsymbol{A}. d\boldsymbol{l} \Rightarrow \Phi = \oint \boldsymbol{A}. d\boldsymbol{l} = n\left(\frac{h}{2e}\right) = n\Phi_0 \quad (1.14)$$

$\Phi_0 = \frac{h}{2e} = 2.07 \times 10^{-15}$ Wb is called magnetic flux quantum.

So, Type II superconductors for applied field H where $H_{c1}<H<H_{c2}$ have mixed state having both normal and superconducting regions. With normal regions having magnetic flux integer multiple of flux quantum passing through it. These regions are called vortices.

To screen the magnetic field that is threading through the sample, a screening current flows at the edge of each vortex. Depending of the direction of applied magnetic field, screening current flowing at the edge of each vortex have same sense. As a result, vortices feel mutual repulsion among themselves. So, to minimise the interaction energy the vortices tend to align periodically in the form of a triangular (hexagonal) lattice having nearest neighbour separation $a_0 = 1.075(\frac{\Phi_0}{B})^{\frac{1}{2}}$. This periodic array of flux lines/ vortices is called flux line lattice (FLL) or Abrikosov vortex lattice[19]. Figure 1.3 shows one such Abrikosov vortex lattice.

Vortex lattice (VL) in a Type II superconductor provides an extremely versatile model system to study the ODT. The density of constituent particles (here the number of vortices) and interaction among them can be varied over several orders of magnitude just by changing the external magnetic field. As a real superconducting system has disorder present in it in the form of crystalline defects, impurities etc., it provides an ideal medium to study the ODT in the presence of disorder.

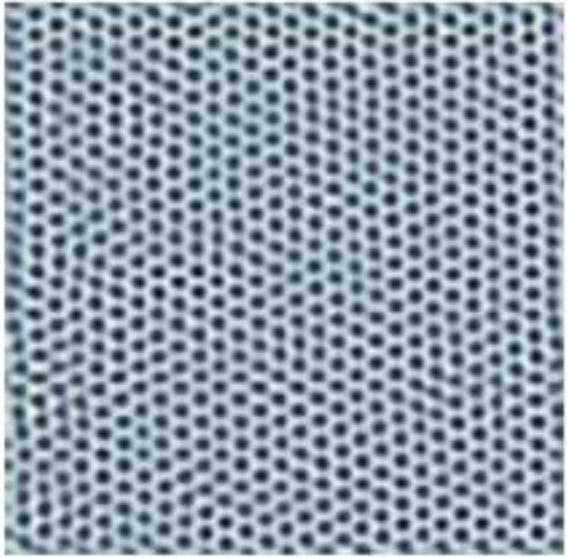

**Figure 1.3** Abrikosov vortex lattice imaged on NbSe$_2$, a typical Type II superconductor. Applied magnetic field: 2 Tesla.

## 1.3.6 BCS theory

To understand the phenomenon of superconductivity from microscopic point of view, Bardeen, Cooper and Schireffer proposed BCS theory in 1957[20]. It describes how a net attractive interaction between two electrons near the Fermi level mediated by phonons can form a bound state known as Cooper pair. These Cooper pairs subsequently form a phase-coherent macroscopic condensate giving rise the superconductivity.

The origin of attractive interaction between electrons can be understood according to Fröhlich (1950) as follows[21]. The first electron with negative charge attracts the positively charged ions and thereby polarizing the medium. The second electrons is then attracted by the excess positive charge created due to lattice distortion giving rise to effective attractive interaction between the two electrons. If this attractive interaction is strong enough compared to the repulsive screened Coulomb interaction a net attractive interaction between the two electrons results.

# Chapter 1

Cooper showed that Fermi sea is unstable against the formation of at least one bound pair[22]. This can be understood by adding two electrons to a Fermi sea at T=0. Assuming they have equal and opposite momenta **k** and –**k** (for the lowest energy state to have zero net angular momentum), we can construct the 2-electron anisymmetric wavefunction as

$$\psi_0 = \sum_{k>k_F} g_k \cos \boldsymbol{k}.(\boldsymbol{r_1} - \boldsymbol{r_2})](\alpha_1\beta_2 - \beta_1\alpha_2) \quad (1.15)$$

Where $\alpha, \beta$ represents the spin up and spin down states respectively. 1, 2 refers to the particles. For the orbital part, cosinusoidal term is chosen as it gives larger probability amplitudes for the electrons to be near each other as we anticipate an attractive interaction.

Inserting this wavefunction into the Schrödinger equation and replacing the summation by a continuous integration we finally end up with the following

$$E = 2E_F - 2\hbar\omega_D e^{-\frac{2}{N(0)V}} \quad (1.16)$$

Where we have assumed $V_{kk'} = \frac{1}{\Omega}\int V(\boldsymbol{r})e^{i(\boldsymbol{k'}-\boldsymbol{k}).\boldsymbol{r}}d\boldsymbol{r} = -V$ ($\Omega$ is the normalization volume) for $\boldsymbol{k}$ states out to a cut off energy $\hbar\omega_D$ away from $E_F$. $N(0)$ is the density of states at the Fermi level for one spin orientation.

We thus have a bound state between electrons having opposite momenta and spin regardless of how small the attractive potential $V$ is.

We have the BCS ground state as

$$|\psi_{BCS}\rangle = \prod_k (u_k + v_k c_{k\uparrow}^\dagger c_{-k\downarrow}^\dagger)|F\rangle \quad (1.17)$$

Where $|v_k|^2, |u_k|^2$ is the probability that the pair $(\boldsymbol{k}\uparrow, -\boldsymbol{k}\downarrow)$ is occupied and unoccupied respectively. $|F\rangle$ represents filled Fermi sea up to $k_F$. Coefficients $u_k$ and $v_k$ are chosen so as to minimize the energy $E = \langle\psi_{BCS}|\mathcal{H}|\psi_{BCS}\rangle$ using the the reduced BCS Hamiltonian,

$$\mathcal{H} = \sum_{k,\sigma} \epsilon_k n_{k\sigma} + \sum_{k,l} V_{kl} c_{k\uparrow}^\dagger c_{-k\downarrow}^\dagger c_{-l\downarrow}^\dagger c_{l\uparrow}^\dagger \quad (1.18)$$

Where the scattering matrix

$$V_{kl} = \begin{cases} -V \ for \ |\xi_k|, |\xi_l| < \hbar\omega_D \\ 0 \ otherwise \end{cases}$$

$\xi_k = \varepsilon_k - \mu$ is the single particle energy relative to Fermi energy

We obtain the following results

### 1.3.6.1 The gap function

$\Delta(0) \approx 2\hbar\omega_D e^{-\frac{1}{N(0)V}}$ (1.20) given $N(0)V \ll 1$ (weak coupling limit)

At non zero temperature, $\Delta$ can be obtained by solving the following equation numerically

$$\frac{1}{N(0)V} = \int_0^{\hbar\omega_D} \frac{\tanh\frac{\beta}{2}(\xi^2+\Delta^2)^{\frac{1}{2}}}{(\xi^2+\Delta^2)^{\frac{1}{2}}} d\xi \quad (1.21)$$

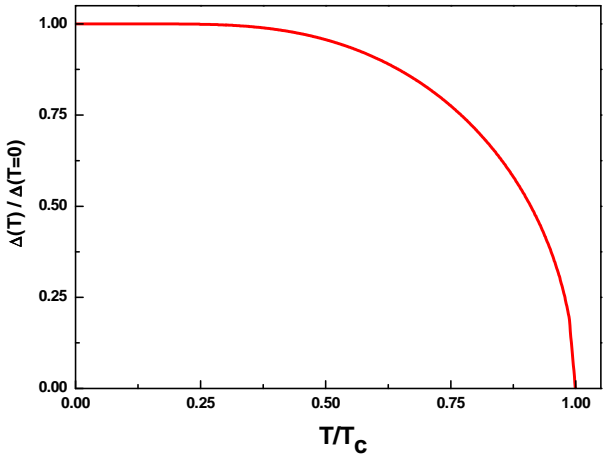

**Figure 1.4** Temperature dependence of energy gap within BCS theory

### 1.3.6.2 Condensation energy

$$E_s(0) - E_n(0) = -\frac{1}{2}N(0)\Delta^2(0) \quad (1.22)$$

$E_s(0)$ and $E_n(0)$ represents free energy densities in superconducting and normal states respectively.

### 1.3.6.3 Single particle density of states

$$\frac{N_S(E)}{N(0)} = \frac{d\xi}{dE} = \begin{cases} \frac{E}{\sqrt{E^2-\Delta^2}} & \text{for } E > \Delta \\ 0 & \text{for } E < \Delta \end{cases} \quad (1.23)$$

So, we have a divergence of density of states at $E = \pm\Delta$ (Fig 1.5). These are called coherence peaks.

For a disordered superconductor, the BCS states are not exact eigenstates as they are broadened and have an associated lifetime. Then the modified density of states can be written as

Chapter 1

$$\frac{N_S(E)}{N(0)} = \begin{cases} Re(\frac{E+i\Gamma}{\sqrt{(E+i\Gamma)^2-\Delta^2}}) \; for \; E > \Delta \\ 0 \qquad\qquad\qquad\quad for \; E < \Delta \end{cases} \quad (1.24)$$

Re corresponds to the real part. $\Gamma$ is called the Dynes' parameter[23] associated with broadening due to finite lifetime of the superconducting quasiparticles.

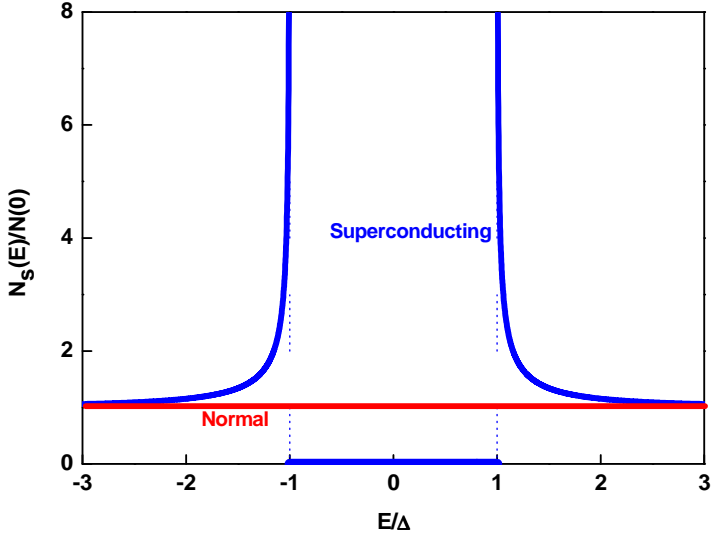

**Figure 1.5** Density of state of superconductor with respect to normal metal.

## 1.3.7 Giever tunnelling and measurement of energy gap

As we can see from the BCS theory, an energy gap $\Delta$ opens up in single particle density of states in the superconducting phase. In 1960, Giaever introduced a method to measure this energy gap using the principle of quantum mechanical tunnelling across a superconductor-insulator-normal metal (SIN) junction[24]. The tunnelling current across such a junction is given by

$$I_{ns} = \frac{G_{nn}}{e} \int_{-\infty}^{\infty} \frac{N_S(E)}{N_n(0)} [f(E) - f(E+eV)] \, dE \quad (1.25)$$

So, the differential tunnelling conductance

$$G_{ns}(E) = \frac{dI_{ns}}{dV} = \frac{G_{nn}}{e} \int_{-\infty}^{\infty} \frac{N_S(E)}{N_n(0)} [-\frac{\partial f(E+eV)}{\partial (eV)}] \, dE \quad (1.26)$$

As T→0, Fermi function mimics a step function and its derivative delta-function yielding

$$G_{ns}|_{T=0} = G_{nn} \frac{N_S(e|V|)}{N_n(0)} \quad (1.27)$$

So, the differential tunnelling conductance mimics density of state of the superconductor. A typical tunnelling conductance spectra along with its BCS density of states fit is shown in Figure 1.6.

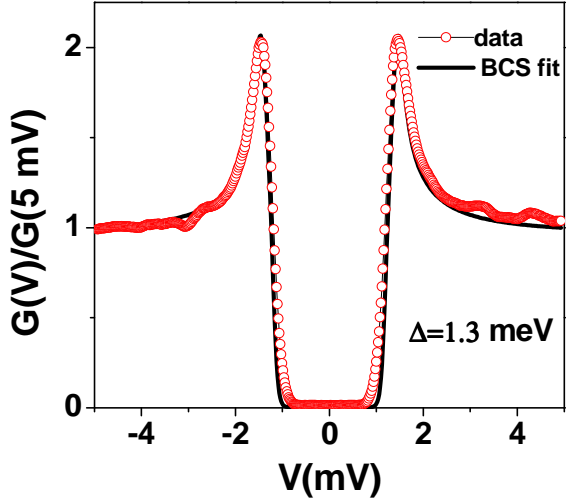

**Figure 1.6** Tunneling spectroscopy on Pb single crystal acquired with Pt-Ir tip at 500 mK. Black line is the BCS fit with the energy gap Δ=1.3 meV

## 1.4 Pinning and critical current in Type II superconductor

Now, if we pass a finite current density J through a Type II superconductor in its mixed state then the Lorentz force density

$$\boldsymbol{F} = \boldsymbol{J} \times \frac{B}{c} \quad (1.28)$$

Due to this force the flux lines will tend to move transverse to the current, say, with velocity **v**. This will induce an electric field of magnitude

$$\boldsymbol{E} = \boldsymbol{B} \times \frac{v}{c} \quad (1.29)$$

This acts parallel to *J*, so a finite power *E.J* is dissipated resulting in non-zero resistance. So, for an ideal Type II superconductor in mixed state, a finite current no matter how small, will drive it to the normal state.

This problem is avoided due to the presence of point disorder (impurities, vacancies etc.) in real systems. These point disorders act as pinning centres for the vortices. To create a vortex, superconductivity is destroyed locally. By passing through pinning centre the energy gained is $\sim \pi \xi^2 L_i (\frac{1}{2} \mu_0 H_c^2)$. Here $L_i$ is the linear extension of the pinning centre. So, the vortices would

Chapter 1

prefer to pass through pinning centres. To move the vortices out of this potential trap, a finite current density is required. It is called critical current density ($J_c$).

## 1.5 Larkin-Ovchinnikov theory of collective pinning

A perfectly periodic and rigid flux-line lattice (FLL) cannot be pinned by any random collection of pinning sites because for any position of the FLL relative to the pinning landscape, there will be an equal number of random pinning sites exerting forces adding to the Lorentz force as opposing it, which sums up to zero net force. It is the elastic response of the FLL to these pinning forces which results in a finite volume pinning force. For most real systems, the pins are either too strong, too extended or too correlated for calculation by statistical theory. In the limit of weak pinning and assuming a random collection of pins, the concept of collective pinning was introduced by Larkin and Ovchinnikov[25]. The interaction between FLL has a range of $\lambda_L$ which is large for high $\kappa$ materials. The concentration of pinning centres $n_p$ is often very large for a real system, resulting $n_p^{-1/3} \ll \lambda_L$. So, the collective treatment of pinning forces is appropriate.

In the collective pinning model, it is assumed that there is no long range order present in the FLL. Instead there is a finite volume $V_c$ within which short range order exists and the FLL is periodic. When a current below the critical value is passed, each of the volumes $V_c$ is displaced independently by the Lorentz force for a distance less than ~$\xi$, so that the pinning force compensates the Lorentz force. The random pinning forces would add in the manner of random walk resulting in a net force ~ $\sqrt{N} \propto \sqrt{V_c}$. Where $N = nV_c$ is the total number of pinning centres in volume $V_c$, n is no. density of pinning centres. For applied current density $J$, the Lorentz force is $JBV_c$. At critical current density, Lorentz force equals maximum pinning force. If f is pinning force per unit volume then we have

$$J_c B V_c = f n^{\frac{1}{2}} V_c^{\frac{1}{2}} \Rightarrow J_c = \frac{1}{B} \frac{f n^{\frac{1}{2}}}{V_c^{\frac{1}{2}}} \quad (1.30)$$

So, we can see that for perfectly ordered vortex lattice $V_c \to \infty$ hence $J_c \to 0$. For disordered VL, $V_c$ is finite and hence $J_c$ is non zero.

If $L_c$ and $R_c$ are correlation lengths along and transverse to the field directions then we have $V_c = R_c^2 L_c$. Since the lattice correlations are lost if the distortion distance ~ $\xi$. So, for distortions $L_c$ and $R_c$ for tilt and shear respectively the strains are of the order $\xi/L_c$ and $\xi/R_c$ respectively. So, the resulting increase in elastic free energy per unit volume,

$$\frac{1}{2}(C_{66}\left(\frac{\xi}{R_c}\right)^2 + C_{44}\left(\frac{\xi}{L_c}\right)^2)$$

$C_{66}$, $C_{44}$ are the elastic moduli for FLL shear and tilt.

The pinning force acts only through a distance $\sim\xi$ before changing randomly. Therefore we have the net free-energy change per unit volume in the presence of pinning sites,

$$\delta F = \frac{1}{2}C_{66}\left(\frac{\xi}{R_c}\right)^2 + \frac{1}{2}C_{44}\left(\frac{\xi}{L_c}\right)^2 - f\xi\frac{n^{\frac{1}{2}}}{V_c^{\frac{1}{2}}} \quad (1.31)$$

Minimising the free energy with respect to $L_c$ and $R_c$,

$$L_c = \frac{2C_{44}C_{66}\xi^2}{nf^2}, R_c = \frac{2^{\frac{1}{2}}C_{44}^{\frac{1}{2}}C_{66}^{\frac{3}{2}}\xi^2}{nf^2}, V_c = \frac{4C_{44}^2 C_{66}^4 \xi^6}{n^3 f^6}$$

$$\delta F_{min} = -\frac{n^2 f^4}{8C_{44}C_{66}^2 \xi^2}, J_c = \frac{1}{B}\frac{n^2 f^4}{2C_{44}C_{66}^2 \xi^3} \quad (1.32)$$

We therefore obtain a larger pinning energy and larger critical current for a softer FLL (smaller elastic moduli).

## 1.6 Vortex lattice melting and peak effect

Similar to melting observed when an ordered crystalline solid at a characteristic temperature and pressure becomes an isotropic liquid, at a characteristic temperature and magnetic field the ordered VL goes into an isotropic vortex liquid phase. This is known as vortex lattice melting.

Phenomenological argument for this VL disordering was given by Pippard[26]. He argued that the disordering happens due to interplay of two opposing interaction energies namely vortex-vortex interaction energy which tries to keep the VL ordered and vortex-impurity interaction energy (pinning energy) which disorders the VL by pinning individual vortices at random positions and thereby creating defects.

These energies are magnetic Gibbs free energy which is a minima at equilibrium when T, H are kept constant. These are obtained by calculating the area under the equilibrium magnetisation curve. Since for a Type II superconductor M falls linearly to zero as $-kh$,

$$g(h) = -\frac{1}{2}kh^2 \quad (1.33)$$

Chapter 1

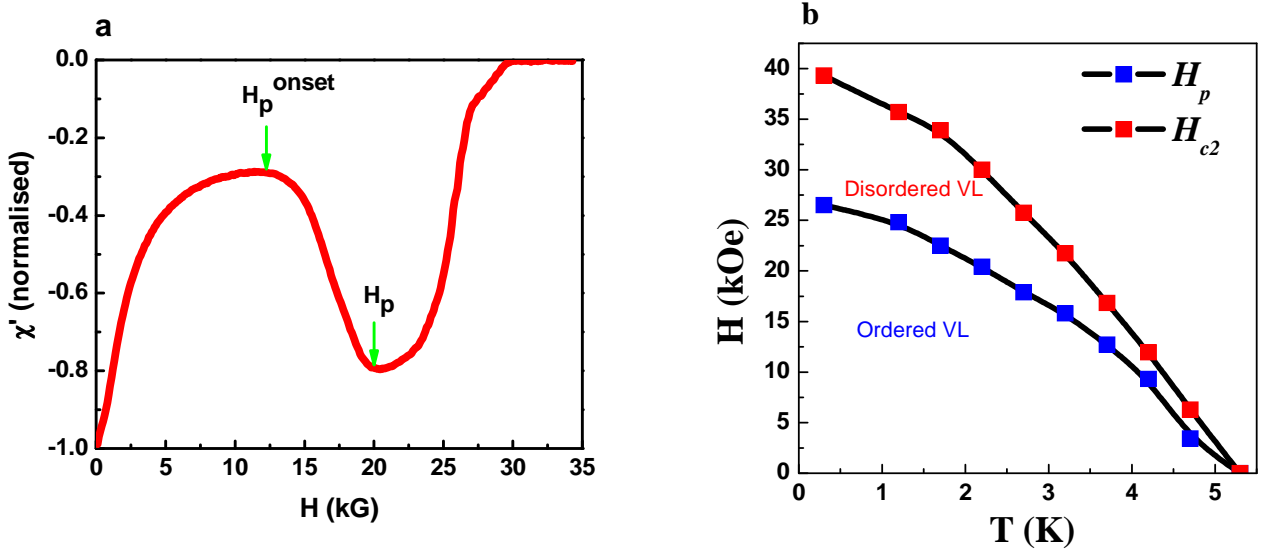

**Figure 1.7** (a) Peak effect in ac magnetic susceptibility of Co-doped NbSe$_2$. Data was taken at 2.2K. $H_p^{onset}$ and $H_p$ represent onset and peak of the peak effect. (b) VL phase diagram demarcating ordered and disordered regimes.

According to computations of Abrikosov (1957) and of Kleiner, Roth and Autler (1964) for square and triangular lattices respectively, the only difference is in the value of *k*. $H_{c2}$ does not depend on any special choice of lattice. The 'structural energy' $\Delta g_1$, which is the difference in g at a given value of *h* for the triangular and square lattices gives rigidity of the lattice. It falls to zero as $h^2$

$$\Delta g_1 = -\frac{1}{2}h^2 \Delta k \quad (1.34)$$

Now if we consider the effect of inhomogeneities, the FLL is then constrained to occupy a particular set of sites. So, we need to compare its behaviour in that position and in a laterally displaced position. Since in the two positions the points at which superconducting condensation is greatest differ in composition, strain etc., we may imagine that the mean value of $H_{c2}$ is changed by the translation. So, the pinning energy $\Delta g_2$ thus falls to zero as h

$$\Delta g_2 = -kh\Delta h \quad (1.35)$$

As we can see vortex-vortex interaction energy varies with applied magnetic field H as (H-$H_{c2}$)$^2$; whereas the vortex impurity interaction goes as (H-$H_{c2}$). So as H→$H_{c2}$, vortex-vortex interaction energy falls faster to zero and at a particular field $H_p^{onset}$, the vortex-impurity energy wins over so the vortices prefer to pass through pinning centres. As a result, the VL starts to disorder.

There are various bulk signatures of this ODT in VL. The most well-known signature is 'Peak effect' in bulk magnetization/critical current density[27,28,29] manifested as a non-monotonic decrease in the bulk magnetisation/ ac susceptibility (increase in critical current density) with external magnetic field/temperature. Figure 1.7(a) demonstrates this peak effect in ac magnetic susceptibility. For magnetic field H< $H_p^{onset}$, the VL is ordered. And for fields above $H_p^{onset}$, the VL starts to disorder. At the peak of the peak effect, $H_P$, the disordering is complete and we have vortex liquid phase. We can conclude about the VL phase diagram from the point of occurrence of the peak effect. In figure 1.7(b), we show one such (H, T) phase diagram indicating the parameter space over which we obtain ordered/ disordered vortex lattice.

## 1.7 Nature of the order-disorder transition in vortex lattice melting

For VL in superconducting thin films, the order-disorder transition can be understood within the framework of BKTHNY theory of 2-dimensional (2-D) melting[30,31,32]. But for VL in a 3-dimensional (3-D) superconductor, the vortex line is rigid only up to a length scale much shorter than sample dimensions. As a result, in a weakly pinned single crystal, the vortex line can bend considerably along the length of the vortex. It is generally accepted that in the presence of weak pinning the Abrikosov VL can transform into a quasi-long range ordered state such as Bragg glass[33] (BG), which retains long-range orientational order of a perfect hexagonal lattice but where the positional order decays algebraically with distance. Theoretically, both the possibility of a direct first order transition from a BG to a vortex glass (VG) state[34,35] (with short range positional and orientational order) as well as transitions through an intermediate state, such as multi-domain glass or a hexatic glass[36,37], have been discussed in the literature. While many experiments find evidence of a first-order order-disorder transition[38,39,40,41], additional continuous transitions and crossovers have been reported in other regions[42,43,44] of the *H-T* parameter space, both in low-$T_c$ conventional superconductors and in layered high-$T_c$ cuprates.

The nature of vortex lattice melting in conventional superconductors has been subjected to a great deal of controversy due contradictory results from various experimental groups. Using small angle neutron scattering (SANS) experiments, thermodynamic signatures of a first-order ODT was found[45,46] in the presence of weak or moderate pinning. However, similar SANS showed no evidence of VL melting till extremely close to $H_{c2}$ for extremely pure Nb single crystals[47]. Also, since signatures of the ODT in conventional superconductors get considerably broadened in the presence of random pinning, it has been suggested by some

Chapter 1

authors that the ODT could be a continuous crossover rather than a phase transition[48,49]. Presence of random pinning potential also prevents establishing perfect crystalline or liquid phase by making the states glassy in nature with very slow kinetics.

## 1.8 Disorder induced melting

For thermal melting of vortex lattice, Lindemann criterion[50] can be described in terms of characteristic energies: the vortex lattice melts when the thermal fluctuation energy becomes equal to the elastic energy:

$$T = \mathcal{E}_{el} \quad (1.36)$$

In the presence of point disorder, the random field induced by defects significantly alters the energy landscape for vortex making it rugged. As a result, these disordered configurations contribute to the entropic part of the vortex free energy. If its characteristic energy scale is $\mathcal{E}_{pin}$, which is the pinning energy. Then one can generalize the Lindemann criterion for a disordered system by including $\mathcal{E}_{pin}$ into the energy balance. And we obtain a quasi-periodic lattice to highly disordered entangled vortex solid transition when

$$\mathcal{E}_{pin} = \mathcal{E}_{el} \quad (1.37)$$

And finally from entangled solid to vortex liquid transition at

$$T = \mathcal{E}_{pin} \quad (1.38)$$

So, the most significant contribution of disorder in VL melting is that it can drive the system into an ODT even when the thermal fluctuations are not enough to destroy the order[51].

In our work, we have introduced a weak random disorder in the form of intercalation of Cobalt atoms in a conventional Type-II superconductor 2H-NbSe$_2$. We then observe the evolution of the long-range order (both positional and orientational) of the Abrikisov vortex lattice using real space imaging as we vary the external magnetic field isothermally across the peak effect. Presence of metastable states by different field and temperature cyclings are demonstrated. We then perform thermal hysteresis measurements to confer about the nature of phase transformations across the peak effect. We finally demonstrate the presence of orientational coupling of the vortex lattice with underlying crystalline lattice and its subsequent effect on the order-disorder transition.

Chapter 1

# Chapter 2

# Experimental details, measurement techniques and sample growth details

## 2.1 The scanning tunnelling microscope (STM)

The scanning tunnelling microscope (STM), invented in 1983 by Binnig and Rohrer[1], has proven to be an extremely powerful and versatile probe to study the surface properties of conducting materials. In addition to having excellent energy resolution, it provides unsurpassed spatial resolution enabling us to determine electronic density of states at atomic length scales and thereby determining the microscopic electronic structure. The first demonstration of its spatial resolution was imaging Si(111) and showing 7×7 surface reconstruction in real space[2]. Since then it has been employed in diverse aspects of surface study and manipulation such as building atomic structure by manipulating atom by atom[3], visualising standing waves formed by electrons[4,5], imaging the spin texture[6] etc.

STM is based on the principle of quantum mechanical tunnelling between a sharp metallic wire (tip) and sample. When the tip is brought very close to the sample (~few nm) and a bias voltage is applied between the tip and sample, tunnelling current flows. This current is typically very small (10 pA – 1 nA). It is therefore amplified and recorded.

If an electron of mass m and energy E is subject to a potential barrier U (>E), then classically it cannot cross the barrier. But, in quantum mechanics its wavefunction can be described by Schrodinger's equation as

$$-\frac{\hbar^2}{2m}\frac{d^2}{dz^2}\psi(z) + U(z)\psi(z) = E\psi(z) \quad (2.1)$$

For *E > U (z)* (classically allowed region), the solution is

$$\psi(z) = \psi(0)e^{\pm ikz} \quad (2.2)$$

Where $k = \sqrt{\frac{2m(E-U)}{\hbar}}$

In the classically forbidden region

$$\psi(z) = \psi(0)e^{-\kappa z} \quad (2.3)$$

Chapter 2

Where $\kappa = \sqrt{\frac{2m(U-E)}{\hbar}}$

So the tunnelling current ($\propto |\psi|^2$) $\sim e^{-2\kappa z}$. As a result, the tunnelling current between tip and sample decays exponentially with distance. This property of tunnelling current is used in imaging the topographic features of a conducting surface.

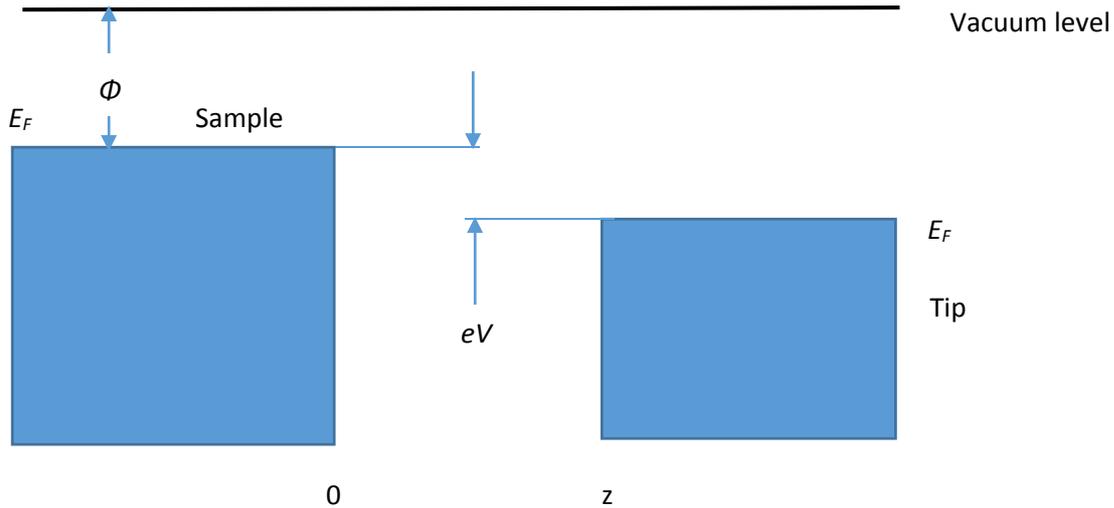

**Figure 2.1** A one-dimensional metal-vacuum-metal tunnelling junction. $\Phi$ is work function of the sample. A bias voltage *V* is applied across sample and tip.

## 2.2 Theory of scanning tunnelling microscopy:

### 2.2.1 Bardeen's approach:

The planner tunnel junction problem is treated by Bardeen[7] as follows.

When two electrodes A and B are far apart, the wavefunction satisfy

$$i\hbar \frac{\partial \Psi}{\partial t} = [-\frac{\hbar^2}{2m}\frac{\partial^2}{\partial z^2} + U_A]\Psi \quad (2.4)$$

Where the stationary state, $\Psi = \psi_\mu e^{-iE_\mu t/\hbar}$. Whose spatial part satisfy the following

$$\left[-\frac{\hbar^2}{2m}\frac{\partial^2}{\partial z^2} + U_A\right]\psi_\mu = E_\mu \psi_\mu \quad (2.5)$$

Similarly, $i\hbar \frac{\partial \Psi}{\partial t} = [-\frac{\hbar^2}{2m}\frac{\partial^2}{\partial z^2} + U_B]\Psi \quad (2.6)$

where $\Psi = \chi_\nu e^{-iE_\nu t/\hbar}$

$$\text{and } \left[-\frac{\hbar^2}{2m}\frac{\partial^2}{\partial z^2} + U_A\right]\chi_\nu = E_\nu\chi_\nu \quad (2.7)$$

$U_A, U_B$ are potential functions of electrodes A, B.

By bringing the two electrodes together, the Schrodinger equation of the combined system becomes

$$i\hbar\frac{\partial \Psi}{\partial t} = \left[-\frac{\hbar^2}{2m}\frac{\partial^2}{\partial z^2} + U_A + U_B\right]\Psi \quad (2.8)$$

With the presence of this combined potential, the states $\psi_\mu$ will have a probability of transferring to the states of electrode B. i.e.,

$$\Psi = \psi_\mu e^{-iE_\mu t/\hbar} + \sum_{\nu=1}^{\infty} c_\nu(t)\chi_\nu e^{-iE_\nu t/\hbar} \quad (2.9)$$

With $c_\nu(t=0)=0$.

Assuming that the sets of wavefunctions $\psi_\mu$ and $\chi_\nu$ are orthogonal to each other, we finally obtain

$$i\hbar\frac{dc_\nu(t)}{dt} = \int_{z>z_0} \psi_\mu U_B \chi_\nu^* d^3r \, e^{-i(E_\mu-E_\nu)t/\hbar} \quad (2.10),$$ since $U_B$ is non-zero only in the volume of electrode B ($z > z_0$).

Defining the tunnelling matrix element as $M_{\mu\nu} = \int_{z>z_0} \psi_\mu U_B \chi_\nu^* d^3r$

$$\text{We have } c_\nu(t) = M_{\mu\nu}\frac{e^{-i(E_\mu-E_\nu)t/\hbar}-1}{E_\mu-E_\nu} \quad (2.11)$$

So, starting with μ-th state of electrode A, the probability of having υ-th state of electrode B at time t is

$$p_{\mu\nu}(t) = |c_\nu(t)|^2 = |M_{\mu\nu}|^2 f(t) \quad (2.12)$$

Where $f(t) = \frac{4\sin^2[(E_\mu-E_\nu)t/2\hbar]}{(E_\mu-E_\nu)^2}$. It reaches maxima for $E_\mu = E_\nu$, for $E_\mu \neq E_\nu$, it approaches zero rapidly. So, the tunnelling current depends on how many states of electrode B are near the energy value of electrode A. Let $\rho_B(E)$ be the density of states of electrode B at energy $E$. Then total probability of tip states that the sample states can tunnel into in time t is

$$p_{\mu\nu}(t) = \frac{2\pi}{\hbar}|M_{\mu\nu}|^2 \rho_B(E_\mu)t \quad (2.13)$$

Chapter 2

Since $\int_{-\infty}^{\infty} \frac{sin^2 ax}{\pi ax^2} dx = 1$. The integrand approached a delta function for large a, i.e. if time of tunnelling is much greater than $\hbar/\Delta E$, where $\Delta E$ is the energy resolution, then Eq. 2.13 Implies a condition of elastic tunnelling: $E_\mu = E_\nu$.

Now, if a bias voltage V is applied across the electrodes, the tunnelling current can be evaluated by summing over all relevant states. At a finite temperature, the electrons in both electrodes follow the Fermi distribution function $f(E) = \frac{1}{1+\exp(\frac{E-E_F}{k_B T})}$. The tunnelling current is

$$I = \frac{4\pi e}{\hbar} \int_{-\infty}^{\infty} [f(E_F - eV + \epsilon) - f(E_F + \epsilon)]\rho_A(E_F - eV + \epsilon)\rho_B(E_F + \epsilon)|M|^2 d\epsilon \quad (2.14)$$

## 2.2.2 Tersoff-Hamann model:

Although the image obtained from the STM is a convolution of tip and sample electronic states, it is difficult to know the tip states. Tersoff and Hamann formulated a model[8] based on Bardeen's theory so that the tip properties can be taken out of the problem. They modelled the tip as a locally spherical potential well with radius of curvature $R$ centred at $\mathbf{r_0}$. Which implies the tip wavefunction is also spherically symmetric. As a result, the tunnelling matrix becomes proportional to the value of sample wavefunction at $x=0$, $y=0$, $z=z_0$, which is the center of curvature of the tip $\mathbf{r_0}$.

$$M \propto \psi(\mathbf{r_0}) \quad (2.15)$$

By summing up all sample states near Fermi level, we have the tunnelling conductance

$$G = \frac{I}{V} \propto |\psi(\mathbf{r_0})|^2 \rho_S(E_F) \quad (2.16)$$

The right hand side is the local density of state of the sample at the Fermi level at the centre of curvature of the tip. i.e.

$$G = \frac{I}{V} \propto \rho_S(E_F, \mathbf{r_0}) \quad (2.17)$$

So, if the tip is spherically symmetric around a point $\mathbf{r_0}$, it is effectively equivalent to a geometrical point at $\mathbf{r_0}$. In this treatment, all other tip wavefunctions except for the *s*-wave tip wavefunction has been neglected. Therefore it is called *s*-wave tip model.

This model is applicable is case the length scale of feature is around 1 nm or more. So, it cannot account for the phenomena observed in atomic length scales.

## 2.3 Construction of low temperature scanning tunnelling microscope:

The scanning tunnelling microscope mainly consists of the following units:

1. STM head containing piezo electric scanning and positioning unit. It consists of (a) course positioner: It brings the tip to a distance from the sample so that tunnelling current is possible to flow. (b) Piezo electric tube: It controls the lateral (x,y) and vertical (z) motion of the tip.

2. Vibration isolation stage: It is employed to get rid of vibrational noise coming from various sources.

2. Control electronics: It is an integrated system controlling the piezo motion through feedback loop, applying bias voltage across sample and tip, amplifying and recording tunnelling current etc. It is also integrated with the data acquisition and storage through computer.

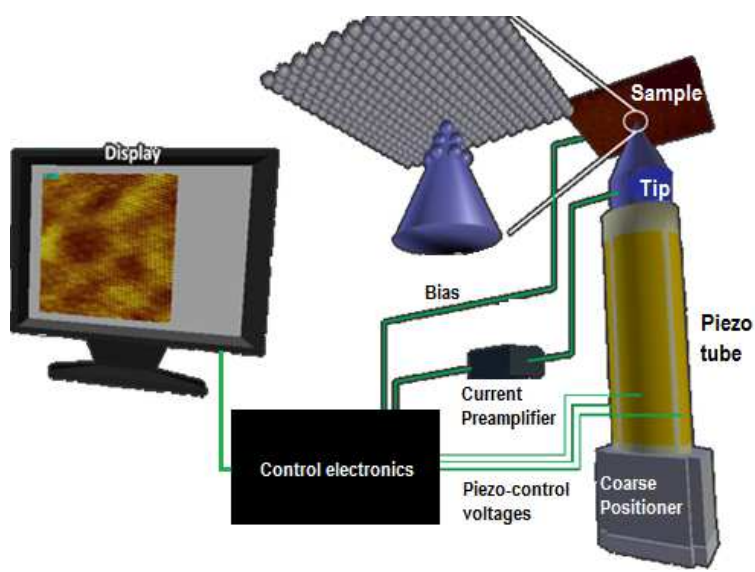

**Figure 2.2** Schematic diagram of scanning tunnelling microscope

The low temperature scanning tunnelling microscope[9] (LT-STM) assembly consists of three primary sub-units: (i) The sample preparation chamber, (ii) the load lock chamber to transfer the sample to the STM and (iii) the $^4$He dewar with 9T magnet housing $^3$He cryostat on which the STM head is attached. The $^4$He dewar hangs from a specially designed vibration isolation table mounted on pneumatic legs. A combination of active and passive vibration isolation systems are used to obtain the required mechanical stability of the tip. Data acquisition is done using the commercial R9 SPM controller from RHK technology[10]. The 3D diagram of our system is shown in Figure 2.3.

Chapter 2

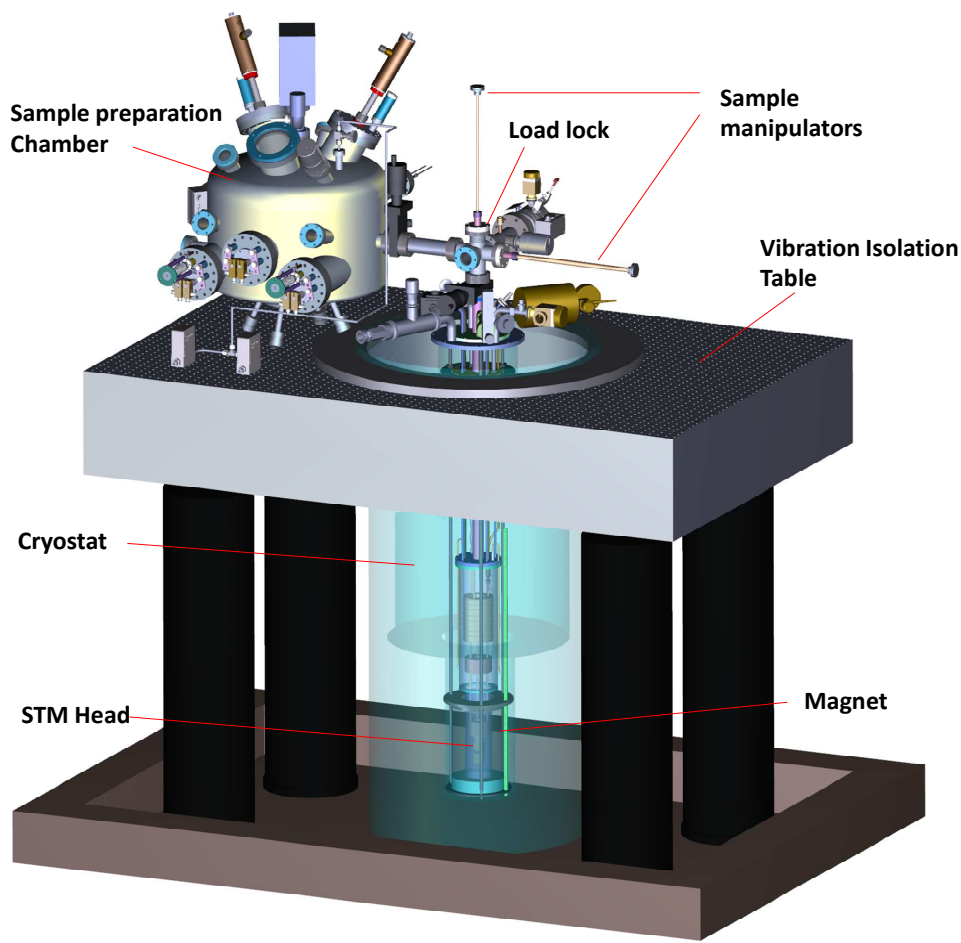

**Figure 2.3**. 3D view of the LT-STM assembly consisting of three primary sub-units: (i) The sample preparation chamber, (ii) the load lock chamber to transfer the sample from the deposition chamber to the STM and (iii) the $^4$He Dewar with 9T magnet housing $^3$He cryostat on which the STM head is attached. The $^4$He Dewar hangs from a specially designed vibration isolation table mounted on pneumatic legs. The Dewar, cryostat and magnet have been made semi-transparent to show the internal construction.

## 2.3.1 STM head

The main body of the STM head is made of single piece of gold plated oxygen free high conductivity (OHFC) copper (Fig 2.4). The sample holder, coming from the top with the sample facing down, engages on a Gold plated Copper part which has 45° conical cut at the top matching with sample holder which gets locked into two nonmagnetic stainless steel studs. The Copper part is electrically isolated from the main body using cylindrical Macor[11] machinable ceramic part. Both these parts are glued together using commercially available low temperature glue Stycast[12]. The copper part also has two phosphor bronze leaf springs which grab the sample holder and hence provide better thermal contact and prevent mechanical vibration of the sample holder. Electrical contact to this copper part is given by soldering a stud which extrude from the lower side.

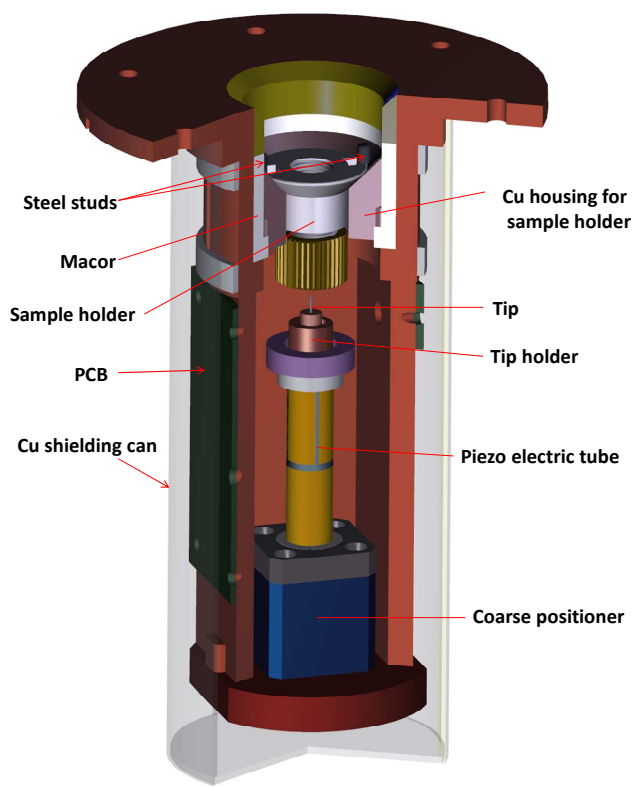

**Figure 2.4**. 3D view showing the construction of the STM head with the coarse positioner, piezoelectric scan-tube mounted, tip holder and sample holder with the sample facing down. PCB mounted on the three sides serves as a connecting stage for electrical connections

Positioning unit is located in the cuboidal cavity in the lower part of STM head. One of the sides of the cavity is open to get access for mounting the positioning unit and changing tip. The positioning unit consists of a coarse approach positioner and a piezoelectric tube on which the STM tip is fixed. Fine positioning and scanning is performed using a 1 inch long piezoelectric tube[13] which has gold plated electrodes inside and outside. The piezo-tube is electrically isolated from coarse positioner at the bottom and the copper tip carrier on the top through Macor pieces which are glued to the tube to reduce differential thermal expansion. The copper tip holder is glued on the upper side of top Macor piece. We use Pt-Ir wire (80-20%) of diameter 300µm as tip which is held frictionally in 400µm bore that is drilled on tip holder. Tip is prepared by cutting the Pt-Ir wire using a sharp scissor at an angle and subsequently field emitted in vacuum at low temperature to achieve the desired sharpness. Printed circuit boards are screwed on the three sides of the cuboid and are used for providing electrical connection to the piezo units, sample and tip. Temperature of STM head is measured using two Cernox[TM] sensors[14] mounted on the bottom plate of the STM as well as on the $^3$He pot. The entire STM head is enclosed in gold plated copper can ensuring temperature homogeneity over the entire

Chapter 2

length of the head. We observe that after achieving a stable temperature for about 10 min the temperature of the STM head and $^3$He pot differ at most by 20mK.

## 2.3.1.1 Coarse positioner

The coarse positioner is fixed to a copper bottom plate using a pair of titanium screws which are in turn screwed to the main body of the STM head. During loading the sample holder to the STM head, the tip is kept at a distance from the sample which is much greater ($\geq 0.5$ mm) than the tunnelling regime. Before start of measurement, we therefore need to bring the tip near the sample within tunnelling regime using coarse positioner also known as piezo walker[15]. It works on the principle of slip-stick motion. Its essential components are, (a) Fixed frame rigidly fixed to the base of STM head, (b) piezoelectric actuator glued to the fixed frame, (c) guiding rod firmly connected to a piezoelectric actuator and (d) sliding block or clamped table which is frictionally held on the guiding rod. Other parts of the positioning unit is screwed on this clamped table.

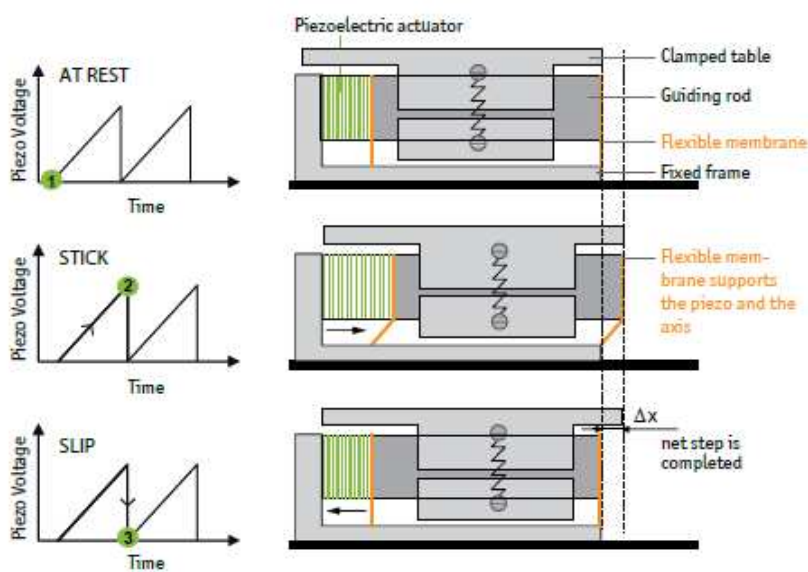

**Figure 2.5**. Schematic explaining working principle of a coarse positioner. Image courtesy Ref. 15.

The piezo expands its length proportional to applied voltages. The slip stick motion is based on the application of sawtooth shaped voltage pulses to the piezo. During the slowly increasing voltage phase the clamped table sticks to the guiding rod and is moved over a distance Δz. The achieved expansion Δz is proportional to the applied maximum voltage. The typical minimum step size for ANP positioners is in the range of 50 nm at ambient conditions and 10 nm at cryogenic temperatures. Subsequently the guiding rod is accelerated very rapidly over a short period of time (typically microseconds) so that the inertia of the clamped table overcomes friction. This way, the clamped table disengages from the accelerated rod and remains nearly stationary. The net step Δz is now completed. By repeating this procedure the table can be moved over large distances with nanometer precision.

## 2.3.1.2 Piezoelectric tube:

In figure 2.6, I have shown the piezoelectric tube used for fine positioning. The piezo tube consists of 6 gold plated electrodes. The upper part contains 4 axially separated electrodes which are 90 degree apart from each other. A pair of the opposite sections of the quartered electrode is referred to as the X, –X electrodes and other pair as Y, –Y electrodes. The bottom part contains Z electrode. The inner electrode is the ground and is wrapped from the top to avoid any static charge build up.

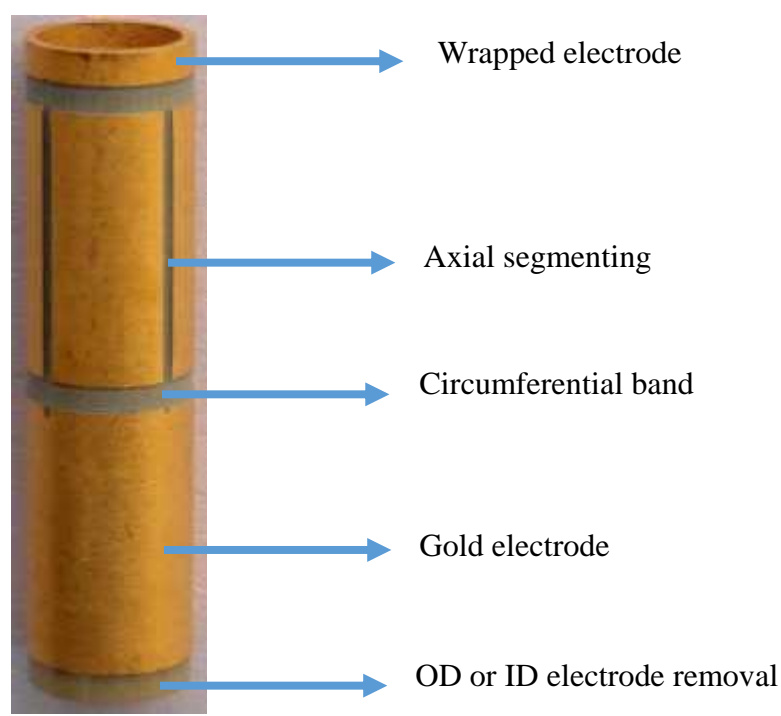

**Figure 2.6** Image of the scan tube used in our STM. Gold plated electrodes have two segments. Lower segment is used for Z motion. Upper segment is subdivided into four quadrants for X-Y movement. Inner electrode is wrapped around at the top which helps to discharge the static charges if any.

The operational principal of the piezo tube is schematically depicted in Fig. 2.7. If an equal voltage is applied in all 4 electrodes with respect to the inner ground electrode, then the piezo tube either expands or contracts depending on the polarity of the applied voltage. If voltage of equal and opposite polarity is applied to the opposing electrode segments, then there is a lateral deformation of the tube. So, by applying appropriate voltages to the different electrodes we can adjust the tip position with extreme precision. Total extension and lateral deformation is given by

Chapter 2

$$\Delta Z = \frac{d_{31}VL}{t}, \Delta X = \Delta Y = \frac{d_{31}VL^2}{d_m t} \quad (2.18)$$

Where $d_{31}$ = Piezoelectric strain constant,

$V$ = applied voltage,

$L$ = Length of the electrode,

$d_m$ = mean diameter of the tube, (OD+ID)/2,

$t$ = thickness of the tube.

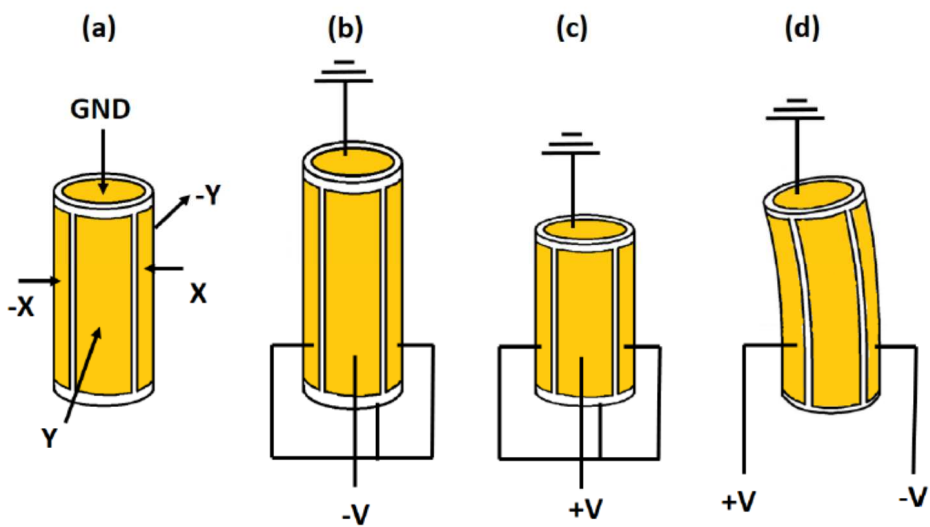

**Figure 2.7** Schematics of the operation of the piezo tube. (b) When a positive voltage is applied to all the four quadrants, the tube contract in z direction. (c) With negative voltages, it extends. (d) When opposite polarity voltages are applied to X, –X (or Y, –Y) then the tube deforms laterally along X (Y) direction.

## 2.3.2 Cryostat and temperature control of the sample

The low temperature stage consists of an internally fitted charcoal sorption pump based $^3$He cryostat from Janis Research Company[16]. We use a custom design with annular shaped sorption pump, 1K pot and $^3$He pot which give us direct line of site access from the top of the cryostat to the STM head mounted below the $^3$He pot. To ensure thermal stability the STM head is bolted to $^3$He pot using 6 screws which ensures good thermal contact between the two. To prevent radiative heating, a radiation plug is inserted in the cryostat after loading the sample using the same vertical manipulator as the one used to insert the sample. The radiation plug (not shown) sits just above the STM head. The $^3$He pot and sorption pump are fitted with resistive heaters. All electrical wires coming from the top are thermally anchored at the 1K pot and the ³He pot. The entire process of cooling the STM from 4.2 K to the base temperature of 350 mK takes about 20 min with a hold time of about 8 hrs. We wait for about 15 min for after the base temperature is reached before starting our measurements. Between the base temperature and 3 K, we control the temperature by controlling the temperature of the sorption pump. For stabilizing above 3K we use the resistive heater fitted on the ³He pot by keeping the sorption pump temperature constant at 30 K.

## 2.3.3 Liquid Helium Dewar

The cryostat is mounted in a 65 liters capacity Al-Fibreglass Dewar with retention time of approximately 5 days. The superconducting magnet with maximum of 9 T aligned along the STM tip hangs from the top flange of the cryostat. Exhaust line of the cryostat is connected with one way valve which maintains a constant pressure slightly above atmosphere. This allows us to flow liquid ⁴He in a capillary wrapped around the sorption pump such that the sorption pump can be cooled without using an external pump.

## 2.3.4 Vibrational and electrical noise reduction:

Most crucial part of any STM design is the vibrational and electrical noise reduction as it is directly reflected in the ultimate noise level in the tunneling current. To obtain reliable topographic images with atomically resolved features, one requires noise amplitude due external disturbances to of the order of 1 pm or less.

### 2.3.4.1 Vibrational noise reduction

We have adopted three isolation schemes to reduce vibrational noise. For sound isolation, the entire setup is located in a sound proof enclosure made of sound proofing perforated foam. To reduce vibrational noise mainly coming from the building, the entire setup rests on a commercial vibration isolation table[17] (Newport SmartTable®) with integrated active and passive stages with horizontal and vertical resonant frequency < 1.7 Hz. Finally, since in our cryostat the 1K pot pump has to be on during STM operation, special precaution has been taken to isolate the system from the pump vibrations which get transmitted in two different ways: (i) Direct pump vibration transmitted through vibration of the connecting bellows is isolated by keeping the pumps on a different floor in the basement and a rigid section of the pumping line is embedded in a heavy concrete block before connecting to the pump. (ii) indirect vibration transmitted through the sound propagated through the ⁴He gas in the pumping

Chapter 2

line. To isolate this source of vibration a special pumping scheme is adopted. The 1K pot is connected to the pump through two alternate pumping lines. While condensing the $^3$He and cooling the STM head from 4.2K to the base temperature, the 1K pot is cooled to 1.6 K by pumping through a 25.4 mm diameter pumping line directly connected to the pump. Once the base temperature of 350 mK is reached on the STM head, the first pumping line is closed and the second pumping line is opened. This line has a 30 cm long 10 cm diameter intermediate section packed with high density polystyrene foam which isolates the STM from the sound generated by the pump. Since the polystyrene foam reduces the pumping speed, the 1K pot warms up to 2.8 K, with no noticeable increase in the temperature of the STM head. During the steady-state operation of the STM at 350 mK the pumping is further reduced by partially closing a valve to keep the 1K pot at a constant temperature of ~ 4 K. While operating in this mode we do not observe any difference in vibration level on the top of the cryostat with the 1K pot pump on or off as shown in Fig. 2.8(a).

### 2.3.4.2 Electrical noise reduction

To reduce the electrical noise coming from the 50Hz line signal, ground connection of all instruments, table and Dewar are made to a separate master ground. RF noise is further reduced by introducing 10 MHz low pass filter before each connection that goes into the STM. The tunneling current is detected using a Femto DLPCA-200 current amplifier with gain of $10^9$ V/A. While the bandwidth of the DLCPA-200 amplifier is 500 kHz, the measurement bandwidth is set digitally restricted to 2.5 kHz in the R9 SPM controller.

### 2.3.4.3 Characterisation of noise

The spectral density (SD) in the current and Z-height signals quantitatively tells us how good our isolation is. We have recorded these signals at 350 mK in actual operating condition. Figure 2.8(b) shows the SD of the current (i) when the tip is out of tunneling range (background noise of electronics), (ii) at a fixed tunneling current with feedback on condition, and (iii) after switching off the feedback for 5 s. The SD with tip out of tunneling range is below 300 fA Hz$^{-1/2}$. At fixed tunneling current (feedback on) additional peaks appear in the SD at 25.5 Hz and 91.5 Hz but the peak signal is only marginally larger than 300 fA Hz$^{-1/2}$. Even after switching off the feedback the peak signal is less than 1 pA Hz$^{-1/2}$. Similarly, the Z-height SD at fixed tunneling current with feedback on (Fig. 2.8(c)) is less than 2 pm Hz$^{-1/2}$ at all frequencies and less than 50 fm Hz$^{-1/2}$ above 150 Hz. The low Z-height and current noise allows us to get very good signal to noise ratio in spectroscopic measurements.

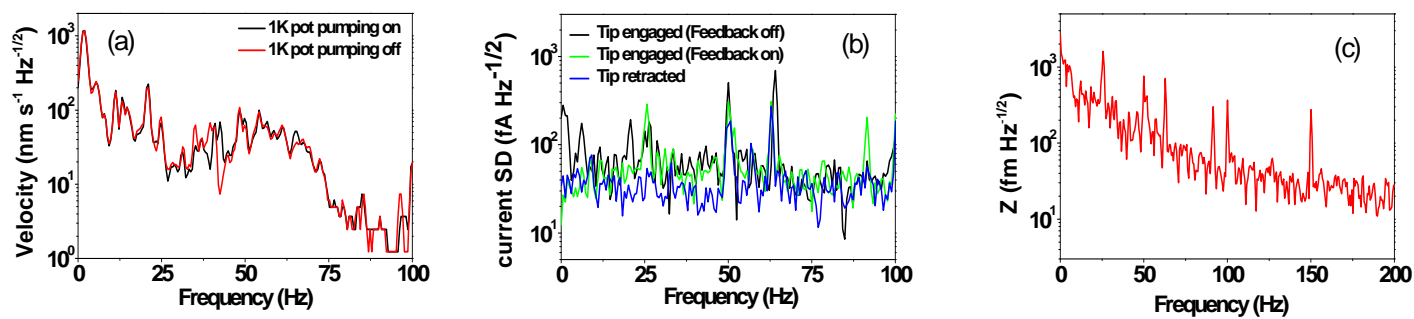

**Figure 2.8.** (a) Spectral density of the velocity vs. frequency on the top of the cryostat measured using an accelerometer. The spectral densities with and without the 1K pot pump on are nearly identical. (b) Spectral density of the tunneling current with the tip out of tunneling range, within tunneling range with feedback on and with feedback off. (c) Spectral density of Z height signal with feedback on. Measurements in (b) and (c) were performed at 350 mK on a NbSe$_2$ single crystal with tunneling current set to 50 pA and bias voltage to 20 mV.

### 2.3.5 Load lock and sample manipulators

The load-lock, located at the top of the $^3$He cryostat, has six CF35 ports and it is connected to sample preparation chamber and STM chamber through gate valves. Typical time to pump the load-lock chamber from ambient pressure to $1\times10^{-6}$ mbar using a turbomolecular pump is about 20 minutes. Sample manipulators (Figure 16) are made of seamless steel tubes (closed at one end) and have matching threads at the end to engage on the corresponding threads on the sample holder. A thermocouple is fitted inside the horizontal sample manipulator to measure the temperature of the sample during deposition.

Chapter 2

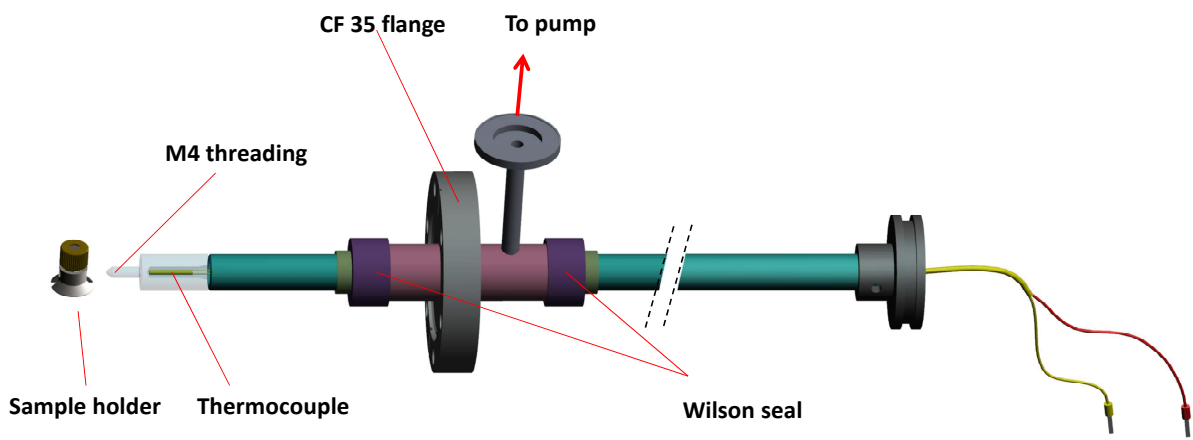

**Figure 2.9.** Design of the horizontal sample manipulator with in-built thermocouple for measuring the temperature during sample deposition. A differential pumping arrangement between two Wilson seals is used to remove any leaked gas during movement. The end of the manipulator is made transparent to show the position of the thermocouple. The vertical manipulator is similar in construction but does not have the thermocouple.

## 2.3.6 Sample holder and cleaving mechanism

The sample holder used in our experiments is made of single piece of Molybdenum (Fig 2.10). The sample holder has threading on side and bottom for holding on the horizontal and vertical manipulators respectively. The sample is placed on top of the holder using Silver epoxy.

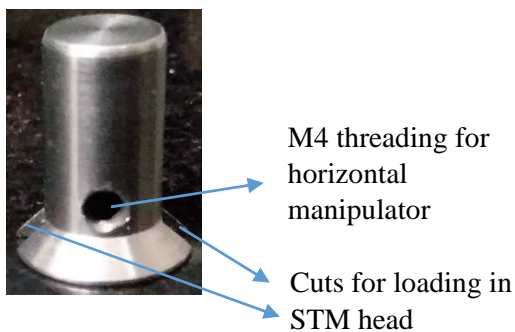

**Figure 2.10.** Actual image of the sample holder. The side threading and the cuts for loading the sample holder are visible

Our sample of interest, NbSe$_2$ can be easily cleaved using a scotch tape. To cleave the sample in-situ, we stick one end of a double sided tape on the sample surface covering it completely and a magnetic cyllinder on its other end. Inside the load-lock cross, after pumping to ~ $1\times10^{-7}$ mbar, the magnetic cyllinder is pulled using a horizontal manipulator with magnet on its end (Figure 2.11). In this way, a freshly cleaved surface in vacuum is formed. This method can be extended for other samples not cleavable by scotch tape also where the magnetic cylinder can be stuck to the sample with any conducting (Ag) or non-conducting (Torsil) epoxy and hitting it with the magnet and thereby cleaving the sample.

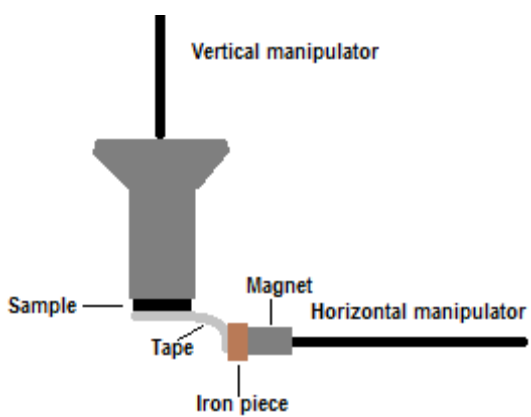

**Figure 2.11.** Schematic of in-situ cleaving assembly

## 2.4 Tip preparation

Commercially available Pt-Ir wire of diameter 300 µm is used as tip. Before loading the tip to the tip holder, it is cut at an angle to make it sharp and pointed. It is further sharpened using field emission. It is a method of electron emission under the influence of strong electric field. The set up for field emission procedure is schematically shown in Figure .Typically (200-250) V is applied between the tip and sample with the tip being at negative bias and the resulting current is being measured by measuring voltage across a voltage divider. As sample, we use commercially bought single crystal of Ag. The feedback is switched off in this case and the current is initially kept between (30-35) µA. As the electrons from the tip are emitted out due to application of high field, the tip becomes sharper. The field emission process is carried out until the current is stable and constant.

Chapter 2

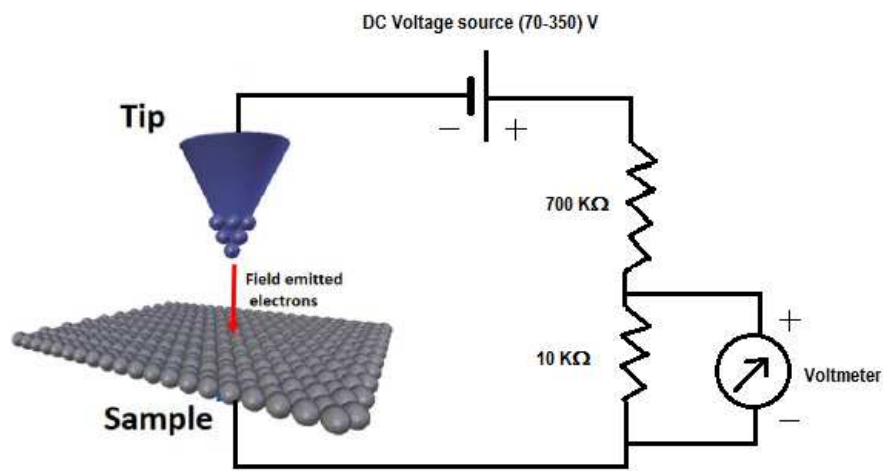

**Figure 2.12.** Circuit diagram of field emission set-up

The stability of the tunnelling current is checked on the Ag single crystal itself. After each field emission, the crystal is cleaned with fine emery paper before loading. As seen from figure , the stability of tunnelling current and conductance spectra improves significantly after field emission.

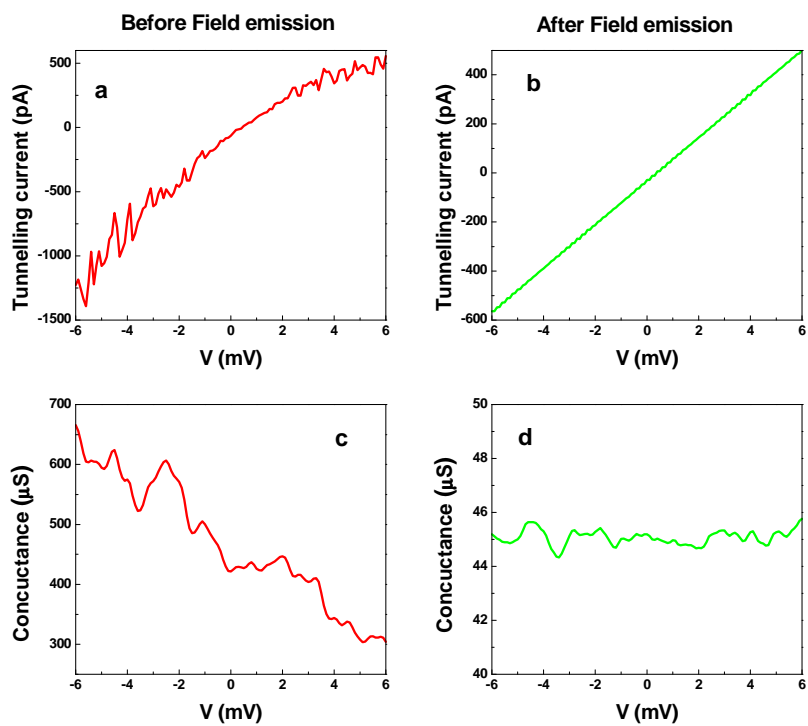

**Figure 2.13.** Tunnelling current and conductance spectra on Ag (a), (c) before field emission. (b), (d) After field emission.

## 2.5 STM modes of operation:

### 2.5.1 Topography

### 2.5.1.1 Constant current mode:

In this mode, the feedback loop controls the piezoelectric tubes in such a way so that the tunnelling current flowing between the electrodes namely tip and the sample surface always remains constant. By recording the voltage which needs to be applied to the piezo in order to maintain the constant tunnelling current, we determine the height of the tip as a function of position z(x, y). In this way, a topographic image is generated. This method of imaging is safe to use for any kind of corrugated surface without destroying or modifying the tip. A disadvantage of this mode of operation is due to the finite response time of the feedback loop which sets an upper limit for maximum scan speed.

To test our system for atomic resolution and in magnetic field, we performed measurements on a 2H-NbSe$_2$ single crystal. Having a hexagonal closed packed layered structure this crystal can be easily cleaved in-plane. We cleaved the crystal in-situ by attaching a tape on the surface and subsequently pulling the tape in vacuum in the load-lock chamber using the sample manipulators. Fig. 2.14 shows the atomic resolution image at 350mK which reveals the hexagonal lattice structure along with the charge density wave modulation. The lattice spacing of 0.33 nm is in good agreement with the lattice constant of NbSe$_2$ known from literature.[18, 19, 20]

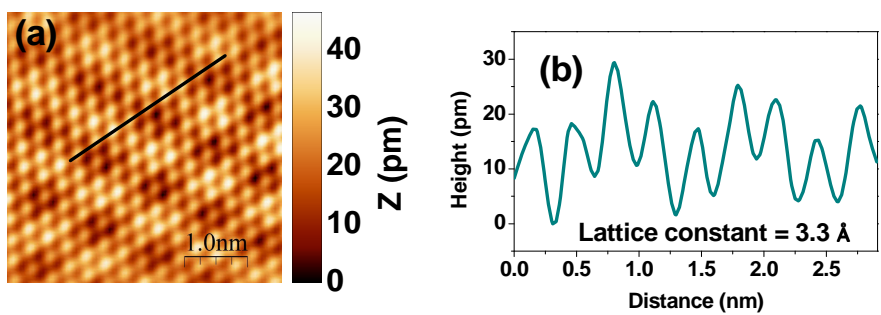

**Figure 2.14**. (a) Atomically resolved topographic image of NbSe$_2$ obtained in constant current mode; the charge density wave modulation is also visible. The tunneling current was set to 150 pA, the bias voltage to 20 mV and the scan speed was 13 nm/s. (b) Line cut along the line shown in (a).

## 2.5.1.2 Constant height mode:

To increase the scan speed, the tip can be scanned at constant height over the sample surface while the feedback loop is switched off. The variation of tunnelling current as function of position will contain the topographic information. But since distance dependence of tunnelling current is not known exactly, extracting topographic information from constant height mode is more difficult. Its main limitation is that it is only applicable to atomically flat surfaces, otherwise the tip might crash into a surface protrusion while scanning.

## 2.6 Scanning tunnelling spectroscopy

We have the tunnelling current flowing between the tip and sample having density of states $\rho_t(\epsilon)$ and $\rho_s(\epsilon)$ respectively

$$I = A \int_{-\infty}^{\infty} |M|^2 \rho_t(\epsilon)\rho_s(\epsilon - eV)[f(\epsilon) - f(\epsilon - eV)]d\epsilon \quad (2.19)$$

Where A is a constant. If both of them are metallic, then

$$I = A|M|^2 \rho_t(0)\rho_s(0) \int_{-\infty}^{\infty}[f(\epsilon) - f(\epsilon - eV)]d\epsilon = A|M|^2 \rho_t(0)\rho_s(0)eV = G_{nn}V \quad (2.20)$$

So, the tunnel conductance between metallic tip and metallic sample is independent of temperature and bias voltage.

If the sample is superconducting, then

$$I = \frac{G_{nn}}{e} \int_{-\infty}^{\infty} \frac{\rho_s(\epsilon)}{\rho_t(0)}[f(\epsilon) - f(\epsilon - eV)]d\epsilon = \frac{G_{nn}}{e} \int_{-\infty}^{\infty} \frac{\rho_s(\epsilon)}{\rho_t(0)}\left[-\frac{\partial f(\epsilon - eV)}{\partial(eV)}\right]d\epsilon \quad (2.21)$$

At sufficiently low temperature, Fermi function becomes a step function and the tunnelling conductance $\frac{dI}{dV}$ becomes proportional to the local density of state of sample at $\epsilon = eV$

In the tunnelling conductance measurement, tip sample distance is fixed by switching off the feedback loop and a small alternating voltage is modulated on the bias. The resultant amplitude of the current modulation as read by the lock-in amplifier is proportional to the dI/dV as can be seen by Taylor expansion of the current,

$$I(V + dV\sin(\omega t)) = I(V) + \left.\frac{dI}{dV}\right|_V \cdot dV\sin(\omega t) + .. \quad (2.22)$$

In our experiments, we keep modulation amplitude of 150 $\mu V$ at a frequency of 2.67 KHz.

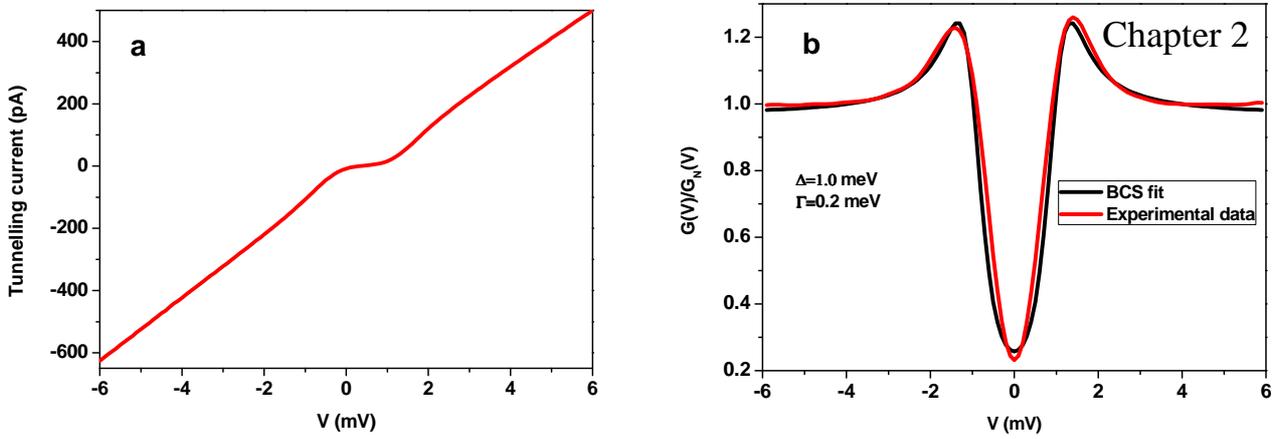

**Figure 2.15** (a) Tunnelling current, (b) conductance spectra along with its BCS fit of NbSe$_2$ at 1.5 K. Γ is the Dynes' parameter defined in chapter 1.

## 2.7 Vortex lattice imaging:

We have first taken full area spectroscopic map over an area of 352 × 352 nm in magnetic field of 200mT at 350mK. In this measurement we recorded the spatially resolved tunneling spectra (*dI/dV vs. V*) at each point of a grid having 64 × 64 pixels by sweeping the bias from 6mV to -6mV. Figure 2.16(a-d) shows intensity plots of the tunneling conductance normalized at 6mV at different bias voltages, showing the hexagonal vortex lattice. The lattice constant, *a* ≈ 109.8 nm is in excellent agreement with the theoretical value expected from Ginzburg Landau theory[21]. For voltages below Δ/*e* the vortices appear as regions with larger conductance whereas for voltages close to the coherence peak the vortices appear as regions with lower conductance. Fig. 2.16(e) shows the line scan sectioned on the line shown in Fig. 2.16(a). Three representative spectra are highlighted in the figure. Spectra 1 and 3 correspond to the superconducting region while the spectrum 2 is at the vortex core and has a zero bias conductance peak which is the signature of Andreev bound state inside the vortex core. In figure 10(f) we show a high resolution (128 × 128) conductance map obtained by measuring *dI/dV* at a fixed bias voltage of 1.4mV (position of coherence peak) while scanning over the same area.

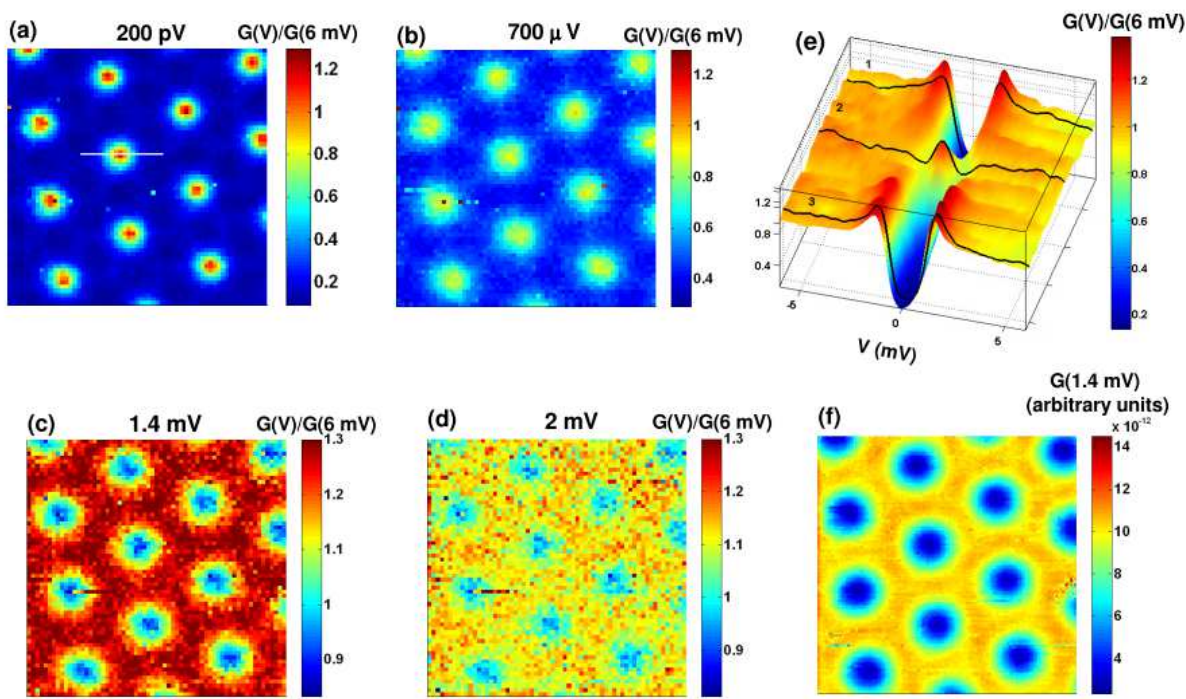

**Figure 2.16**. Vortex imaging on NbSe$_2$. (a)-(d) 64 × 64 conductance maps over 352 × 352 nm area at different voltages at 350 mK in an applied field of 200 mT. The maps are obtained from full spectroscopic scans from -6 mV to 6 mV at each pixel. (e) Line scan of the tunneling spectra along the white line marked in panel (a). Three spectra inside (2) and outside (1 & 3) vortex cores are highlighted in black. (f) High resolution conductance map acquired over the same area by scanning at fixed bias of V=1.4 mV; the tunneling current was set to 50 pA and modulation voltage was set to 150 µV with frequency of 2.3 kHz.

## 2.8 Filtering the STS conductance maps

For better visual depiction, the conductance maps obtained from STS are digitally filtered[22] to remove the noise and scan lines which arise from the raster motion of the tip. The filtering procedure is depicted in Fig. 2.17. Fig. 2.17(a) shows the raw conductance map obtained at 24 kOe. To filter the image we first obtain the 2D Fourier transform (FT) of the image Fig. 2.17(b). In addition to six bright spots corresponding to the Bragg peaks we observe a diffuse intensity at small $k$ corresponding to the random noise and a horizontal line corresponding to the scan lines. We first remove the noise and scan line contribution from the FT by suppressing the intensity along the horizontal line and the diffuse intensity within a circle at small $k$ (Fig. 2.17(c)). The filtered image shown in Fig. 2.17(d) is obtained by taking a reverse FT Fig. 2.17(c).

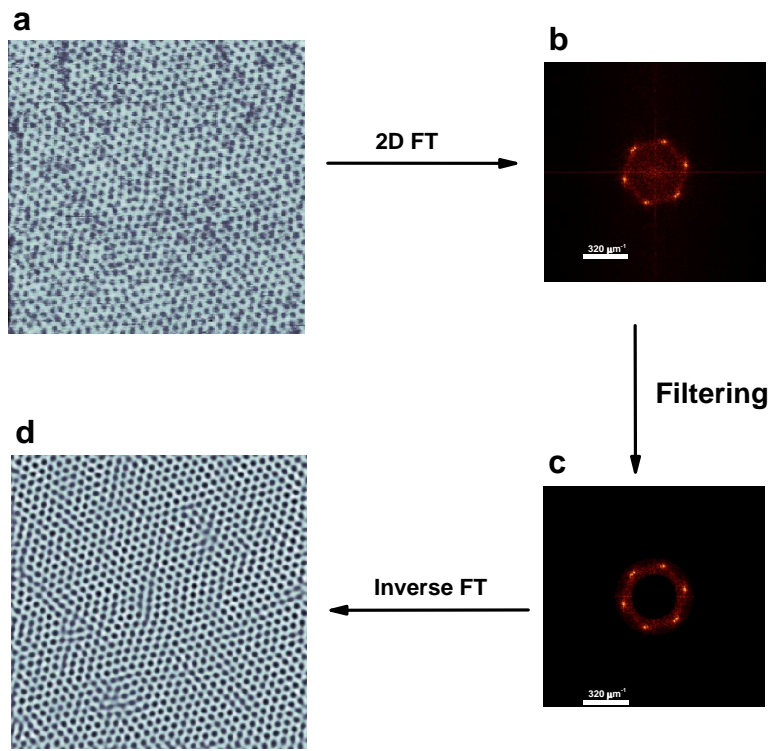

**Figure 2.17.** Filtering the conductance map. (a) Raw conductance map recorded at 24 kOe, 350 mK over an area of 1 µm × 1 µm; (b) 2D FT of (a); (c) 2D FT after removing the contribution from random noise at small $k$ and scan lines; (d) Filtered image obtained from the inverse FT of (c).

## 2.9 Determining the position of the vortices and Delaunay triangulation

To identify topological defects in the VL we need to determine the nearest neighbor coordination of each vortex. For this first the position of individual vortices are obtained from the local minima of the filtered image, generating a map of the center of each vortex (Fig. 2.18(a)-(b)). The position of each vortex is thus determined within the resolution of 1 pixel which for our images (256 × 256 pixel) is ~ 3.9 nm. Subsequently, the center map is Delaunay triangulated (Fig. 2.18(c)-(d)) such that there are no points inside the circumcircle of each triangle. This process generates a unique set of bonds connecting the center of vortices. Topological defects are identified from lattice points that are connected to smaller or larger than 6 nearest neighbors (normally 5 and 7 respectively).

An artifact of the Delaunay triangulation process is to generate some spurious bonds at the edge of the image as seen in Fig. 2.18(c). Thus in our analysis, we ignore all the bonds that are at the edge of the image.

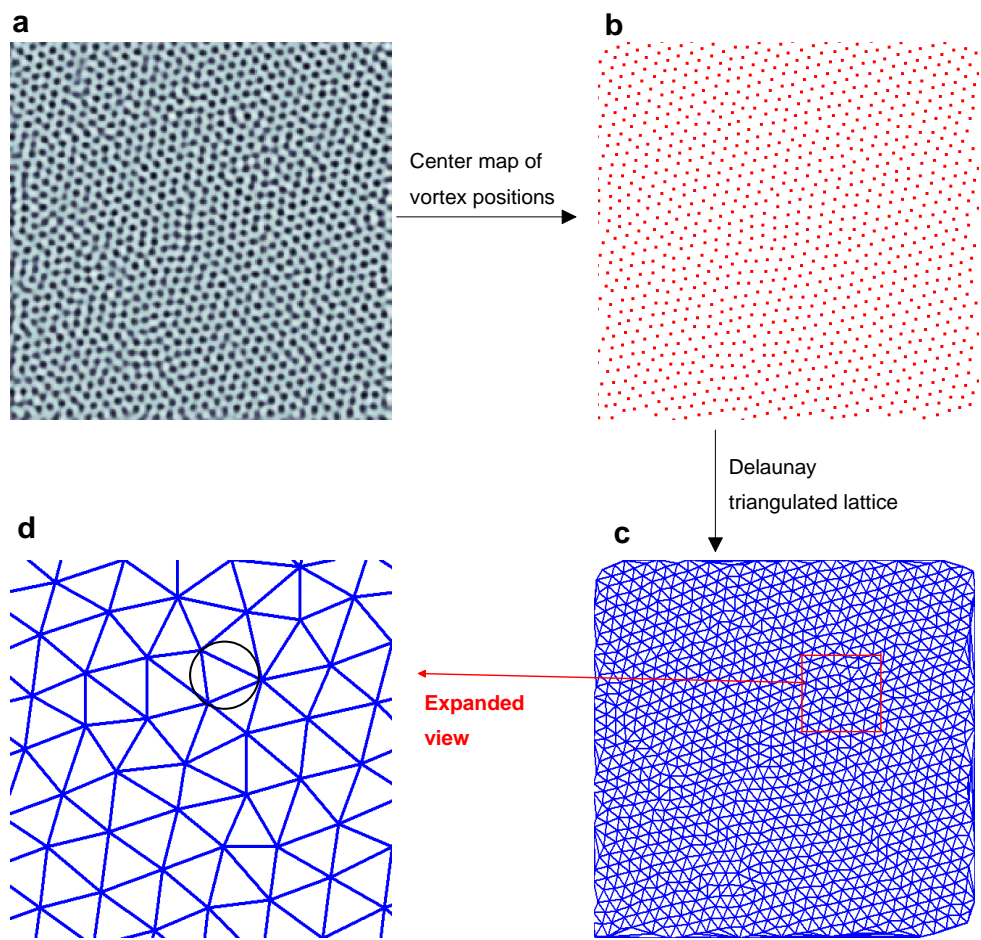

**Figure 2.18** Delaunay triangulation of the VL. (a) Filtered conductance map showing the ZFC VL at 24 kOe, 350 mK; (b) position of the center of each vortex obtained from the local minima of the conductance map; the size of the red dots is adjusted to reflect the uncertainty is position resulting for finite pixel resolution of our image. (c) Delaunay triangulation of the center map; (d) Expanded view of the Delaunay triangulated lattice showing a representative circumcircle enclosing a triangle.

## 2.10 Two coil ac susceptometer

The complex ac magnetic susceptibility is measured using two coil mutual inductance technique[23,24]. In the measurement set up, the sample is placed between a quadrupole primary coil and a dipole secondary coil. A small ac excitation current $I_d$ of frequency ω is passed through the primary coil and the induced voltage $V_p$ at the secondary coil is measured using lock-in amplifier. The mutual inductance, $M = M_1 + iM_2 = \frac{V_p}{\omega I_d}$ is measured as a function of

temperature/ magnetic field. The in phase part of the mutual inductance corresponds to the inductive coupling and the out of phase part corresponds to the resistive coupling between the coils. Close to $T_c$, the resistive part contributes and hence the response becomes complex (Fig. 2.20).

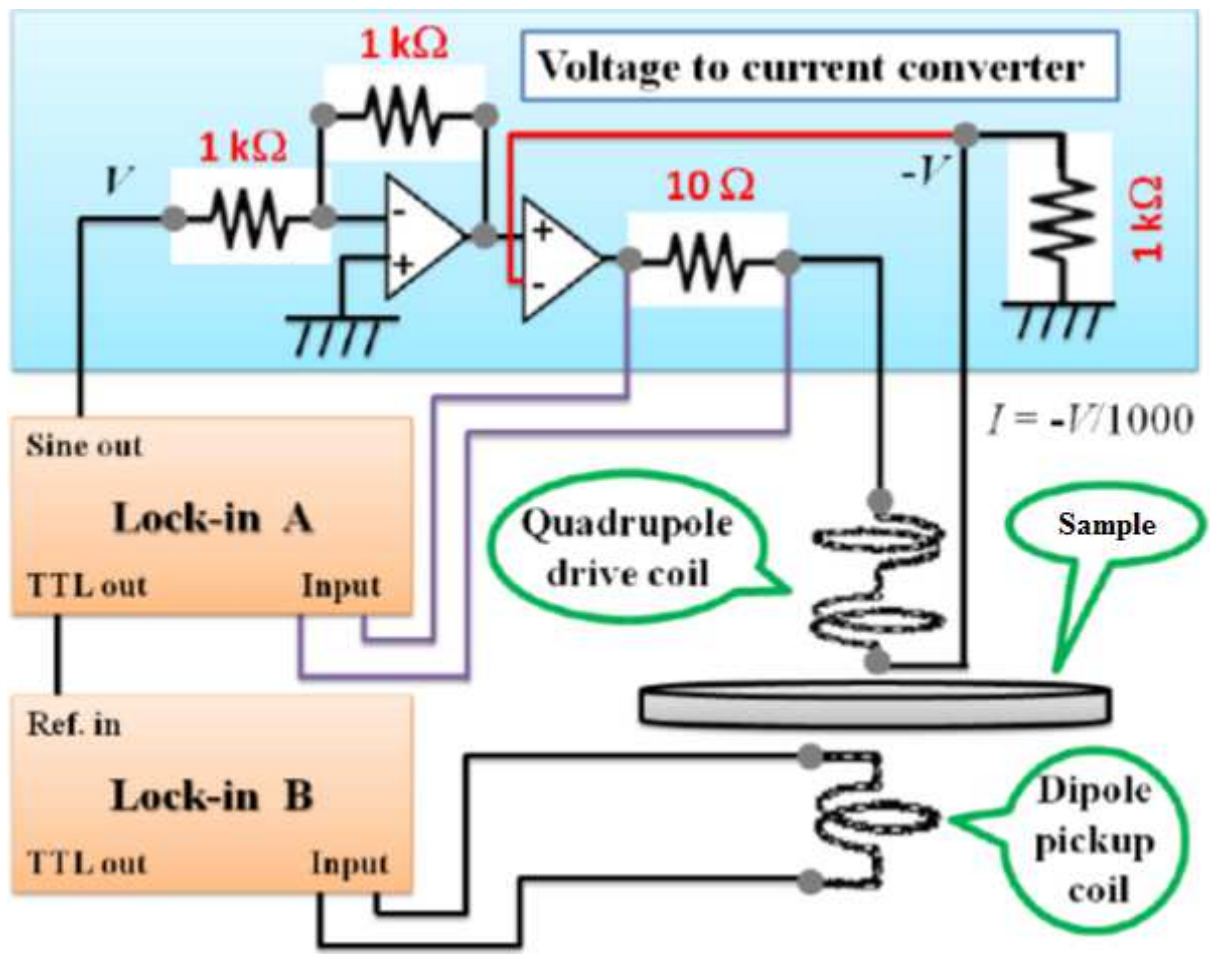

**Figure 2.19.** Schematic diagram of our two coil mutual inductance setup. Image courtesy[25].

The ac current is supplied from a current source followed by a lock in amplifier (Lock-in A in Fig. 2.19). The drive current is measured with a 10 Ω resistance in series with the primary coil and the pickup voltage is measured directly by another Lock-in (Lock-in B in Fig. 2.19). In our coil setup, the primary drive coil is attached with the cylindrical Macor sample housing with the help of thread arrangement. The secondary coil is pressed from bottom using a spring loaded piston made of Macor. The spring arrangement help us to hold the sample coaxially placed in between the drive and pickup coils and also takes care the thermal contraction of bobbins. The whole coil assembly is kept inside a heater can made of Cu in He gas environment which provides very good thermal equilibrium.

The primary drive coil (quadrupole) is coaxially centered above the film, has 28 turns in one layer. The secondary pickup coil (dipole) is coaxially centered below the film, has 120 turns in 4 layers. In the primary coil, the half of the coil nearer to the film is wound in one direction and the farther half is wound in the opposite direction. The radiuses of both drive and pickup coils are ~ 1.0 mm.

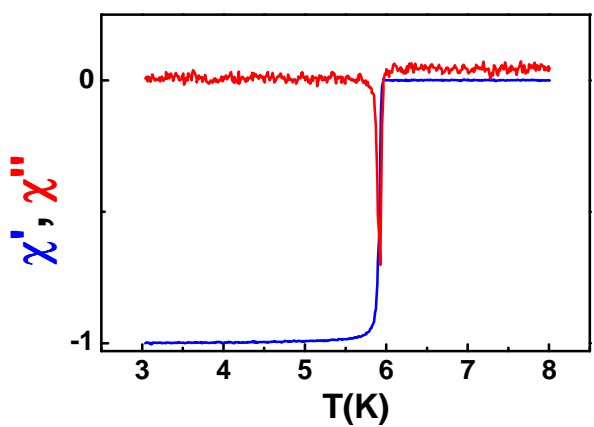

**Figure 2.20.** Temperature dependence of real and imaginary parts of a.c. susceptibility response of NbSe$_2$ in normalized to -1 and 0 in superconducting and normal states respectively. Near the superconducting transition, the response becomes complex.

## 2.10.1 a.c. susceptibility response as a function of excitation field

The a.c. susceptibility reported in Fig. 2.20 was performed with an a.c. excitation amplitude of 3.5 mOe at frequency 31 kHz. Since at large amplitudes, the a.c. drive can significantly modify the susceptibility response of the VL through large scale rearrangement of vortices, we performed several measurements with different a.c. excitation amplitudes to determine the range of a.c. field over which the χ' is independent of excitation field. We observe (Fig. 2.21) that below 3K, χ' shows significant dependence on the magnitude of the a.c. excitation only above 14 mOe.

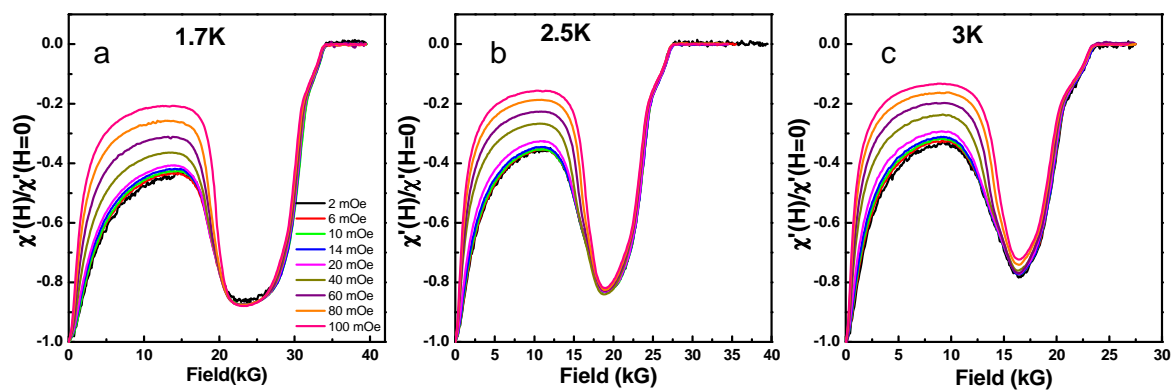

**Figure 2.21** Isothermal χ'-H scans for different a.c. excitations. The susceptibility response shows a strong amplitude dependence above 14 mOe.

## 2.11 Sample preparation and characterisation:

The samples used in this study consist of pure and Co-intercalated 2H-NbSe$_2$ single crystals. NbSe$_2$ is a conventional Type-II superconductor. The random intercalation of Co provides us a handle to control the degree of pinning. Earlier work[26] showed intercalated Co atoms apart from generating random pinning centres for the vortices, reduce anisotropy in the upper critical field compared to undoped NbSe$_2$, thereby making the vortex lines stiffer and hence less susceptible to bending. The Co$_{0.0075}$NbSe$_2$ single crystal was grown by iodine vapor transport method. Stoichiometric amounts of pure Nb, Se and Co together with iodine as the transport agent were mixed and placed in one end of a quartz tube, which was then evacuated and sealed. The sealed quartz tube was heated up in a two zone furnace between 5 to 10 days, with the charge-zone and growth-zone temperatures kept at, 800 °C and 720 °C respectively. We obtained single crystals with lateral size (in the a-b plane) of 2-5 mm and typical thickness varying between 60-150 µm. For the first set of crystals, we started with a nominal composition Co$_{0.0075}$NbSe$_2$ and the growth was continued for 5 days. We obtained single crystals with narrow distribution of $T_c$, in the range 5.3 – 5.93 K. The second set of crystals were also grown with the same nominal composition, but the growth was continued for 10 days. Here we obtained crystals with $T_c$ varying in the range 5.3 – 6.2 K. We conjecture that this larger variation of $T_c$ results from Co gradually depleting from the source such that crystal grown in later periods of the growth run have a slightly lower Co concentration. The third set of samples were pure 2H-NbSe$_2$ single crystal with $T_c$ ranging from 6.9 – 7.2 K.

Compositional analysis of few representative crystals from each batch was performed using energy dispersive x-ray analysis (EDX). We obtained a Co concentration of 0.45 atomic % for sample having $T_c$=5.88 K from the first set and 0.31 atomic % for sample having $T_c$=6.18 K from the second set. These compositions are marginally higher than the ones reported in ref.[25]. Since these measurements are close to the resolution limit of our EDX machine (below 1 atomic %), precise determination of the absolute value is difficult. However, measurements at various

Chapter 2

points on the crystals revealed uniform composition. This is also manifested by the sharp superconducting transitions observed from a.c. susceptibility in these crystals (Fig. 2.22).

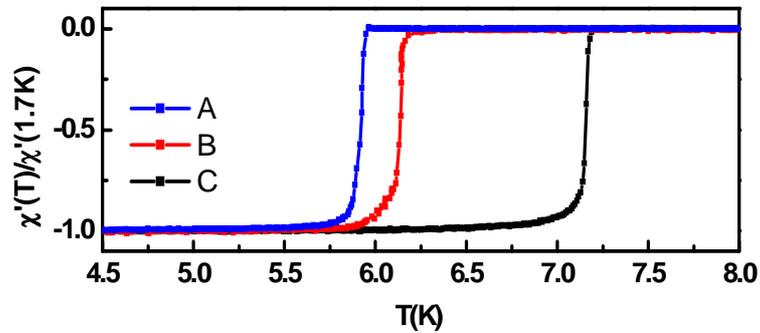

**Figure 2.22**. Temperature variation of χ' in a zero applied dc magnetic field for 3 samples having $T_c$=5.9 K (A), 6.18 K (B) and 7.2 K (C); χ' is normalised to -1 in superconducting state and 0 in normal state.

Chapter 2

---

[26] M. Iavarone, R. Di Capua, G. Karapetrov, A. E. Koshelev, D. Rosenmann, H. Claus, C. D. Malliakas, M. G. Kanatzidis, T. Nishizaki, and N. Kobayashi, Phys. Rev. B **78** 174518 (2008).

# Chapter 3

# Two step disordering of vortex lattice in weakly pinned Co -intercalated NbSe$_2$, presence of metastability

## 3.1 Introduction

In this chapter we shall establish the phase diagram of vortex lattice (VL) in a weakly pinned Type-II superconductor by studying its structural evolution in real space across the magnetic field driven peak effect. Our system of interest is Co-intercalated NbSe$_2$. NbSe$_2$ is very well known in the vortex physics community and is widely used for studying the VL phase diagram. It is extremely important to understand the structural evolution of the VL in a weakly pinned Type II superconductor since it determines superconducting properties that are directly relevant for applications like critical current and the onset of electrical resistance. In the past, there have been numerous attempts to understand the nature of the order-disorder transition of the VL with temperature or magnetic field[1,2,3]. It is generally accepted that in a clean system the hexagonal VL realised at low temperature and magnetic field, can transform to vortex liquid above a characteristic temperature (*T*) and magnetic field (*H*). Crystalline imperfections in a real superconducting system in the form of vacancies, impurty atoms etc. give rise to random pinning. It has been argued that in presence of this random pinning, the system can no longer sustain true long-range order and both the ordered and the disordered state become glassy[4,5], characterised by different degree of positional and orientational order. In addition, the VL can exist in a variety of non-equilibrium metastable states[6,7] depending on the thermomagnetic history of the sample.

Theoretically, for a 3D VL both the possibility of a direct first order transition from a BG to a vortex glass (VG) state[8,9] (with short range positional and orientational order) as well as transitions through an intermediate state, such as multi-domain glass or a hexatic glass[10,11], have been discussed in the literature. While many experiments find evidence of a first-order order-disorder transition[12,13,14,15], additional continuous transitions and crossovers have been reported in other regions[16,17,18] of the *H-T* parameter space, both in low-*T$_c$* conventional superconductors and in layered high-*T$_c$* cuprates.

Chapter 3

Experimentally, the most common technique to study the ODT in 3-D superconductors has been through bulk measurements, such as critical current[19], ac susceptibility[20,21] and dc magnetisation[22,23]. As a consequence of the presence of random pinning centres, the VL gets more strongly pinned to the crystal lattice as the perfect hexagonal order of the VL is relaxed[24], and hence ODT manifests as sudden non-monotonic enhancement of bulk pinning[25]. It is manifested as "peak effect" in the critical current and the diamagnetic response in ac susceptibility measurements. These bulk measurements are essential for establishing the phase diagram of type II superconductors. But, they do not reveal the evolution of the microscopic structure of the VL across the ODT. So in our work, we have used direct imaging of the VL using scanning tunnelling spectroscopy[26,27,28,29,30] (STS). The main challenge in this technique is to get large area images that are representative of the VL in the bulk crystal.

In this chapter, we will track the evolution of the equilibrium state of the VL across the magnetic field driven peak effect at low temperature using direct imaging of the VL using STS in an NbSe$_2$ single crystal, intercalated with 0.75% of Co. It has been shown[31] that the intercalated Co atoms have two competing effects: 1. generating random pinning centres for the vortices, 2. reducing the anisotropy in the upper critical field compared to undoped NbSe$_2$, thereby making the vortex lines (for $H \| c$ ) stiffer and hence less adaptable to bending to accommodate the point pinning centers of Co. Consequently, in weakly Co intercalated NbSe$_2$ single crystals the VL lattice becomes very weakly pinned as manifested by nearly vanishing critical current[31] at low magnetic fields and a pronounced peak effect close to $H_{c2}$.

## 3.2. Sample characterisation

The crystal on which the measurements were performed was characterised using 4-probe resistivity measurements from 286 K to 4 K (Fig. 3.1). The superconducting transition temperature, defined as the temperature where the resistance goes below our measurable limit is 5.3 K. The transition width, defined as the difference between temperatures where the resistance is 90% and 10% of the normal state resistance respectively is ~ 200 mK. The same crystal was used for both a.c. susceptibility and STS measurements.

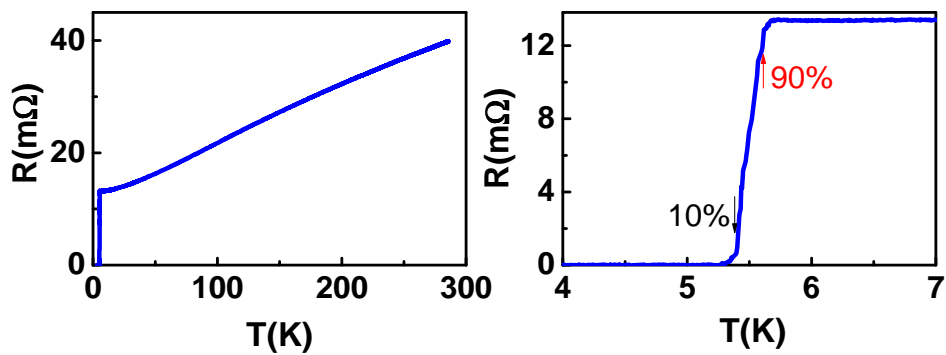

**Figure 3.1** Resistance vs. temperature for the $Co_{0.0075}NbSe_2$ sample. The left panel shows the resistivity from 286 - 4 K. The right panel shows an expanded view close to the superconducting transition. The arrows mark the positions where the resistance is 90% and 10% of the normal state value respectively.

## 3.3 Bulk pinning properties (a.c. suspectibility measurements)

a.c. susceptibility measurements were performed down to 350 mK using a home-built susceptometer, in a $^3$He cryostat fitted with a superconducting solenoid. Both ac and dc magnetic field were applied perpendicular to the Nb planes. The dc magnetic field was varied between 0-60 kOe. The ac excitation field was kept at 10 mOe where the susceptibility is in the linear response regime.

Fig. 3.2 (a) shows the bulk pinning response of the VL at 350 mK, measured from the real part of the linear ac susceptibility ($\chi'$) when the sample is cycled through different thermomagnetic histories. The $\chi'$-$H$ for the zero field cooled (ZFC) state (red line) is obtained while ramping up the magnetic field after cooling the sample to 350 mK in zero magnetic field. The "peak effect" manifests as a sudden increase in the diamagnetic response between 16 kOe ($H_p^{on}$) to 25 kOe ($H_p$) after which $\chi'$ monotonically increases up to $H_{c2} \sim 38$ kOe. When the magnetic field is ramped down after reaching a value $H > H_{c2}$ (black line, henceforth referred as the ramp down branch), we observe a hysteresis starting below $H_p$ and extending well below $H_p^{on}$. A much more disordered state of the VL with stronger diamagnetic response is obtained when the sample is field cooled (FC), by applying a field at 7 K and cooling the sample to 350 mK in the presence of the field (solid squares). This is however a non-equilibrium state: When the magnetic field is ramped up or ramped down from the pristine FC state, $\chi'$ merges with the ZFC branch or the ramp down branch respectively. In contrast, $\chi'$ for the ZFC state is reversible with magnetic field cycling up to $H_p^{on}$, suggesting that it is the more stable state of the system. Fig.

Chapter 3

3.2 (b) shows the phase diagram with $H_p^{on}$, $H_p$ and $H_{c2}$, obtained from isothermal χ'-H scans at different temperatures.

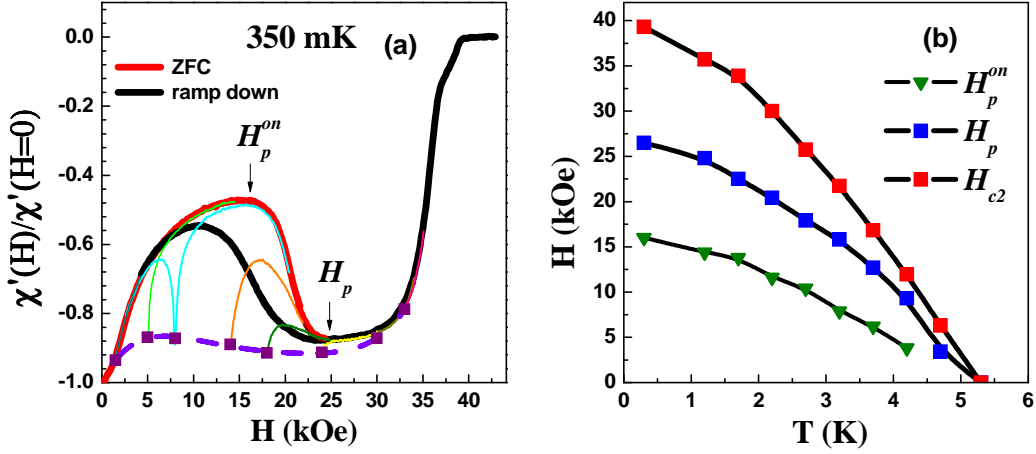

**Figure 3.2** (a) Magnetic field *(H)* dependence of the real part of linear ac susceptibility (χ') (normalised to its value in zero field) at 350 mK for the VL prepared using different thermomagnetic cycling. The red line is χ'-H when the magnetic field is slowly ramped up after cooling the sample in zero field (ZFC state). The black line is χ'-H when the magnetic field is ramped down from a value higher than $H_{c2}$. The square symbols stand for the χ' for the FC states obtained by cooling the sample from $T > T_c$ in the corresponding field; the dashed line shows the locus of these FC states created at different *H*. The thin lines starting from the square symbols show the evolution of χ' when the magnetic field is ramped up or ramped down (ramped down segment shown only for 0.8 T), after preparing the VL in the FC state. We observe that the FC state is extremely unstable to any perturbation in magnetic field. χ' is normalised to the zero field value for the ZFC state. (b) Phase diagram showing the temperature evolution of $H_p^{on}$, $H_p$ and $H_{c2}$ as a function of temperature.

## 3.4 Real space imaging of the VL using scanning tunneling spectroscopy

The VL was imaged using a home-built scanning tunneling microscope[32] (STM) operating down to 350 mK and fitted with an axial 90 kOe superconducting solenoid. Prior to STM measurements, the crystal is cleaved in-situ in vacuum, giving atomically smooth facets larger than 1 µm × 1 µm. Well resolved images of the VL are obtained by measuring the tunneling conductance (*G(V) = dI/dV*) over the surface at a fixed bias voltage (*V*~ 1.2 mV) close to the superconducting energy gap, such that each vortex core manifests as a local minimum in *G(V)*. Each image was acquired over 75 minutes after waiting for 15 minutes after stabilizing to the magnetic field. The precise position of the vortices are obtained from the images after digitally removing scan lines (see supplementary material) and finding the local minima in *G(V)* using WSxM software[33]. To identify topological defects, we Delaunay triangulated the VL and determined the nearest neighbor coordination for each flux lines. Topological defects in the hexagonal lattice manifest as points with 5-fold or 7-fold coordination number. Since, the Delaunay triangulation procedure gives some spurious bonds at the edge of the image we ignore the edge bonds while calculating the average lattice constants and identifying the topological defects. Unless otherwise mentioned VL images are taken over an area of 1 µm × 1 µm. The correlation functions $G_6$ and $G_K$ are calculated using individual images.

The VL is imaged using STS over a 1 µm × 1 µm area close to the center of the cleaved crystal surface. Figure 3.3, 3.4, 3.5, 3.6 show the conductance maps superposed with the Delaunay triangulated VL for 6 representative fields, when the magnetic field is ramped up at 350 mK in the ZFC state. The Fourier transforms (FT) corresponding to the unfiltered images are also shown. We identify 3 distinct regimes. For $H < H_p^{on}$, i.e. before the onset of peak effect where the shielding response monotonically increases, i.e. the sample becomes progressively less diamagnetic. In this region we have two representative points at H=10 KOe (not shown in the figure) and H=15 KOe (Fig 3) the VL is free from topological defects and the FT show 6 bright spots characteristic of a hexagonal lattice.

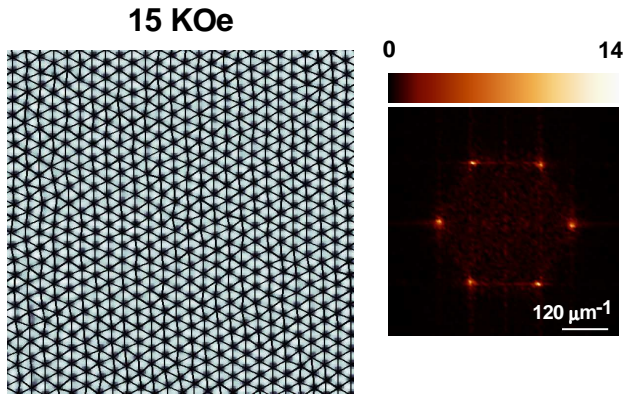

**Figure 3.3.** STS conductance maps showing real space ZFC vortex lattice image at 350 mK along with their Fourier transforms. Delaunay triangulation of the VL are shown as solid lines joining the vortices. Image is acquired over 1 µm × 1µm area, and have been zoomed to show around 600 vortices for clarity. The Fourier transform correspond to the unfiltered images; the colour scale is in arbitrary units.

Between $H_p^{on} < H \leq H_P$, i.e. at the region where the shielding response decreases non-monotonically. Here we have real space VL image at 20, 24 and 25 KOe (Fig 3.4(a), (b), (c)). At 20 KOe we observe a single vortex with 5-fold coordination and its nearest neighbour vortex having 7-fold coordination within our field of view. This defect-pair is called a dislocation. We

Chapter 3

have few more dislocations at 24 KOe. The number of dislocations increases with increasing magnetic field. This is reflected in the Fourier transform of the VL image also. As the field is increased the individual Fourier spots get broader. The dislocations do not completely destroy the orientational order which can be seen from FTs which continue to display a six-fold symmetry. We call this state an orientational glass (OG).

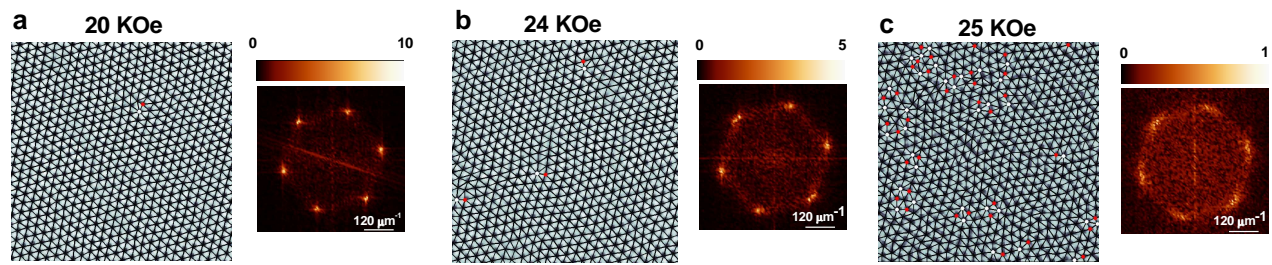

**Fig 3.4** (a)-(c) STS conductance maps showing real space ZFC vortex lattice image at 350 mK along with their Fourier transforms. Delaunay triangulation of the VL are shown as solid lines joining the vortices and sites with 5-fold and 7-fold coordination are shown as red and white dots respectively. Images are acquired over 1 µm × 1µm area, images shown here have been zoomed to show around 600 vortices for clarity. The Fourier transforms correspond to the unfiltered images; the color scales are in arbitrary units.

For $H > H_p$ at 26 KOe (Fig 3.5(a)) we observe that apart from 5-fold and 7-fold coordinated vortices appearing as nearest neighbours, there are individual 5-fold defects which have no corresponding 7-fold defect at nearest neighbour position and vice-versa. These individual defects are called disclination. Appearance of disclination reflect on the Fourier transform as it becomes isotropic with no clear maxima indicating broken orientational order. The number of disclinations also increase with increasing field as seen in 30 KOe data (Fig 3.5(b)). The disclinations proliferate in the system driving the VL into an isotropic VG. The FT shows a ring, characteristic of an isotropic disordered state. We observe a significant range of phase coexistence[34], where both large patches with dislocations coexist with isolated disclinations.

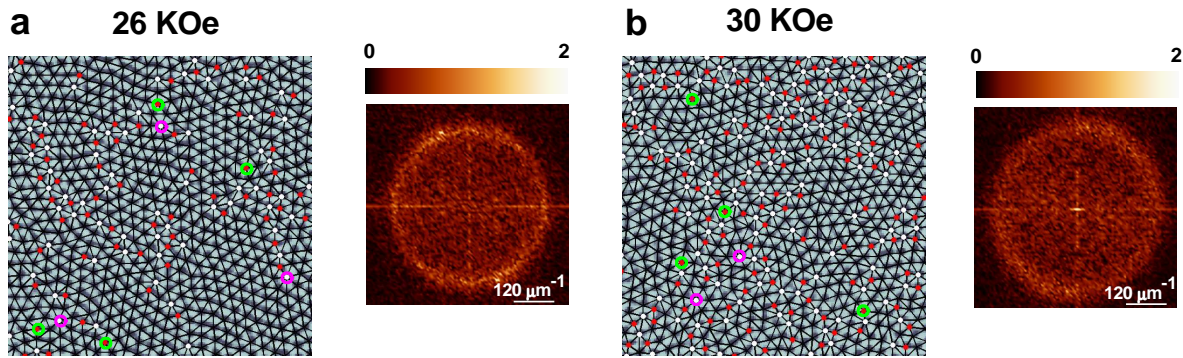

**Figure 3.5** (a), (b) STS conductance maps showing real space ZFC vortex lattice image at 350 mK along with their Fourier transforms. Delaunay triangulation of the VL are shown as solid lines joining the vortices and sites with 5-fold and 7-fold coordination are shown as red and white dots respectively. The disclinations (unpaired 5-fold or 7-fold coordination sites) observed at 26 and 30 kOe are highlighted with green and purple circles. While all images are acquired over 1 µm × 1µm area, images shown here have been zoomed to show around 600 vortices for clarity. The Fourier transforms correspond to the unfiltered images; the color scales are in arbitrary units.

Going to higher fields, 32 and 34 kOe (Fig. 3.6 (a)-(b)) we observe that the VL gets gradually blurred to form a randomly oriented linear structures, where the vortex lines start moving along preferred directions determined by the local surrounding. This indicates a softening of the VG and a gradual evolution towards the liquid state close to $H_{c2}$.

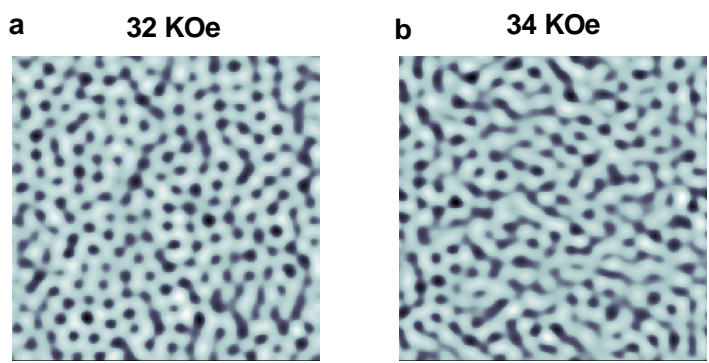

**Figure 3.6** (a), (b) STS conductance maps showing real space ZFC vortex lattice image at 350 mK. VL images are taken over (400 nm × 400 nm). At these fields the VL image becomes blurred due to the motion of vortices.

## 3.4.1 Correlation functions

Further quantitative information on this sequence of disordering is obtained from the orientational and positional correlation functions, $G_6(\bar{r})$ and $G_{\bar{K}}(\bar{r})$, which measure the degree of misalignment of the lattice vectors and the relative displacement between two vortices

Chapter 3

separated by distance $r$ respectively, with respect to the lattice vectors of an ideal hexagonal lattice. The orientational correlation function is defined as[28],

$$G_6(r) = (1/n(r,\Delta r)) \left( \sum_{i,j} \Theta\left(\frac{\Delta r}{2} - \left|r - \left|\bar{r}_i - \bar{r}_j\right|\right|\right) \cos 6(\theta(\bar{r}_i) - \theta(\bar{r}_j)) \right),$$ where $\Theta(r)$ is the Heaviside step function, $\theta(\bar{r}_i) - \theta(\bar{r}_j)$ is the angle between the bonds located at $\bar{r}_i$ and the bond located at $\bar{r}_j$, $n(r,\Delta r) = \sum_{i,j} \Theta\left(\frac{\Delta r}{2} - \left|r - \left|\bar{r}_i - \bar{r}_j\right|\right|\right)$, $\Delta r$ defines a small window of the size of the pixel around $r$ and the sums run over all the bonds. We define the position of each bond as the coordinate of the mid-point of the bond. Similarly, the spatial correlation function,

$$G_{\bar{K}}(r) = (1/N(r,\Delta r)) \left( \sum_{i,j} \Theta\left(\frac{\Delta r}{2} - \left|r - \left|\bar{R}_i - \bar{R}_j\right|\right|\right) \cos \bar{K} \cdot (\bar{R}_i - \bar{R}_j) \right),$$ where $\bm{K}$ is the reciprocal lattice vector obtained from the Fourier transform, $R_i$ is the position of the $i$-th vortex, $N(r,\Delta r) = \sum_{i,j} \Theta\left(\frac{\Delta r}{2} - \left|r - \left|\bar{R}_i - \bar{R}_j\right|\right|\right)$ and the sum runs over all lattice points. We restrict the range of $r$ to half the lateral size (1 µm) of each image, which corresponds to 11$a_0$ (where $a_0$ is the average lattice constant) at 10 kOe and 17$a_0$ for 30 kOe. For an ideal hexagonal lattice, $G_6(r)$ and $G_{\bar{K}}(r)$ shows sharp peaks with unity amplitude around 1$^{st}$, 2$^{nd}$, 3$^{rd}$ etc… nearest neighbour distance for the bonds and the lattice points respectively. As the lattice disorder increases, the amplitude of the peaks decay with distance and neighbouring peaks at large $r$ merge with each other.

Figure 3.7(a) and 3.7(b) show the $G_{\bar{K}}(r)$ (averaged over the 3 principal $\bm{K}$ directions) and $G_6(\bar{r})$, calculated from individual VL images, as a function of $r/a_0$ for different fields.

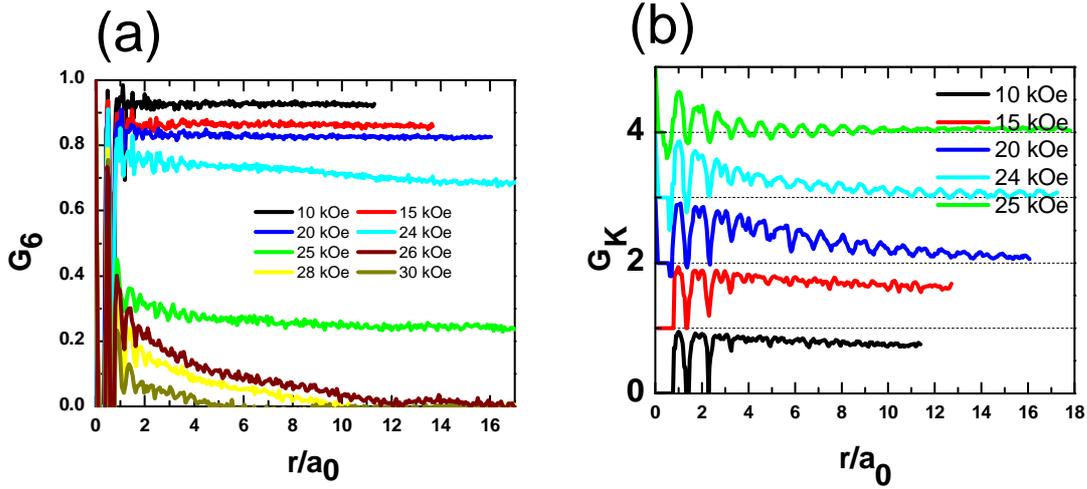

**Figure 3.7** (a) Orientational correlation function, $G_6$ and (b) and positional correlation function, $G_K$ (averaged over the principal symmetry directions) as a function of $r/a_0$ for the ZFC state at various fields. $a_0$ is calculated by averaging over all the bonds after Delaunay triangulating the image. In panel (b), $G_K$ for each successive fields are separated by adding a multiple of one for clarity.

At 10 kOe and 15 kOe, $G_6(r)$ saturates to a constant value of ~0.93 and ~0.86 respectively after 2-3 lattice constants, indicating that the presence of long-range orientational order. The envelope of $G_{\bar{K}}(r)$ decays slowly but almost linearly with $r$. Since the linear decay cannot continue for large $r$, this reflects our inability to capture the asymptotic behaviour at large $r$ at low fields due to limited field of view. While we cannot ascertain whether $G_{\bar{K}}(r)$ decays as a power-law for large $r$ as predicted for a BG, the slow decay of $G_{\bar{K}}(r)$ combined with the long-range orientational order is indicative of quasi long-range positional order (QLRPO). In the OG state (20-25 kOe), $G_6(r)$ decays slowly with increasing $r$. The decay of $G_6(r)$ with $r$ is consistent with a power-law ($G_6(r) \propto 1/r^\eta$), characteristic of quasi-long-range orientational order (Fig. 3.8(a)). $G_{\bar{K}}(r)$, on the other hand displays a more complex behaviour. At 20 kOe, within our field of view the $G_{\bar{K}}(r)$ envelope decays exponentially with positional decay length, $\xi_p \sim 6.7$. However for 24 and 25 kOe where the initial decay is faster, we observe that the exponential decay is actually restricted to small values of $r/a_0$ (Fig. 3.8(b)), whereas at higher values $G_{\bar{K}}(r)$ decays as a power-law (Fig. 3.8(c)). The OG state thus differs from the QLRPO state in that it does not have a true long-range orientational order. It also differs from the hexatic

Chapter 3

state in 2-D systems, where $G_{\bar{K}}(r)$ is expected to decay exponentially at large distance. Similar variation of $G_{\bar{K}}(r)$ has earlier been reported at intermediate fields in the VL of a neutron irradiated NbSe$_2$ single crystal[28], although the data in that case did not extend to the VG state.

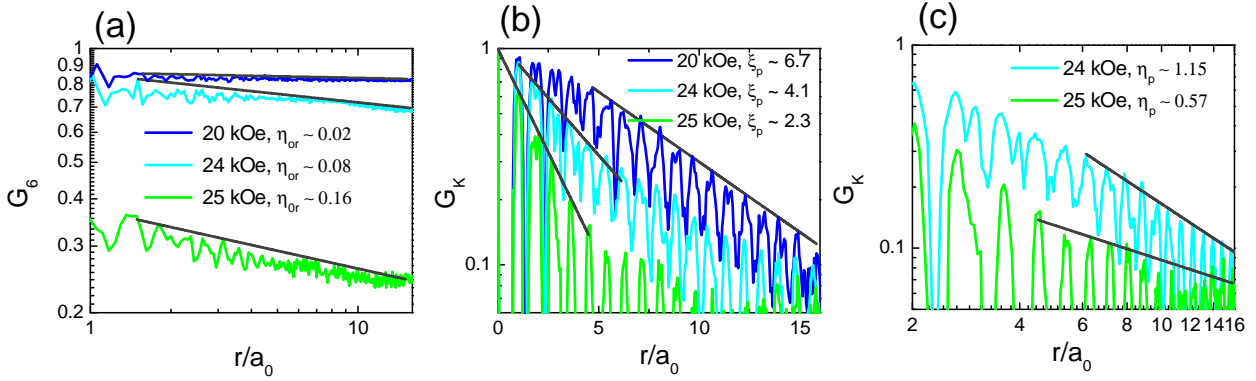

**Figure 3.8** (a) G$_6$ plotted in log-log scale for 20 kOe, 24 kOe and 25 kOe, along with fits (grey lines) of the power law decay of the envelope; η$_{or}$ are the exponents for power-law decay of G$_6$. (b) G$_K$ for 20 kOe, 24 kOe and 25 kOe plotted in semi-log scale along with the fits (grey lines) to the exponential decay of the envelope at short distance, with characteristic decay length, ξ$_P$. (c) G$_K$ for 24 and 25 kOe log-log scale along with fits (grey lines) to the power-law decay of the envelope at large distance; η$_P$ are the exponents for power-law decay of G$_K$.

Finally, above 26 kOe, $G_6(r) \propto e^{-r/\xi_{or}}$ (ξ$_{or}$ is the decay length of orientational order), giving rise to regular amorphous VG state with short-range positional and orientational order (Fig. 3.9).

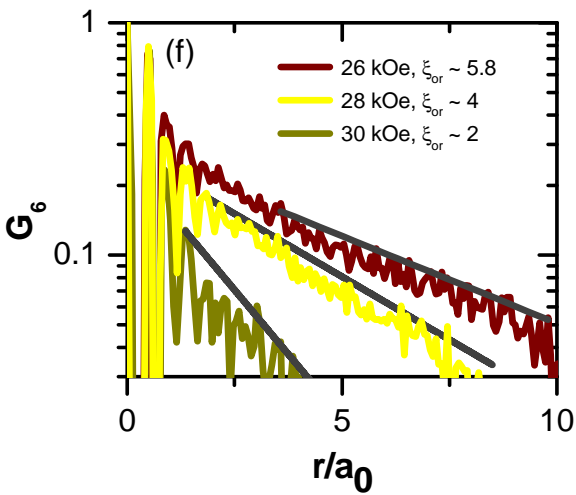

**Figure 3.9**. G$_6$ plotted in semi-log scale for 26 kOe, 28 kOe, 30 kOe along with the fits (grey lines) to the exponential decay of the envelope with characteristic decay length

## 3.4.2 Calculation of correlation lengths from Fourier transform

The possibility of a state with hexatic correlations between the onset and the peak of the "peak effect" has earlier been suggested from the field variation of the positional correlation length of the VL parallel ($\xi^{\|}$) and perpendicular ($\xi^{\perp}$) to the reciprocal lattice vector **K,** from neutron scattering studies in Nb single crystal[35]. In that measurement, $\xi^{\|}$ and $\xi^{\perp}$ was inferred from the radial width and azimuthal width of the six first order scattering in the Ewald sphere, projected on the plane of the detector[36]. While the relation between these correlation lengths and the ones obtained from the decay of $G_{\bar{K}}(r)$ or $G_6(r)$ is not straightforward, it is nevertheless instructive to compare the corresponding correlation lengths obtained from our data. For an infinite lattice the correlation lengths along and perpendicular to the reciprocal lattice vectors **K** can be obtained from the width of the first order Bragg peaks (BP) of the reciprocal lattice of each VL (obtained by taking the Fourier transform of a binary map constructed using the position of the vortices) using the relations, $\xi^{\|} = 1/\Delta k_{\|}$ and $\xi^{\perp} = 1/\Delta k_{\perp}$, where $\Delta k_{\|}$ and $\Delta k_{\perp}$ are the width of the first order Bragg peaks parallel and perpendicular to **K**. The 2D Fourier transform of binary image reveals the first order Bragg peaks which appear as six symmetric bright spots as well as higher order peaks (Fig. 3.10 (b)). The position of the first order peaks with respect to the origin correspond to the reciprocal lattice vector, **K**. We determine the peak profile along the azimuthal direction by taking the cut along a circle going through the center of the six first order Bragg spots (Fig. 3.10 (c)). Similarly, the profile along the radial direction is determined by taking a line cut along the direction of the reciprocal lattice vector (Fig. 3.10 (d)).

Three factors contribute to the peak width determined in this process: (i) The intrinsic disorder in the lattice, (ii) the finite size of the image and (iii) the positional uncertainty arising from the finite pixel resolution of our images. For each field we construct the binary image of an ideal hexagonal lattice of the same size with the same density of lattice points, where the position of each lattice point is rounded off to same accuracy as the pixel resolution of our image (256 × 256). The radial and the azimuthal width of the first order Bragg spots for this ideal lattice is determined using the same procedure as before (see Fig. 3.10(e)-(h)). In principle the experimental peak width is a convolution of intrinsic and extrinsic factors, and to correct for the contributions arising from extrinsic factors one needs to follow an elaborate deconvolution procedure. However, when the peak can be fitted with a pure Lorentzian function, the situation is simpler and the peak widths arising from different contributions are

Chapter 3

additive. We tried to fit peak profile averaged over the six symmetric BPs with both Lorentzian and Gaussian functions. The goodness of fit can be estimated from the residual sum of squares ($\chi$-square) and the coefficient of determination[37] (adjusted $R^2$) which should be 1 when the fit is perfect. We observe that the Lorentzian function gives a much better fit with smaller $\chi$-square and $R^2$ value close to unity. Since the Bragg peaks fit well to a Lorentzian profile, we correct for the peak broadening arising from the finite size of the images and the position uncertainly arising from the finite pixel size of the images by subtracting the peak width of the Bragg spots of the ideal lattice from the peak width obtained from the actual image to obtain $\Delta k_\perp$ and $\Delta k_{//}$ arising from the lattice disorder alone (Fig. 3.10(e), (f)).

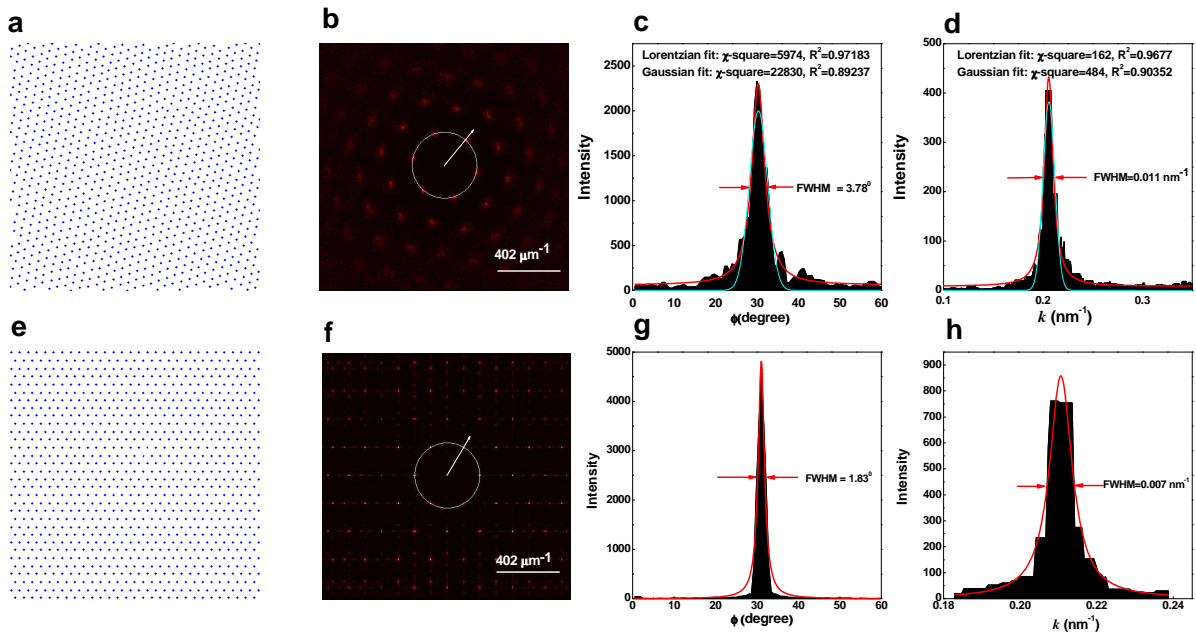

**Figure. 3.10** (a) Binary image of the vortex positions constructed from the VL image at 20 kOe. (b) 2D Fourier transform of (a) showing one reciprocal lattice vector, ***K***. (c) Azimuthal profile of the first order Bragg peak averaged over the six symmetric peak along the circle shown in (b); $\varphi$ is the azimuthal angle with respect to an arbitrary axis roughly midway between two Bragg peaks. (d) Radial profile of the First order Bragg peak along the reciprocal lattice vector averaged over the six symmetric peaks; $k_{//}$ is the magnitude of wave-vector along the reciprocal lattice vector. The lines show the fit to the Lorentzian (red) and the Gaussian (indigo) functions; the corresponding $\chi$-square and $R^2$ values are shown in the panel. (e)-(h) Same as (a)-(d) for the ideal hexagonal lattice of same size with the same density of lattice points.

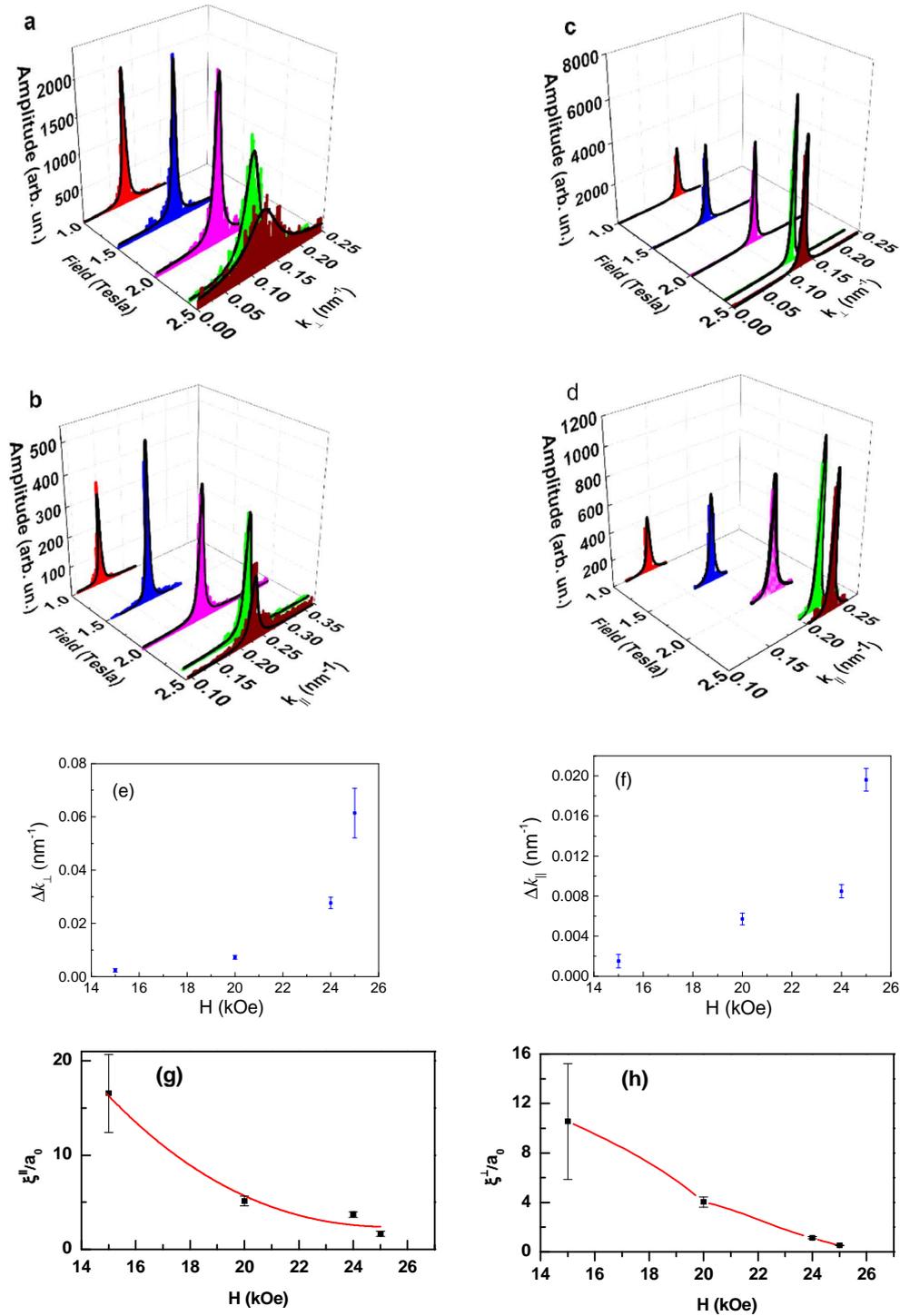

**Figure. 3.11** (a)-(b) Profiles of the first order Bragg peaks (averaged over the six symmetric peaks) along the radial and azimuthal directions at different fields obtained from the VL images. $k_\perp = \varphi_0$ (in radian) $\times |\mathbf{K}|$ is the distance in reciprocal space measured along the azimuthal direction. (c)-(d) Corresponding peak profiles for an ideal hexagonal lattice of the same size and same density of lattice points. The black lines are the fits to a Lorentzian function. (e)-(f) variation of $\Delta k_\perp$ and $\Delta k_\parallel$ with magnetic field. (g)-(h) Variation of the VL correlations lengths $\xi^\parallel$ and $\xi^\perp$ with magnetic field; below 15 kOe the correlation lengths reach the size of the image.

Fig. 3.11 (g)-(h) show the variation of $\xi^{\parallel} \sim 1/\Delta k_{\parallel}$ and $\xi^{\perp} \sim 1/\Delta k_{\perp}$ with magnetic field. We note that the evaluation of the correlations lengths is only valid up to the size of the image. Consequently, for our data $\xi^{\parallel}$ and $\xi^{\perp}$ are evaluated only for fields of 15 kOe and higher. We observe that $\xi^{\perp}/a_0$ and $\xi^{\parallel}/a_0$ decreases rapidly between $H_p^{on}$ and $H_p$ signalling a progressive decrease in orientational and positional order. At $H_P$, both $\xi^{\perp}/a_0$, $\xi^{\parallel}/a_0 \sim 1$, showing both positional and orientational order are completely lost. Both the magnitude and field dependence of $\xi^{\perp}$ and $\xi^{\parallel}$ are qualitatively consistent with ref. 35 even though the measurements here are performed at a much lower temperature.

## 3.5 Ramp down branch

We now discuss the ramp down branch focussing on the hysteresis region in the $\chi'$-$H$ measurements. Fig. 3.12 (a)-(c) shows the VL configurations for the ramp down branch at 25, 20 and 15 kOe. The VL structures for the ramp down branch are similar to ZFC: At 25 and 20 kOe the VL shows the presence of dislocations and at 15 kOe it is topologically ordered. In Fig. 3.12 (d)-(f) we compare $G_6(r)$ and $G_{\bar{K}}(r)$ calculated for the ZFC and the ramp down branch. At 25 kOe $\approx H_P$, we observe that $G_{\bar{K}}(r)$ for ZFC and ramp down branch are similar whereas $G_6(r)$ decays faster for the ramp down branch. However, analysis of the data shows that in both cases $G_6(r)$ decays as a power-law (Fig. 3.12 (f)) characteristic of the OG state. At 15 kOe, which is just below $H_p^{on}$, both ZFC and ramp down branch show long-range orientational order, while $G_{\bar{K}}(r)$ decays marginally faster for the ramp down branch. Thus, while the VL in the ramp down branch is more disordered, our data do not provide any evidence of supercooling across either QLRPO→OG or OG→VG transitions as expected for a first order phase transition. Therefore, we attribute the hysteresis to the inability of the VL to fully relax below the transition in the ramp down branch.

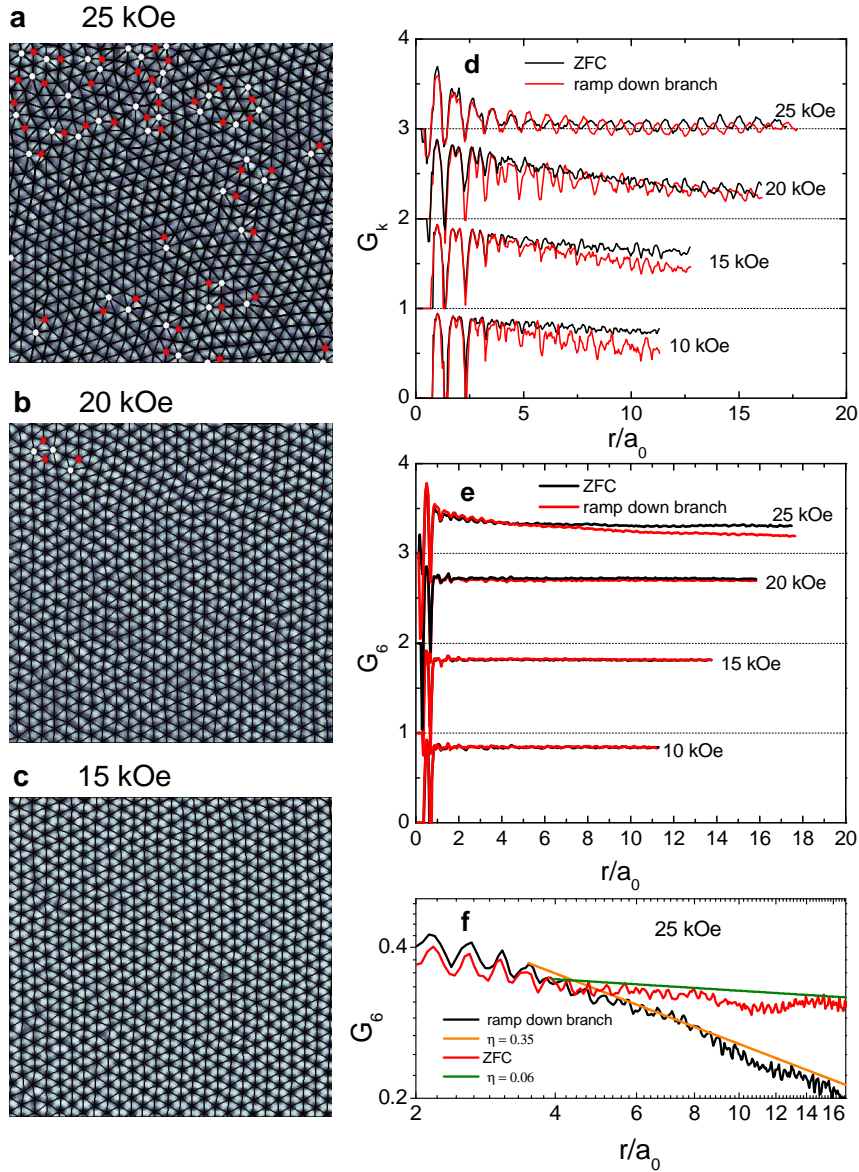

**Figure 3.12** (a)-(c) VL images at 350 mK along with Delaunay triangulation for the ramp down branch at 25, 20 and 15 kOe. Sites with 5-fold and 7-fold coordination are shown as red and white dots respectively. At 25 and 20 kOe we observe dislocations in the VL. At 15 kOe the VL is topologically ordered. While all images are acquired over 1 μm × 1μm area, images shown here have been zoomed to show around 600 vortices for clarity. (d) $G_6$ and (e) $G_K$ calculated from the VL images at different field when the magnetic field is ramped down from $H > H_{c2}$ at 350 mK. The curves for each successive field are separated by adding a multiple of one for clarity. (f) $G_6$ at 25 kOe for ZFC and ramp down branch along with the fit (solid lines) to the power-law decay of the corresponding envelope; the exponents for the power-law fits, η, are shown in the legend.

Chapter 3

## 3.6 The field cooled branch

We now follow the magnetic field evolution of the FC state (Fig. 3.13). The FC state is created by cooling the sample from 7K (>$T_c$) in the presence of applied field. The real space images reveal the FC states to be much more disordered than the ZFC states at corresponding fields. The FC state show an OG at 10 kOe (not shown) and 15 kOe with free dislocations (Fig. 3.13(a)), and a VG above 20 kOe with free disclinations (Fig. 3.13 (b)-(d)). The FC OG state is however extremely unstable. This is readily seen by applying a small magnetic pulse (by ramping up the field by a small amount and ramping back), which annihilates the dislocations in the FC OG (Fig. 3.14) eventually causing a dynamic transition to the QLRPO state. It is interesting to note that metastability of the VL persists above $H_p$ where the ZFC state is a VG. The FC state is more disordered with a faster decay in $G_6(r)$ (Fig. 3.13 (e), (f)), and consequently is more strongly pinned than the ZFC state.

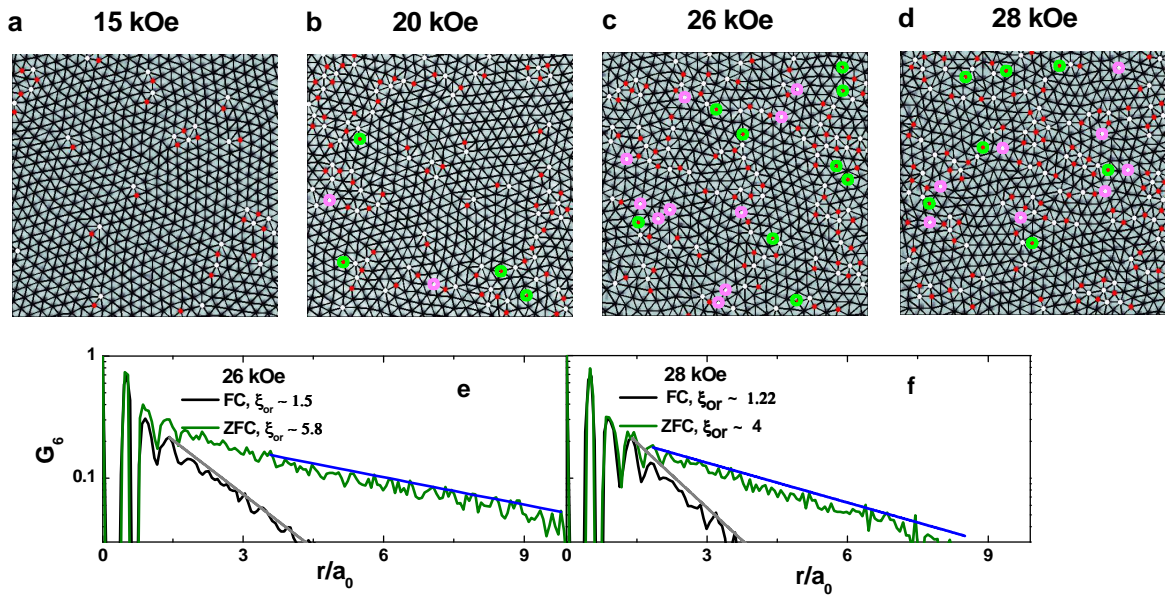

**Figure 3.13 (a)-(d)** Field Cooled VL images (upper panels) at 15 kOe, 20 kOe, 26 kOe, 28 kOe at 350 mK. Delaunay triangulation of the VL are shown as solid lines joining the vortices and sites with 5-fold and 7-fold coordination are shown as red and white dots respectively. The disclinations are highlighted with green and purple circles. **(e), (f)** the variation of $G_6$ for the FC and ZFC states at 26 kOe and 28 kOe along with the corresponding fits for exponential decay (solid lines); the decay lengths for the orientational order, $\xi_{or}$ are shown in the legend. Images have been zoomed to show around 600 vortices for clarity.

## 3.6.1 Metastability of the FC state

Here we demonstrate the metastability of the FC state by applying a dc magnetic perturbation and observing the resulting state. VL is created in FC protocol at 15 KOe at 350 mK. The VL is in OG phase with free dislocations (Fig. 3.14 (a)). Now, the VL is perturbed at 350 mK with a magnetic field pulse of 300 Oe by ramping up the field to 15.3 kOe over 8 sec followed by a dwell time of 10 sec and ramping down over 8 sec to its original value (15 KOe). We observe that the resultant state still has free dislocations but the number of free dislocations has reduced greatly (Fig. 3.14 (b)). If we apply a larger field pulse of 900 Oe, we observe that the resultant state has become completely ordered with no free dislocations (Fig. 3.14 (c)).

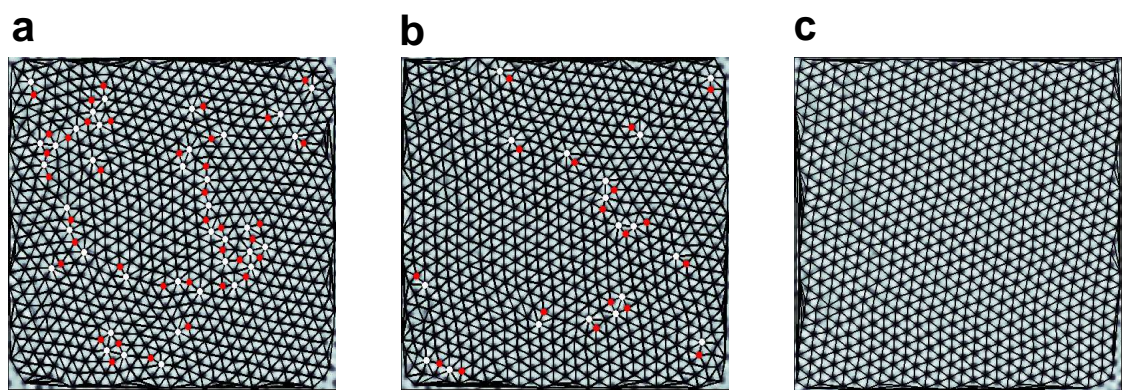

**Figure 3.14** Annihilation of dislocations in a FC vortex lattice at 15 kOe with application of magnetic field pulse. (a) FC VL at 350 mK; the same VL after a applying a magnetic field pulse of (b) 0.3 kOe and (c) 0.9 kOe. Dislocations in the VL are shown as pairs of adjacent points with five-fold (red) and seven-fold (white) coordination. In (b) many of the dislocations are annihilated, whereas in (c) all dislocations are annihilated. Delaunay triangulation of the VL are shown as solid lines joining the vortices. Images have been zoomed to show around 600 vortices for clarity.

## 3.7 Discussion

To summarise, our real space VL images demonstrate that across the peak effect, the VL in a weakly pinned Type II superconductor disorders through gradual destruction of positional and orientational order. The sequence of disordering observed in our experiment is reminiscent of the two-step melting observed in 2-D systems[38], where a hexatic fluid exists as

# Chapter 3

an intermediate state between the solid and the isotropic liquid. However, the situation for the weakly pinned 3-D VL is more complex. Here, the reduced influence of thermal fluctuations prevents from establishing a fluid state in the presence of a pinning potential. Thus, the OG and the VG state are both glassy states with very slow kinetics, which appear completely frozen within the timescales of our experiment, till very close to $H_{c2}$. The glassy nature of the VL also manifests by producing a number of non-equilibrium states, such as the OG below $H_p^{on}$ and the VG state below $H_p$ when the sample is field cooled.

The sequence of disordering observed in our experiment also bears some similarity with the phase diagram proposed in ref. 36 where the BG disorders to a VG through an intermediate "multidomain glass" (MG) state. However, in a MG the spatial correlations are similar to a BG up to a characteristic length-scale followed by a more rapid decay at larger distance. In contrast, in the intermediate OG state observed in our experiment, $G_{\bar{K}}(r)$ decays rapidly at short distance followed by a more gradual behaviour. The possibility of the BG transforming to an intermediate state that retains hexatic correlations, which would be closer to our experimental findings, has also been proposed as a possibility[10,39], but has not been explored in detail.

Our isothermal field hysteresis measurement across the phase boundaries did not provide any evidence for superheating/ supercooling. It might be due to the fact that as field ramping changes the density of vortices and thereby driving the VL into equilibrium state, we are not able to observe metastable superheated/ supercooled states. So, we need to perform thermal hysteresis measurements at a constant field across the phase boundaries. This is discussed in detail in chapter 4. Finally, the gradual softening of the VG as the magnetic field approaches $H_{c2}$ supports the view[35] that the VG and the vortex liquid are thermodynamically identical states in two different limits of viscosity.

One relevent question in this context, is the role of the coupling of the VL the crystal lattice, on the emergence of OG state. When the symmetry of the crystal lattice is not hexagonal such coupling has been shown to give rise to non-trivial distortion of the hexagonal vortex lattice[40]. In our case however, the crystal lattice has a hexagonal symmetry in the plane perpendicular to the magnetic field. It is therefore possible that orientational pinning of the VL to the crystal lattice is an important factor contributing to the stability of the orientational order in OG state and has to be explored in detail. This is explored in detail in chapter 5.

Chapter 3

# Chapter 4

# Structural evidence of disorder induced two-step melting of vortex matter in Co-intercalated NbSe$_2$ single crystals from superheating and supercooling

## 4.1. Introduction

In chapter 3, I have discussed about STS imaging of the VL across the ODT in a Co$_{0.0075}$NbSe$_2$ single crystal[1]. It revealed that the hexagonal ordered state (OS) of the VL disorders in two steps, reminiscent of the Berezinski-Kosterlitz-Thouless-Halperin-Nelson-Young (BKTHNY) transition[2] in two-dimensional solids. At a fixed temperature as the magnetic field is increased, dislocations, in the form of nearest-neighbor pairs with 5-fold and 7-fold coordination, first proliferate in the VL. We call this state an orientational glass (OG). At a higher field dislocations dissociate into isolated disclinations driving the VL into an amorphous vortex glass (VG). These three states are characterized by their positional and orientational order. In the OS, the VL has long-range or quasi long-range positional order and long-range orientational order. The OG is characterized by a rapid decay of positional order, but a (quasi)-long range orientational order analogous to the hexatic state in 2-dimensional (2D) solids. In the VG both positional and orientational order are short range. A somewhat different two-step disordering sequence has also been reported in neutron irradiated NbSe$_2$ single crystals[3], though in that case the study was restricted to low fields. In the presence of weak or moderate pinning several studies find signatures of a thermodynamic first-order ODT[4,5,6,7,8]. However, experimental investigations on extremely pure Nb single crystals did not find evidence[9,10] of VL melting below the upper critical field ($H_{c2}$). Since for our system, which is a 3-dimensional VL BKTHNY transition is not expected, it is important to investigate whether these two transformations correspond to two distinct phase transitions, or a gradual cross-over as suggested by some authors[11,12,13].

An alternative viewpoint to the thermal route to melting is the disorder induced ODT originally proposed by Vinokur et al.[14]. Here it was speculated that in the presence of weak pinning the transition can be driven by point disorder rather than temperature. In this scenario topological defects proliferate in the VL through the local tilt of vortices caused by point

Chapter 4

disorder, creating an "entangled solid" of vortex lines. The key difference with conventional thermal melting is that here, the positional entropy generates instability in the ordered VL driving it into a disordered state, even when thermal excitation alone is not sufficient to induce a phase transition. Recently, similar notions have also been extended to the melting of vortex lattice in Bose-Einstein condensates formed of ultra-cold atoms in a disordered optical potential[15]. There have also been direct evidence of melting induced by random pinning potential in two-dimensional colloidal crystal[16].

In this chapter, I shall address the ODT in Co-intercalated $NbSe_2$ by tracking the structural evolution of the VL, imaged using STS. Our data provides structural evidence of superheating and supercooling across both OS-OG and OG-VG transitions. Furthermore, comparing crystals with different degree of pinning, we show that these two transitions come closer to each other when pinning is reduced, suggesting that they are fundamentally associated with the random background potential created by random pinning.

## 4.2. Sample characterization

We have performed bulk measurements using 2-coil ac susceptibility and bulk magnetization measurements (SQUID). The samples used in this study consist of pure and Co-intercalated $NbSe_2$ single crystals[17] grown through iodine vapor transport in sealed quartz ampules. The random intercalation of Co provides us a handle to control the degree of pinning[18]. For the first crystal on which all the STS measurements were carried out, we started with a nominal composition $Co_{0.0075}NbSe_2$ and the growth was continued for 5 days. We obtained single crystals with narrow distribution of $T_c$, in the range 5.3 – 5.93 K. The crystal chosen for our studies had a $T_c$ ~ 5.88 K (sample A). The second crystal was also grown starting from the same nominal composition (in a different ampule), but the growth was continued for 10 days. Here we obtained crystals with $T_c$ varying in the range 5.3 – 6.2 K. We conjecture that this larger variation of $T_c$ results from Co gradually depleting from the source such that crystal grown in later periods of the growth run have a lower Co concentration. However, over the 2mm × 2mm crystal chosen from this growth run ($T_c$ ~ 6.18 K) (sample B), we did not see significant compositional variation. The third sample was a pure $NbSe_2$ single crystal with $T_c$ ~ 7.25 K (sample C). Compositional analysis of the three crystals was performed using energy dispersive x-ray analysis (EDX). We obtained a Co concentration of 0.45 atomic % for A and 0.31 atomic % for B. While these Co concentrations are marginally higher than the ones reported in ref. 18, we note that these measurements are close to the resolution limit of our EDX machine, where precise determination of the absolute value is difficult. However, measurements at various points on the crystals revealed the composition to be uniform, which is also validated by the sharp superconducting transitions observed from a.c. susceptibility in these crystals (Fig. 4.1).

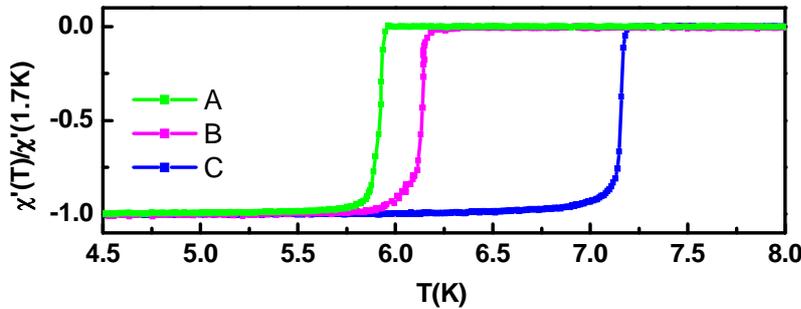

**Figure 4.1.** Temperature variation of $\chi'$ in zero applied d.c magnetic field for samples A, B and C.

## 4.2.1 Critical current density measurement using SQUID magnetometer

The bulk pinning properties at low temperature are characterized through d.c. magnetization measurements using a Quantum Design SQUID magnetometer. The magnetic field in all measurements reported in this paper is applied along the *c*-axis of the hexagonal crystal structure. Figure 4.2 (a) shows the representative hysteresis-loop in the d.c. magnetization (*M-H*) measured at 1.8 K for the sample A. The hysteresis first collapses below our resolution limit at fields above 5 kOe and then opens up showing a bubble close to $H_{c2}$. This reopening of the hysteresis curve signals a sudden anomalous increase in the critical current, the "peak effect" [19], which is associated with the ODT of the VL. We estimate the critical current ($J_c$) as a function of magnetic field, using the critical state model[20] which relates the width of the hysteresis loop ($\Delta M$) to $J_c$ of the superconductor, through the relation, $J_c \approx 20\Delta M/d$ where $J_c$ is in units of A/cm$^2$, $\Delta M$ is in units of emu/cm$^3$ and *d* is the lateral dimension perpendicular to the applied magnetic field of the crystal in cm. Fig. 4.2 (b) shows $J_c(H)$ at 1.8 K for the three crystals under investigation. While the absolute value is likely to have some error due to demagnetization effects since the magnetic field is applied along the short dimension of the crystal, we observe that the peak in $J_c$ progressively increases for samples with lower $T_c$ showing that the pinning becomes progressively stronger as Co is introduced. However, we note that even our most disordered sample (A) is in the weak pinning limit, which is

Chapter 4

functionally defined as the pinning range where a topologically ordered vortex lattice is realized at low temperatures and low fields.

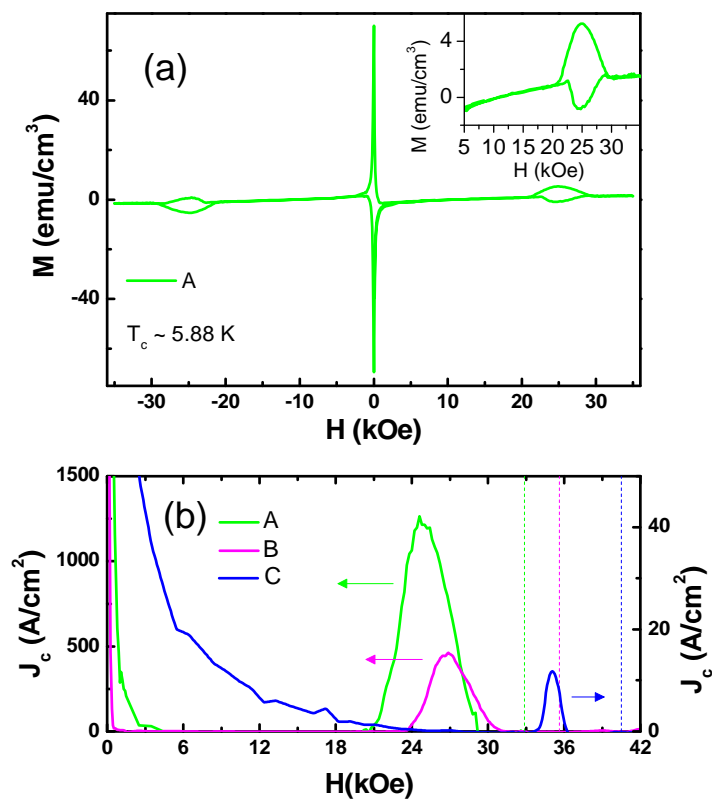

**Figure 4.2.** (a) 5-quadrant *M-H* loop for the sample A at 1.8 K; the inset shows the expanded view of the peak effect. (b) Variation of $J_c$ with magnetic field at 1.8 K for the crystals A, B, C ; the vertical dashed lines show the upper critical field for the 3 crystals, at the same temperature.

## 4.3 Phase diagram

The phase diagram for the three crystals is established from variation of the real part of a.c. susceptibility with magnetic field ($\chi'$-*H*) at different temperatures in a home built a.c. susceptometer (Fig. 4.3). We have earlier shown in chapter 2 that above 14 mOe of a.c. excitation field the $\chi'$ response becomes non-linear. Here we also observed that in the non-linear regime the onset of the peak effect moves to higher fields with increase in excitation. We believe that this is a consequence of cycling the sample through minor hysteresis loops resulting from the oscillatory a.c. excitation. Consequently in all these measurements we fix the a.c. excitation field to 3.5 mOe (frequency of 31 kHz) which is well within the linear regime. The sample is first cooled in zero magnetic field across the superconducting transition (the zero field cooled (ZFC) protocol) before the magnetic field is applied. $\chi'$-*H* (*inset* Fig. 3 (a-c)) shows a characteristic drop with magnetic field below $H_{c2}$, the peak effect, signaling the ODT of the VL. Correlating with real space STS images, we earlier identified two characteristic fields[1]: The field at which χ' starts to decrease (defined as the onset of peak effect, $H_p^{on}$), where dislocations start proliferating into the VL and the peak of the peak effect ($H_p$) where disclinations start proliferating in the VL. Tracking the loci of $H_p^{on}$ and $H_p$ obtained from isothermal χ'-$H$ scans at different temperatures, we obtain two lines in the $H$-$T$ parameter space demarcating the regions of existence of an OS, the OG and the VG.

We observe that region in the $H$-$T$ space where we observe the OG gradually shrinks as pinning is reduced. In this context we would like to note that the onset of the peak effect in χ' is lower than the corresponding onset of the peak effect from d.c. magnetization measurements, a feature which has also been reported in other weakly pinned Type II superconductors[21]. While the reason for this difference is still not completely understood, one likely reason is the inhomogeneity in the superconducting magnet. Unlike χ' measurements, where the sample is held at a fixed position, in conventional d.c. magnetization measurements, the sample in moved over a distance inside the pick-up coils. In this process the sample is cycled through minor hysteresis loops dictated by the inhomogeneity in magnetic field in the superconducting magnet which tends to collapse the magnetic hysteresis when $J_c$ is small (for a scan length of 3 cm used in the Quantum Design SQUID, we estimate the magnetic field inhomogeneity experienced by the sample at 15 kOe to be of the order of 9 Oe)[22,23]. $H_p$ on the other hand is the same in both measurements, since the critical current is large at this field and the magnet inhomogeneity has very little effect.

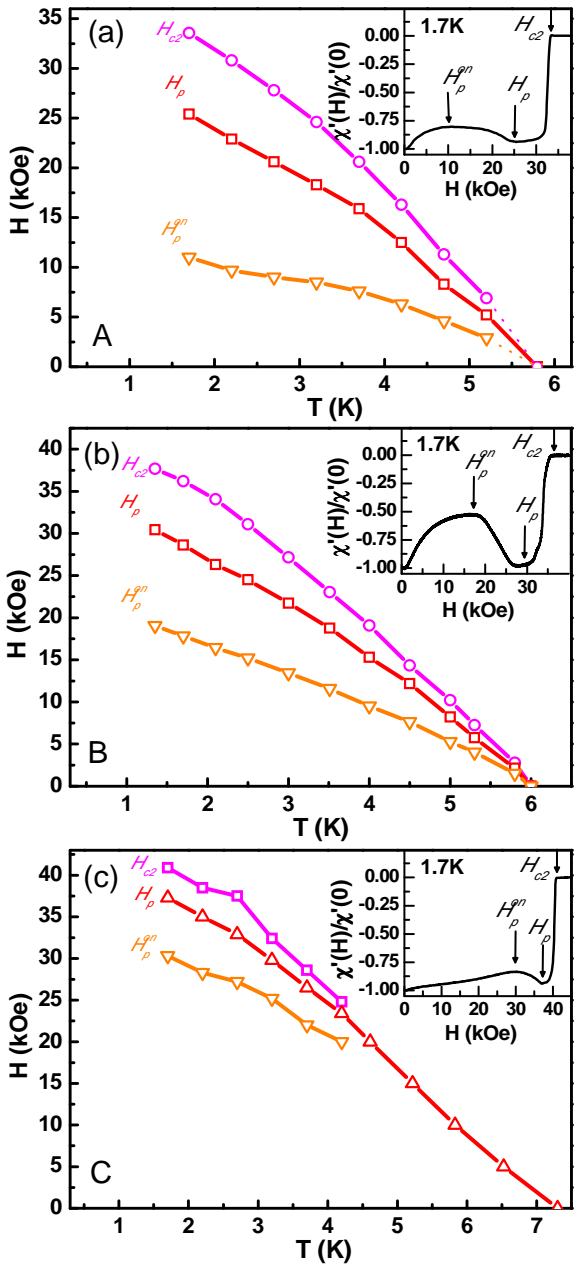

**Figure 4.3.** (a)-(c) Phase diagrams of pure and Co intercalated NbSe$_2$ crystals (A, B, C), showing the variation of $H_p^{on}$, $H_p$ and $H_{c2}$ as a function of temperature; the solid lines connecting the points are guides to the eye. Below $H_p^{on}$ the VL is in a topologically ordered state. Between $H_p^{on}$ and $H_p$ the dislocations proliferate in the VL. Between $H_p$ and $H_{c2}$ disclinations proliferate in the VL. The *insets* show χ'(H)/χ'(0) as a function of *H* at 1.7 K; the onset of the peak effect $H_p^{on}$, the peak of the peak effect $H_p$ and the upper critical field $H_{c2}$ are marked with arrows.

## 4.4 Real space investigations of the VL from STS imaging

We now investigate the thermal history dependence of the VL by imaging the VL using STS imaging across the OS→OG and OG→VG transitions. We focus on the crystal A for which these two boundaries are well separated in the *H-T* parameter space. STS measurements are performed using our home built low temperature scanning tunneling microscope[24] operating down to 350 mK and fitted with a 90 kOe superconducting solenoid. Prior to STS measurements the crystal is cleaved in-situ exposing atomically smooth facets in the *a-b* plane, several microns in size. The VL is imaged by measuring the tunneling conductance over the sample surface $\left(G(V) = \left.\dfrac{dI}{dV}\right|_V\right)$ at fixed bias voltage ($V = 1.2$ mV) close to the superconducting energy gap, such that each vortex manifest as a local minimum in the tunneling conductance. Topological defects in the VL are identified by first Delaunay triangulating the VL and finding the nearest neighbor coordination for each point. The magnetic field is applied along the six-fold symmetric *c*-axis of the hexagonal NbSe$_2$ crystal.

## 4.4.1 Thermal hysteresis across the OS➔OG phase boundary

We first concentrate on the boundary separating the OS and OG. Figure 4.4 indicates the thermal hysteresis path. The upper panels of Fig. 4.5 show the images of the VL acquired at different points during the temperature cycling and the lower panels show the corresponding autocorrelation functions defined as, $G(\bar{r}) = \sum_{r'} f(\bar{r} + \bar{r}')f(\bar{r})$ where $f(\bar{r})$ is the image matrix. A faster radial decay of the auto-correlation function implies a more disordered state. Figure 4.5 (a) shows the VL in at 8 kOe and 420 mK (corresponding to point **A** in Fig. 4.4) prepared using ZFC protocol. The VL is in the topological defect free OS. We now heat the sample to 4.14 K (point **B** in Fig. 4.4) without changing the field, thereby crossing the OS-OG boundary in the phase diagram.

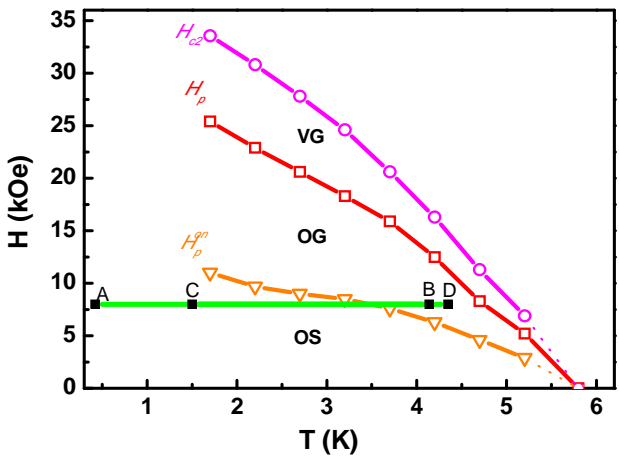

**Figure 4.4** Path along which thermal hysteresis is measured across OS➔OG phase boundary

However, the VL (Fig. 4.5 (b)) continues to remain topologically ordered. To demonstrate that this is actually a metastable superheated state, we apply a small magnetic field perturbation in the form of a pulse, by ramping up the field by 300 Oe over 8 sec followed by a dwell time of 5 sec and then ramping down over 8 sec to its original value. Since such a.c. or

Chapter 4

d.c. magnetic[1,4] (or current[25,26]) perturbation helps to overcome the local potential barriers causing a dynamic transition from a metastable state towards equilibrium state of the VL. In this case after application of a field pulse, dislocations proliferate in the VL driving it into the OG state (Fig. 4.5(c)).

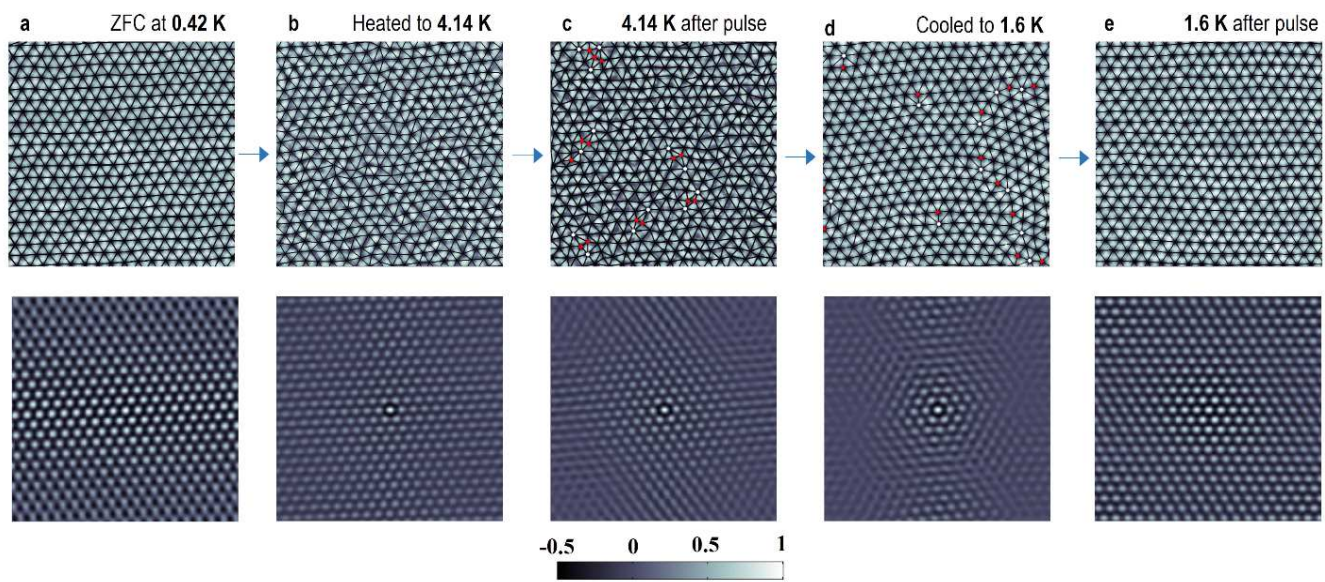

**Figure 4.5.** Hysteresis of the VL across the OS-OG boundary. Conductance maps (upper panel) and the corresponding autocorrelation function (lower panel) showing (a) the ZFC VL created at 0.42 K in a field of 8 kOe; (b) the VL after heating the crystal to 4.14 K keeping the field constant and (c) after applying a magnetic field pulse of 300 Oe at the same temperature; (d) the VL at 1.6 K after the crystal is heated to 4.35 K and cooled to 1.6 K; (e) VL at 1.6 K after applying a magnetic field pulse of 300 Oe. In the upper panels, Delaunay triangulation of the VL is shown with black lines and sites with 5-fold and 7-fold coordination are shown with red and white dots respectively. The color-scale of the autocorrelation functions is shown in the bottom.

Subsequent application of field pulse does not alter the state anymore, showing that this is the equilibrium state of the VL. To demonstrate supercooling across the OS-OG transition we heat the crystal to 4.35 K (point **D** in Fig. 4.4) and cool it back in the same field to 1.6 K (point **C** in Fig. 4.4). The additional heating is to ensure that the VL completely relaxes in OG state. The supercooled VL continues to have dislocations, characteristic of the OG state (Fig. 4.5(d)). However, after applying a 300 Oe magnetic field pulse the dislocations annihilate driving the VL into the equilibrium OS (Fig. 4.5(e)).

## 4.4.2 Thermal hysteresis across the OG→VG phase boundary

We now focus on the OG-VG boundary. Figure 4.6 indicates the thermal hysteresis path. We prepare the VL in a field of 24 kOe at 420 mK (point **E** in Fig. 4.6) using ZFC protocol. The VL shown in Fig. 4.7(a) contains dislocations as expected in the OG state. Fig. 4.7(b) shows the VL when the crystal is heated above OG-VG boundary to 2.2 K keeping the field unchanged (point **F** in Fig. 4.6). Here, the number of dislocations greatly increase. In addition to dislocations comprising of nearest neighbor pairs of 5-fold and 7-fold coordinated vortices, we also observe dislocations comprising of nearest neighbor pairs with 4-fold and 8-fold coordination, 8-fold coordinated site with two adjacent 5-fold coordinated site, and 4-fold with two adjacent 7-fold coordinated site. In addition, we also observe a small number of disclinations in the field of view. To determine the nature of this state, we examine the 2-D Fourier transform (FT) of the VL image. The FT of the image shows 6 diffuse spots showing that the orientational order is present in the VL. This is not unexpected since a small number of disclinations do not necessarily destroy the long-range orientational order[27]. This state is thus a superheated OG state. (Further evidence of OG is obtained from the orientational correlation function, $G_6(|r|)$ discussed later.) However, when a magnetic field pulse of 300 Oe is applied large number of disclinations proliferate the VL (Fig. 4.7 (c)) and the FT shows an isotropic ring, corresponding to an amorphous VG. When the crystal is subsequently cooled to 1.5 K (point **G** in Fig. 4.6) the FT shows an isotropic ring corresponding to a VG. This state is the supercooled VG (Fig. 4.7 (d)). When a magnetic field pulse of 300 Oe is applied at this temperature, the disclinations disappear driving the VL into its equilibrium OG state where the FT recovers the clear six-fold pattern (Fig. 4.7 (e)).

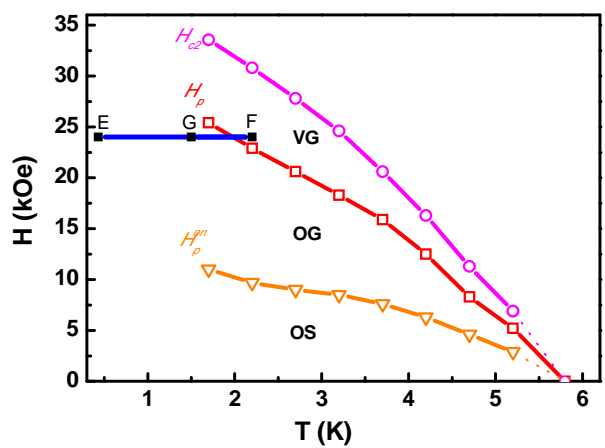

**Figure 4.6** Path along which thermal hysteresis is measured across OG→VG phase boundary

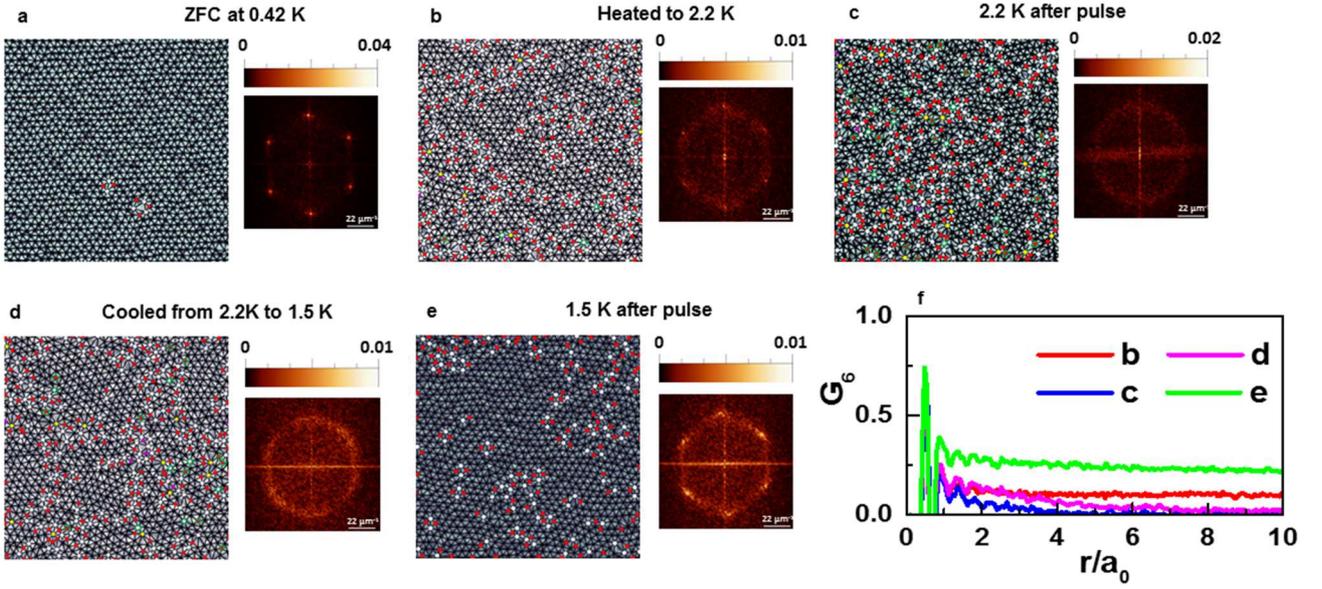

**Figure 4.7.** Hysteresis of the VL across the OG-VG boundary. Conductance map showing (a) the ZFC VL created at 0.42 K in a field of 24 kOe; (b) the VL after heating the crystal to 2.2 K keeping the field constant and (c) after applying a magnetic field pulse of 300 Oe at the same temperature; (d) the VL after the crystal is subsequently cooled to 1.5 K. (e) VL at 1.5 K after applying a magnetic field pulse of 300 Oe. The right hand panels next to each VL image show the 2D Fourier transform of the image; the color bars are in arbitrary units. Delaunay triangulation of the VL is shown with black lines, sites with 5-fold and 7-fold coordination are shown with red and white dots respectively and sites with 4-fold and 8-fold coordination are shown with purple and yellow dots respectively. The disclinations are circled in green. (f) Variation of $G_6$ as a function of $r/a_0$ (where $a_0$ is the average lattice constant) for the VL shown the panels (b)-(e).

Figure 4.7 (f) shows the orientational correlation functions, $G_6(r) = \langle \Psi_6(0)\Psi_6^*(r) \rangle$, which measure the spatial variation of the orientational order parameter, $\Psi_6(r) = \exp[6i\theta(r)]$, where $\theta(r)$ is the angle of a bond between two nearest neighbor points on the lattice located at position $r$, with respect to an arbitrary reference axis. For the superheated OG state at 2.2 K and the equilibrium OG state at 1.5 K, $G_6(r)$ tends towards a constant value for large $r$ showing long range orientational order. On the other hand for the supercooled VG state at 1.5 K and the equilibrium VG state at 2.2 K, $G_6(r)$ tends towards zero for large $r$, characteristic of an isotropic amorphous state.

In principle, the OS-OG and OG-VG phase boundaries can also be crossed by isothermal field ramping. In chapter 3, field ramping measurements[1] performed at 350 mK did not provide unambiguous evidence of superheating and supercooling though a significant hysteresis was observed between the field ramp up and ramp down branch. The most likely reason is that field ramping which changes the density of vortices involves large scale movement of vortices, which provides the activation energy to drive the VL into its equilibrium state. In contrast temperature sweeping does not significantly perturb the vortex lattice owing to the low operating temperatures, and makes these metastable states observable.

### 4.4.3 Thermal hysteresis within the phase boundary

In this context it is important to note that for a glassy system the presence of thermal hysteresis alone does not necessarily imply a phase transition, since due to random pinning the VL might not be able to relax to its equilibrium configuration with change in temperature even if we do not cross any phase boundary. We have also observed this kind of metastable states in our experiments. However, the key difference with this kind of metastable states and the superheated/supercooled states is that in this case the difference is merely in the number of topological defects. For example, ramping up the temperature at fixed field within the OG state creates such metastable states which vary from the corresponding equilibrium state only in the number of dislocations (and in the asymptotic value of $G_6(r)$) though both states have long-range orientational order. In contrast, the superheated and supercooled states are distinct from the corresponding equilibrium states both in the nature of topological defects and consequently in their symmetry properties.

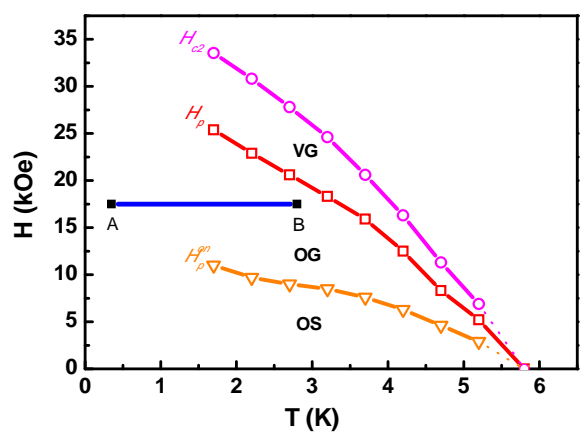

**Figure 4.8** Thermal hysteresis path within the phase boundary

Chapter 4

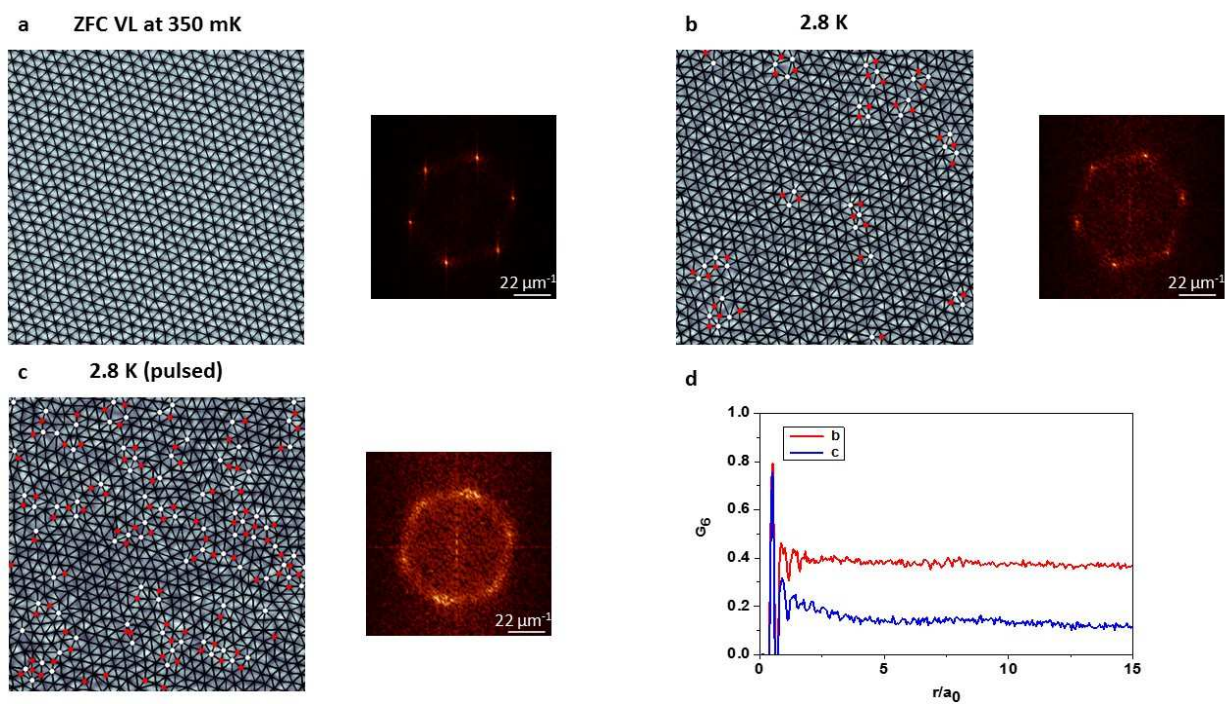

**Figure. 4.9** (a) ZFC VL created at 350 mK in a field of 17.5 kOe. (b) The VL after warming up the sample to 2.8 K without changing the magnetic field. (c) VL after applying a magnetic field pulse of 300 Oe. The panels on the right show the Fourier transform of the image. (d) $G_6$ as a function of $r/a_0$ corresponding to the VL in panels (b) and (c) respectively.

Fig. 4.9 (a) shows the ZFC VL created in a field of 350 mK in a field of 17.5 kOe (Point A in Fig. 4.8). When the crystal is heated to 2.8 K without changing the field (Point B in Fig. 4.8) we observe a finite density of dislocations (Fig. 4.9 (b)). However, when a magnetic field pulse of 300 Oe is applied on the crystal at the same temperature the density of dislocations greatly increases (Fig. 4.9 (c)), showing that the warmed up ZFC state was an unrelaxed metastable state. However, $G_6(r)$ calculated for the state before and after pulsing shows that both these states have long-range orientational order characteristic of the OG state. Thus the metastable and equilibrium state of the VL here have the same symmetry, while the difference is in the degree of long range orientational order. In contrast the superheated/supercooled states have different symmetry with respect to the corresponding equilibrium states.

## 4.5 Evidence of disorder induced phase transition

Our experiments provide structural evidence of superheating and supercooling across both OS-OG and OG-VG phase boundaries, a hallmark of thermodynamic phase transition. BKTHNY mechanism of two-step melting is not applicable in this system since it requires logarithmic interaction between vortices which is not realized in a 3-D VL. In our case, the two-step disordering is essentially induced by the presence of quenched random disorder in the crystalline lattice, which provides random pinning sites for the vortices.

Further evidence for this is obtained by comparing $\chi'$ as a function of reduced magnetic field, $h = H / H_{c2}$, at 1.7 K for A, B and C. Fig. 4.10 (a)-(c) show $\chi'$-$h$ for the three crystals. As the pinning gets weaker the difference $\Delta h = h_p - h_p^{on}$ decreases thereby shrinking the phase space over which the OG state is observed (Fig. 4.3). We speculate that in the limit of infinitesimal small pinning, $\Delta h \to 0$, thereby merging the two transitions into a single first order transition possibly very close to $H_{c2}$.

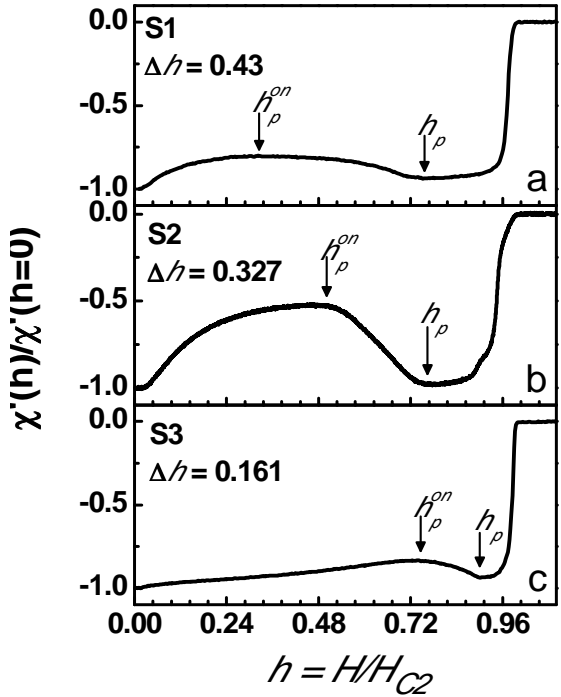

**Figure 4.10.** (a)-(c) Variation of $\chi'(h)/\chi'(0)$ as a function of reduced magnetic field, $h = H/H_{c2}$, at 1.7 K for three crystals S1, S2 and S3. $T_c$ and $\Delta h$ of the crystals are shown in the legend and $h_p^{on}$ and $h_p$ are marked with arrows.

## 4.6 Discussion

We have provided conclusive structural evidence that in the presence of weak pinning, the VL in a 3-D superconductor disorders through two thermodynamic topological phase transitions. In this context, we can also compare our results with the neutron scattering experiments reported in refs. 6 and 11 which studied the thermal history dependence of the VL across the ODT. The results presented there have strong similarities with our results, with the important

Chapter 4

difference that the two distinct steps accompanying the ODT was not identified. This could be for two reasons. First, the samples used in those studies could have weaker pinning. More importantly, the experiments there were performed in a relatively low field ($H \ll H_{c2}$) where the OG state is observed over a very narrow range of temperatures and is difficult to resolve unless measurements are performed at very small temperature intervals. Evidence of superheating and supercooling of the VL has also been observed from bulk transport measurements[28], though such measurements cannot discriminate between the OG from the VG.

# Chapter 5

# Orientational coupling between the vortex lattice and the crystalline lattice in a weakly pinned Co-doped NbSe$_2$ single crystal

## 5.1. Introduction

Until now, most theoretical descriptions of the vortex state in homogeneous 3D superconductors takes into account the vortex-vortex interaction, which stabilizes a hexagonal Abrikosov vortex lattice, and random pinning of vortices by crystalline defects, which tend to destroy this order by pinning the vortices at random positions[1,2,3,4]. For weakly pinned Type II superconductors, these theories predict that the topological defect free VL undergoes an order to disorder transition (ODT) through proliferation of topological defects[5] (TD) as one approaches the superconductor-normal metal phase boundary. These TD relax the hexagonal order which make it easier for vortices to accommodate the random pinning potential thereby enhancing the effective pinning manifested as "peak effect"[6] widely studied in weakly pinned superconducting crystals[7].

In principle, the VL can also couple with symmetry of the underlying substrate. It has been shown that in superconductors with artificially engineered periodic pinning, this coupling gives rise to interesting matching effects. In those cases, the VL gets oriented in specific direction with respect to the pinning potential when the lattice constant is commensurate with the pinning potential[8]. In single crystals of conventional superconductors, barring one exception[9], most theories dealing with the vortex phase diagram[1,2,3,4] consider the VL to be decoupled from the crystal lattice (CL) except for the random pinning potential created by defects which hinders the relative motion between two. However, in cubic and tetragonal systems, it has been theoretically[10] and experimentally[11,12] shown that non-local corrections to the vortex-vortex interaction can carry the imprint of crystal symmetry. Recent neutron diffraction experiments[13] on Nb single crystal also show that the structure of the VL varies depending on the symmetry of the crystalline axis along which the magnetic field is applied. Therefore the influence of the symmetry of the CL on the VL and consequently its effect on the ODT needs to be explored further.

Chapter 5

In this chapter, I will describe the coupling between the symmetry of the VL and CL in $Co_{0.0075}NbSe_2$, a Type II superconductor with hexagonal crystal structure. Our crystal is in the weak-pinning limit, which we functionally define as the pinning range where a topological defect free hexagonal ground state of the VL is realized at low temperatures and low fields. By simultaneously imaging the VL and CL using STM down to 350 mK, we shall investigate the orientation of the VL with respect to the CL. We shall also investigate whether the orientational coupling of the VL with the crystal lattice affects the ODT of the VL.

## 5.2. Sample details:

The single crystal with used in these set of STM and bulk measurements had superconducting transition temperature $T_c \sim 5.88$ K (Fig 5.1). It was characterized using a home built a.c. susceptometer operating at 60 kHz. The amplitude of the a.c. signal was fixed at 3.5 mOe which lies within the linear response regime. Energy dispersive X-ray (EDX) analysis was performed on the single crystal to confirm the presence of Co atoms. The EDX spectra showed a Co concentration slightly smaller than the nominal one, varying between 0.5-0.55 atomic %, on different points of the single crystal.

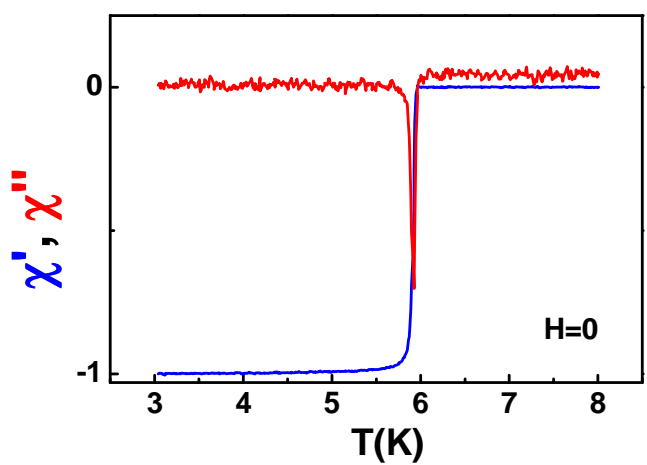

**Figure 5.1.** Real ($\chi'$) and imaginary ($\chi''$) part of a.c. susceptibility of the $Co_{0.0075}NbSe_2$ crystal in zero field. Both curves are normalized to the $\chi'$ value at 3 K.

## 5.3 Simultaneous imaging of CL and VL

### 5.3.1 Atomic resolution using STM

The VL and the CL were imaged using a home-built low temperature STM[14] fitted with a superconducting solenoid operating down to 350 mK. Prior to STM measurements the crystal is cleaved *in-situ* at room temperature, in a vacuum better than $1 \times 10^{-7}$ mbar. $NbSe_2$ has a layered hexagonal crystal structure with each unit cell consisting of two sandwiches of hexagonal Se-Nb-Se layers. Thus the crystal cleaves between the weakly coupled neighboring

Se layers exposing the hexagonal Se terminated surface. STM topographic images were captured at various locations to identify an atomically smooth surface. Fig. 5.2 (a) shows one such region (with surface height variation < 2 Å) over an area of 1.5 μm × 1.5 μm close to the center of the crystal. Atomic resolution images were subsequently captured at various points within this area to determine the unique orientation of the in-plane crystallographic axes (Fig. 5.2 (b)-(d)). The 3 × 3 charge density wave (CDW) modulation is also visible in the atomic resolution images, though it is blurred due to the presence of Co dopant atoms. The Fourier transform of the atomic resolution image (Fig. 5.2 (e)) shows 6 symmetric sharp Bragg spots corresponding to the Se lattice and 6 diffuse spots corresponding to the CDW. All VL images other than those shown in Fig. 5.3 (b) and (c) were performed within the area shown in Fig. 5.2. All the topographic images were acquired in constant current mode with set tunneling current 150 pA and a applied bias voltage of 10 mV.

## 5.3.2 VL imaging using STS

To image the VL, the differential tunneling conductance, $G(V) = dI/dV$, between the tip and the sample is measured as the tip scans the surface at a fixed bias voltage (in constant current mode), $V \sim 1.2$ mV (position of coherence peak in tunneling conductance spectra), with a d.c. tunneling current of 50 pA. $G(V)$ is measured by adding a small a.c. modulation voltage ($V_{mod}$) with frequency 2.3 kHz and amplitude 150 μV to the bias voltage and detecting the resulting modulation in the tunneling current ($I_{mod}$) using a lock-in amplifier, such that, $G(V) \approx I_{mod}/V_{mod}$. The $G(V)$-$V$ curves away from the vortex core consist of well resolved coherence peaks close to the superconducting energy gap and a minimum at zero bias characteristic of a superconductor, whereas inside the normal core they are flat and featureless. Since while acquiring the VL images the bias voltage ($V \sim 1.2$ mV) is kept close to the coherence peak, each vortex manifests as local minimum in the conductance map.

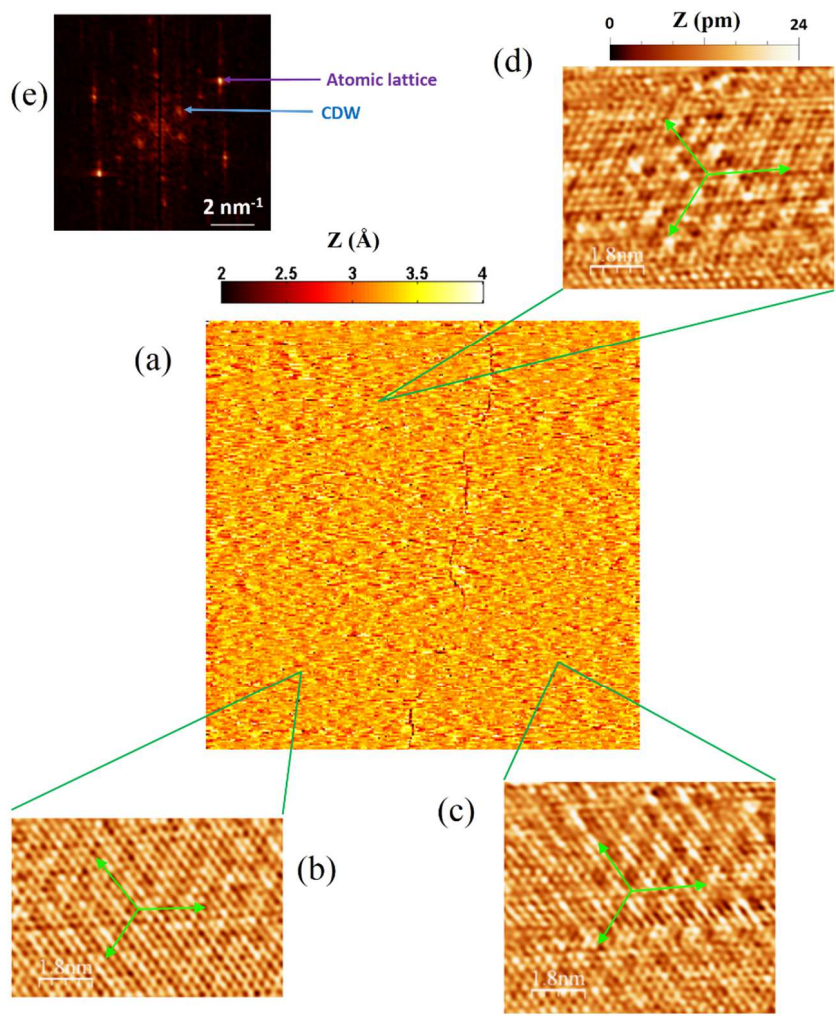

**Figure 5.2.** (a) The central panel shows topographic image of an atomically smooth surface on the Co$_{0.0075}$NbSe$_2$ single crystal. The surrounding panels (b)-(d) show representative atomic resolution images on three different areas within this region; the direction of the basis vectors in the hexagonal crystal lattice plane comprising of selenium atoms are shown with green arrows. Panel (e) shows the Fourier transform corresponding to (b); the Bragg spots corresponding to the atomic lattice and CDW in the FT and shown by arrows. We use the convention (length)$^{-1}$ for the scale-bar on the FT.

### 5.3.3 Domain formation at low fields

We first focus on the VL created at a relatively low field of 2.5 kOe in the zero field cooled state (ZFC) where the magnetic field is applied after cooling the sample to the lowest temperature. Fig. 5.3 (a)-(c) show representative images of the ZFC VL at 350 mK imaged over 1.5 μm × 1.5 μm on three different areas of the surface. While Fig. 5.3 (a) was imaged on the same area as that shown in Fig. 5.2 (a), Fig. 5.3 (b) and (c) were performed on two other similar atomically smooth areas on the crystal. We observe that the VL is oriented along different directions at different locations with no apparent relation with the orientation of the crystalline lattice. In addition, in Fig. 3(c) we observe a domain boundary created by a line of dislocations, with the VL having different orientation on two sides of the boundary. These observations are consistent with earlier Bitter decoration experiments[15] where large area images of the ZFC VL revealed large randomly oriented domains, albeit at much lower fields.

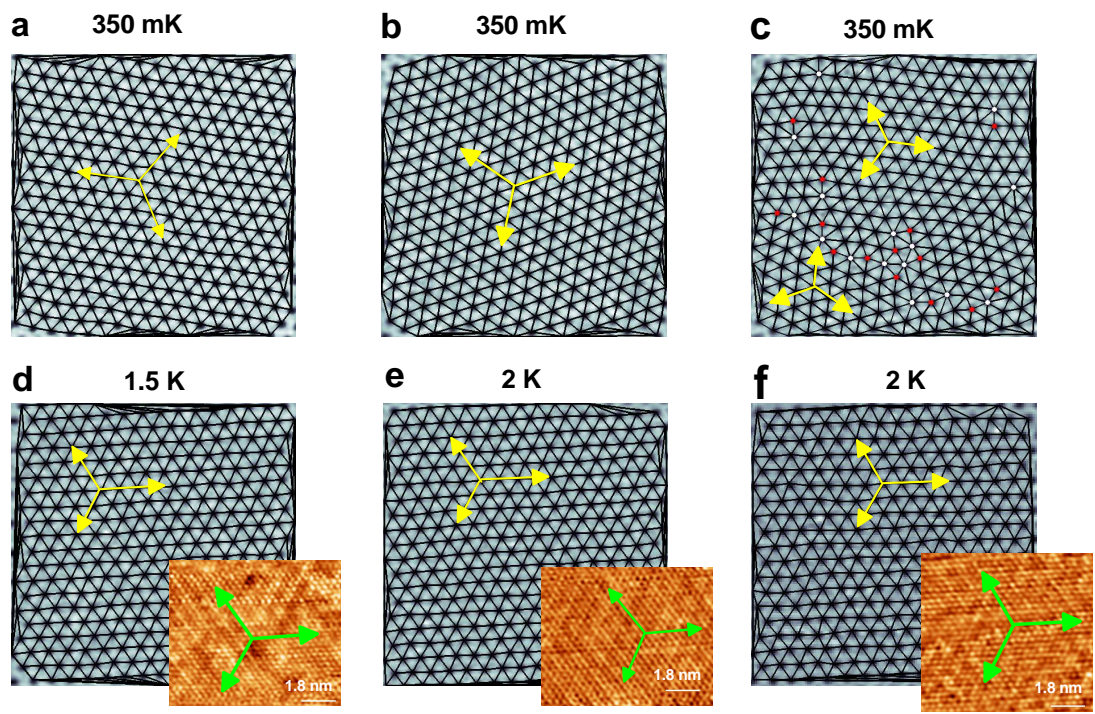

**Figure 5.3.** (a)-(c) Differential conductance maps showing the ZFC VL images (1.5 μm × 1.5 μm) recorded at 350 mK, 2.5 kOe at three different places on the crystal surface. (d)-(f) VL images at the same places as (a)-(c) respectively after heating the crystal to 1.5 K (for (d)) or 2 K ((e) and (f)) and applying a magnetic pulse of 300 Oe. Solid lines joining the vortices show the Delaunay triangulation of the VL and sites with 5-fold and 7-fold coordination are shown as red and white dots respectively. The direction of the basis vectors of the VL are shown by yellow arrows. In figure (c) a line of dislocations separate the VL into two domains with different orientations. The right inset in (d)-(f) show the orientation of the lattice, imaged within the area where the VL is imaged.

We now show that such randomly oriented domains do not represent the equilibrium state of the VL. It has been shown that shaking the VL through a small magnetic perturbation, forces the system out of metastable states causing a dynamic transition to its equilibrium configuration[16,17,18]. In our case, we did not observe any change when the ZFC VL is perturbed at 350 mK with a magnetic field pulse of 300 Oe by ramping up the field to 2.8 kOe over 8 sec followed by a dwell time of 10 sec and ramping down over 8 sec to its original value. However,

Chapter 5

when the pulse is applied after heating the sample to a temperature higher than 1.5 K the VL gets oriented along the orientation of the crystal lattice (Fig. 5.3(d)-(f)). For the VL in Fig. 5.3(f), in addition, this process annihilates the domain boundary.

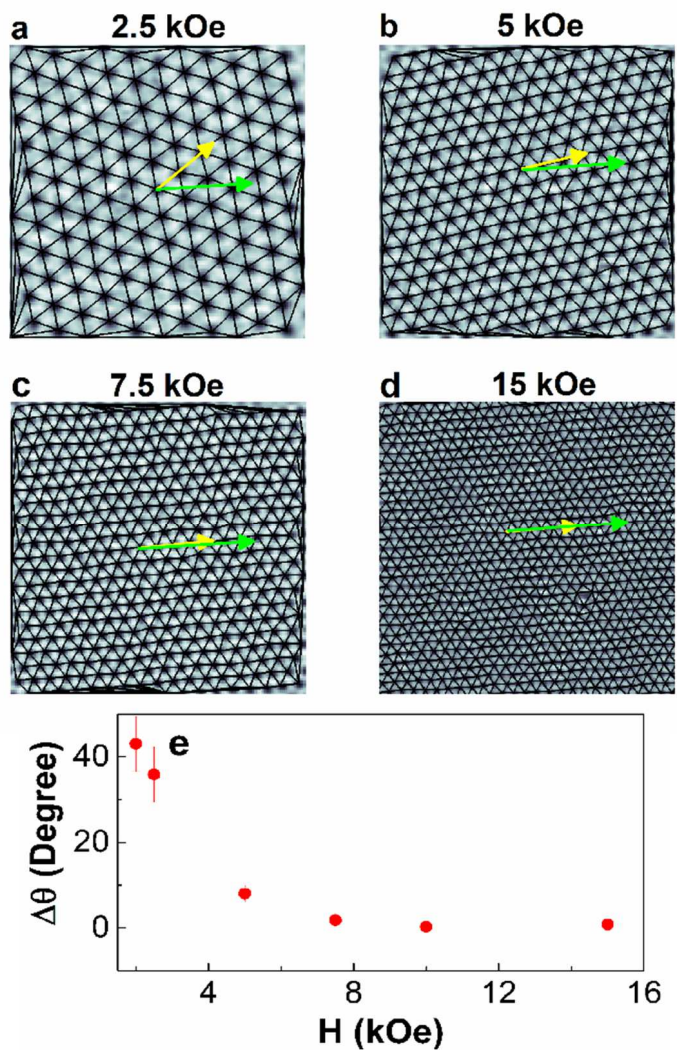

**Figure 5.4.** VL images at the same location on the sample surface imaged at (a) 2.5 kOe, (b) 5 kOe, (c) 7.5 kOe and (d) 15 kOe; the images are recorded while ramping up the field after cooling the sample to 350 mK in zero magnetic field. Solid lines joining the vortices show the Delaunay triangulation of the VL. The yellow arrow shows the direction of one basis vector of the hexagonal VL. For reference, the direction of the corresponding basis vector of the crystal lattice is shown by green arrow. While all images were acquired over 1 μm × 1 μm, the image at 15 kOe has been zoomed to show around 600 vortices for clarity. (e) Angle between the orientation of the VL

We now investigate the orientation of a single domain as the magnetic field is increased. Fig. 5.4 (a)-(d) shows the VL at the same position for 4 different fields when the magnetic field is ramped up from 2.5 kOe to 15 kOe at 350 mK. We observe that the ZFC VL gradually orients towards the crystal lattice with increase in field and above 10 kOe becomes fully oriented along the crystal lattice (Fig 5.4(e)). At 10 kOe, images taken at various locations of the sample surface shows that this orientation is global and no domain boundaries are observed in the ZFC VL till the onset of the ODT.

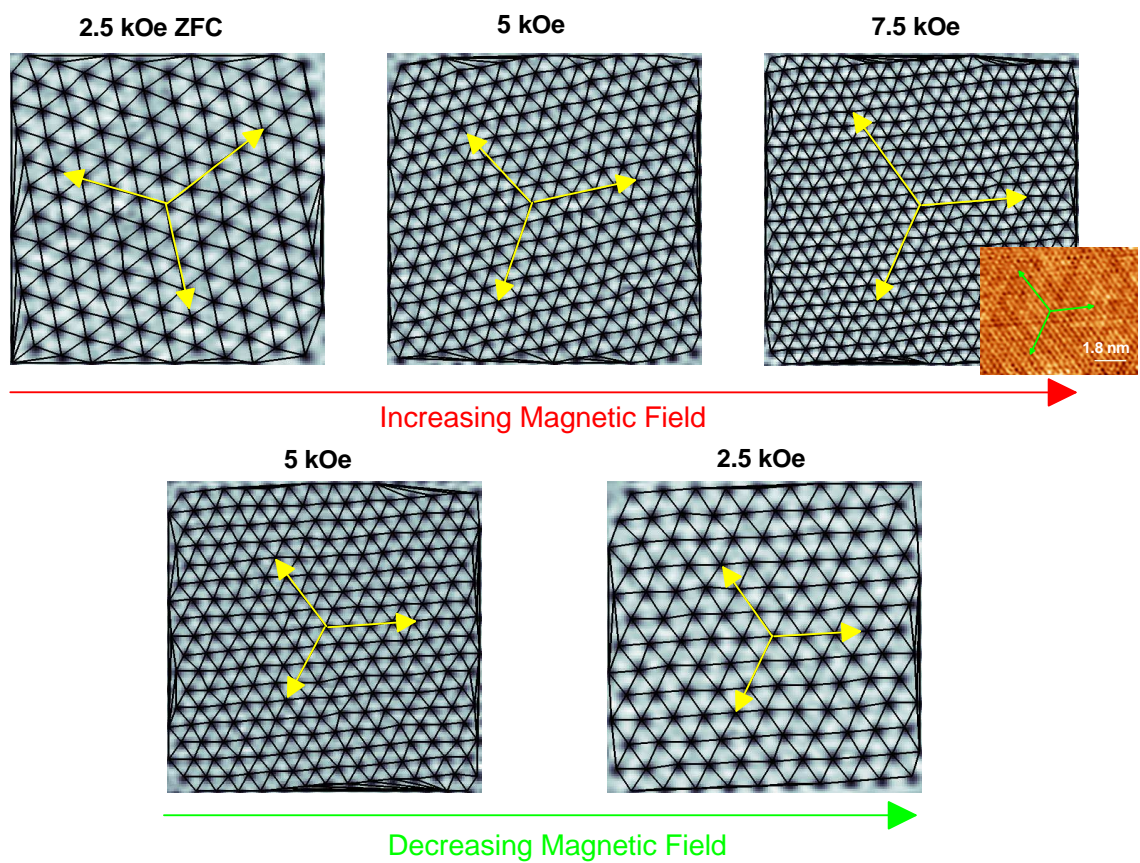

**Figure 5.5.** Differential conductance maps showing the VL at 350 mK at the same location on the sample surface imaged while cycling the magnetic field. The upper panels show the VL as the field in increased from 2.5 kOe to 7.5 kOe after preparing the VL in the ZFC state at 2.5 kOe. The two lower panels show the VL as the field is decreased from 7.5 kOe. Solid lines joining the vortices show the Delaunay triangulation of the VL. The yellow arrows show the direction of the basis vectors of the hexagonal VL. The inset in the lower right of the VL at 7.5 kOe shows representative topographic image of the atomic lattice within this area. The green arrows show the basis vector of the CL. All VL images were acquired over 1 μm × 1 μm.

At 350 mK we need to cycle the VL up to a much higher field compared to higher temperatures in order to orient the domains along the CL. In Fig. 5.5, we show the VL at the same position as the magnetic field is increased from 2.5 kOe to 7.5 kOe and then decreased to 2.5 kOe. We observe that the VL orientation gradually rotates with increasing field and at 7.5 kOe the VL is completely oriented along the CL. Upon decreasing the field, the VL maintains its orientation and remains oriented along the CL. We also observed that the orientation of the VL does not change when the crystal is heated up to 4.5 K without applying any magnetic field

Chapter 5

perturbation. Therefore, the domain structure in the ZFC VL corresponds to a metastable state, where different parts of the VL get locked in different orientations.

Since we do not observe a reorientation of the ZFC vortex lattice at 2.5 kOe at 350 mK when a magnetic field pulse of 300 Oe is applied, it is important to ensure that the applied magnetic field pulse indeed penetrates the sample. To verify this we have imaged the ZFC VL at the same place first at 350 mK first in a field of 2.5 kOe and then after ramping up the field to 2.8 kOe (Fig. 5.6). The average lattice constant from both images is calculated from the nearest neighbor bond length distribution of the Delaunay triangulated lattice. Fitting the histogram of the bond lengths to Gaussian distribution, we observe that the average lattice constant decreases by 5% after the magnetic field is ramped up by 300 Oe. This is very close to the value of 5.5% expected if the field fully penetrates in the superconductor, showing that our magnetic field pulse indeed penetrates the sample.

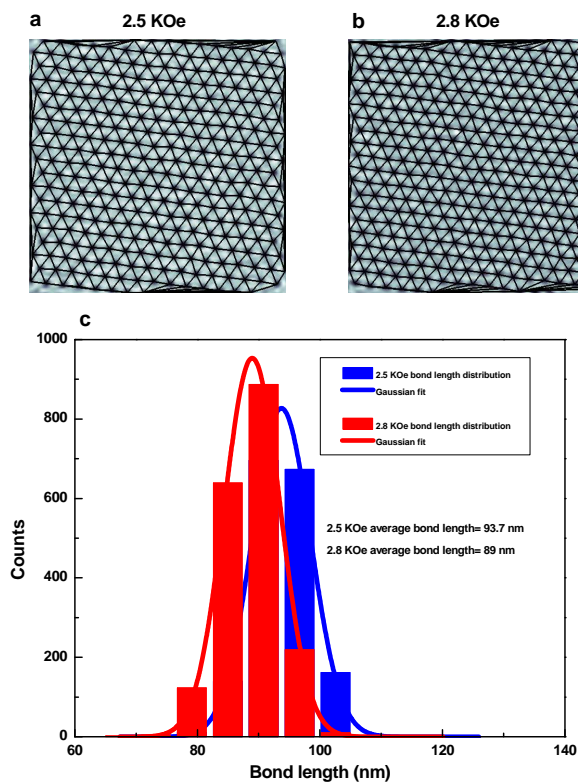

**Figure 5.6** VL image at (a) 2.5 kOe and (b) 2.8 kOe (right). The solid lines show the Delaunay triangulation of the VL. (c) Histograms showing the nearest neighbour bond length distributions at 2.5 and 2.8 kOe. The solid lines are the Gaussian fits. Spurious bonds at the edge of the images arising from the Delaunay triangulation procedure are ignored in this analysis.

## 5.3.4 Effect on bulk properties

To explore if this orientational ordering leaves its signature on the bulk pinning properties of the VL we performed ac susceptibility measurements on the same crystal (Fig. 5.7) using three protocols: In the first two protocols, the VL is prepared in the field cooled (FC) and ZFC state respectively (at 2.5 kOe) and the real part of susceptibility ($\chi'$) is measured while increasing the temperature; in the third protocol the vortex lattice is prepared at the lowest temperature in the ZFC state and the $\chi'$–T is measured while a magnetic field pulse of 300 Oe is applied at regular intervals of 100 mK while warming up. As expected the disordered FC state has a stronger diamagnetic shielding response than the ZFC state representing stronger bulk pinning. For the pulsed-ZFC state, $\chi'$–T gradually diverges from the ZFC warmed up state and shows a weaker diamagnetic shielding response and exhibits a more pronounced dip at the peak effect. Both these show that the pulsed-ZFC state is more ordered than the ZFC warmed up state, consistent with the annihilation of domain walls with magnetic field pulse.

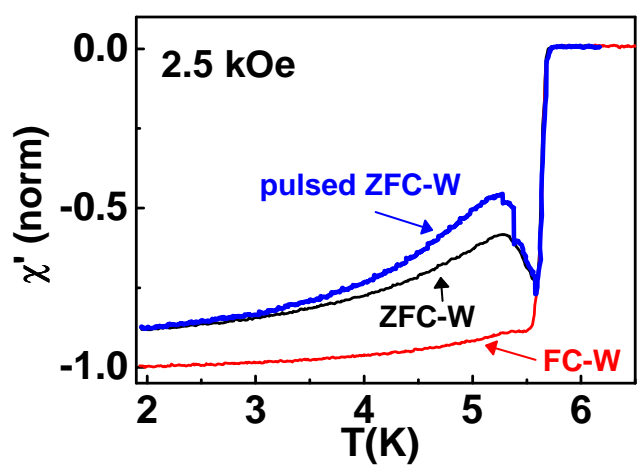

**Figure 5.7.** Susceptibility ($\chi'$) as a function of temperature (T) measured at 2.5 kOe while warming up the sample from the lowest temperature. The three curves correspond to $\chi'$-T measured after field cooling the sample (FC-W), after zero field cooling the sample (ZFC-W) and zero field cooling the sample and then applying a magnetic pulse of 300 Oe at temperature intervals of 0.1 K while warming up (pulsed ZFC-W). The y-axis is normalized to the FC-W $\chi'$ at 1.9 K. The measurements are done at 60 kHz using an a.c. excitation field of 10 mOe.

Chapter 5

## 5.4 Effect of orientational coupling on the ODT of the VL

We now investigate the impact of this orientational coupling on the ODT of the VL. As the field is increased further from the ordering field between CL and VL, the VL remains topologically ordered up to 24 kOe (Fig. 5.8). At 26 kOe dislocations proliferate in the VL, in the form of neighboring sites with 5-fold and 7-fold coordination (Fig 5.8 (b)). At 28 kOe, we observe that the disclinations proliferate into the lattice (Fig 5.8 (c)). However, the corresponding FT show a six-fold symmetry all through the sequence of disordering of the VL. Comparing the orientation of the principal reciprocal lattice vectors with the corresponding ones from the FT of crystal lattice, we observe that the VL is always oriented along the crystal lattice direction.

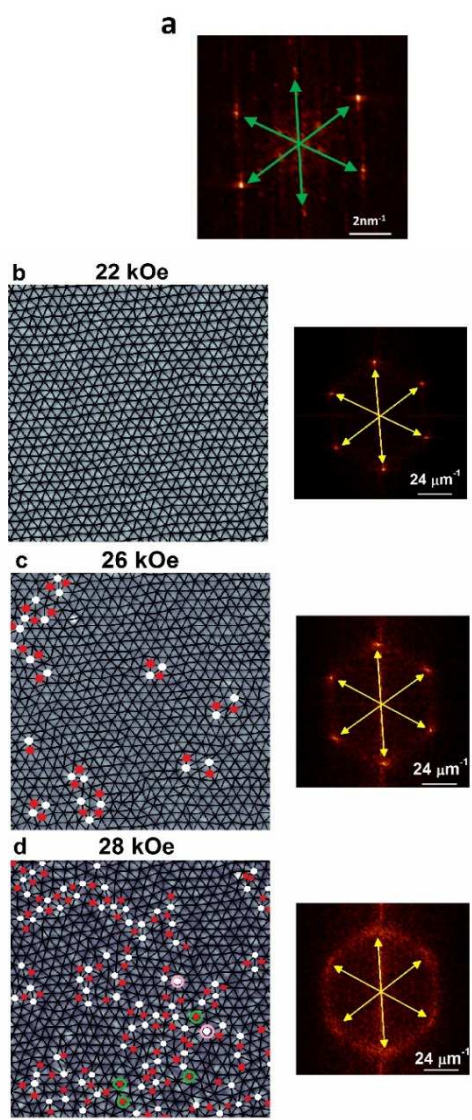

**Figure 5.8.** (a) FT of the crystal lattice; the directions of the principal reciprocal lattice vectors are shown with green arrow. (b)-(d) VL lattice images at 22, 26 and 28 kOe (left) along with their FT (right); Delaunay triangulation of the VL are shown as solid lines joining the vortices and sites with 5-fold and 7-fold coordination are shown as red and white dots respectively. The disclinations (unpaired 5-fold or 7-fold coordination sites) observed at 26 and 28 kOe are highlighted with green and purple circles. The directions of the principal reciprocal lattice vectors are shown with yellow arrows.

In contrast to our 2-step disordering observation in chapter 3, here an isotropic amorphous vortex glass phase is not realized even after the disclinations have proliferated in the system. This difference reflects the weaker defect pinning in the present crystal[Error! Bookmark not defined.], which enhances the effect of orientational coupling in maintaining the orientation of the VL along the crystal lattice.

## 5.5 Exploring the origin of orientational coupling

Now, the question that arrises from our experiments is, what is the origin of this orientational coupling? Conventional pinning cannot explain these observations, since it requires a modulation of the superconducting order parameter over a length scale of the order of the size of the vortex core, which is an order of magnitude larger that the interatomic separation and the CDW modulation in NbSe$_2$. The likely origin of orientational coupling is from anisotropic vortex cores whose orientation is locked along a specific direction of the crystal lattice. In unconventional superconductors (e.g. (La,Sr)CuO$_4$, CeCoIn$_5$) [19,20] such anisotropic cores could naturally arise from the symmetry of the gap function, which has nodes along specific directions. However even in an s-wave superconductor such as YNi$_2$B$_2$C vortex cores with 4-fold anisotropy has been observed, and has been attributed to the anisotropy in the superconducting energy gap resulting from Fermi surface anisotropy[21]. As the magnetic field is increased, the vortices come closer to each other and start feeling the shape of neighboring vortices. The interaction between the vortices in such a situation would also be anisotropic, possessing the same symmetry as that of the vortex core. Thus the interaction energy would get minimized for a specific orientation of the VL with respect to the CL.

### 5.5.1 High resolution spectroscopic imaging of single vortex core

To explore this possibility we performed high resolution imaging of a single vortex core at 350 mK. To minimize the influence of neighboring vortices we first created a ZFC vortex lattice at 350 mK (Fig. 5.9(a)) in a field of 700 Oe, for which the inter vortex separation (177 nm) is much larger than the coherence length. As expected at this low field the VL is not aligned with the CL (Fig. 5.9(e)). We then chose a square area enclosing a single vortex and measured the full $G(V)$-$V$ curve from 3mV to -3mV at every point on a $64 \times 64$ grid. In Fig 5.9(b)-(d) we plot the normalized conductance $G(V)/G(3\ mV)$ at 3 bias voltages. The normalized conductance images reveal a hexagonal star shape pattern consistent with previous measurements[22,23] in undoped NbSe$_2$ single crystals. Atomic resolution images captured within the same area (Fig. 5.5(e)) reveals that the arms of the star shape in oriented along the principal

Chapter 5

directions of the CL. We observe that the star shape is not specifically oriented along any of the principal directions of the VL, which rules out the possibility that the shape arises from the interaction of supercurrents surrounding each vortex core.

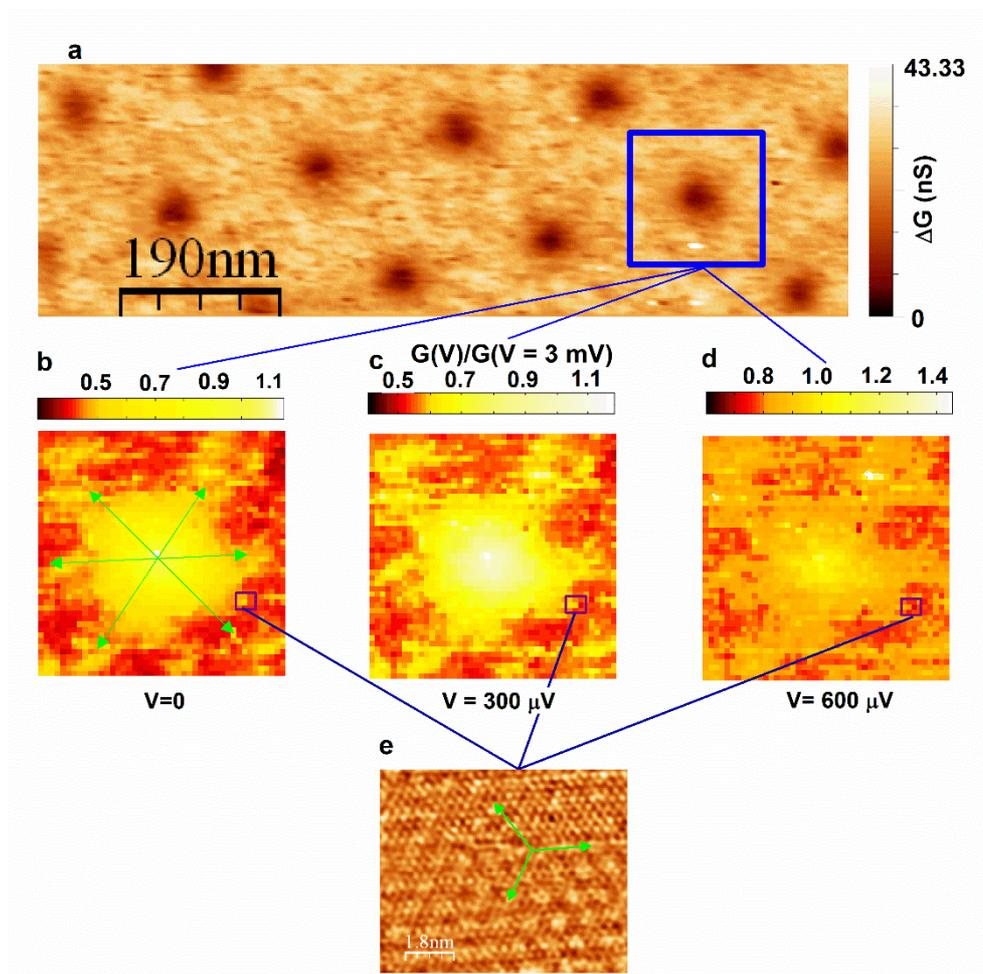

**Figure 5.9.** (a) Differential conductance map showing the ZFC VL at 700 Oe and 350 mK. (b) High resolution image of the single vortex (114 nm ×114 nm) highlighted in the blue box in panel (a) obtained from the normalized conductance maps ( G(V)/G(V = 3mV) ) at 3 different bias voltages. The vortex core shows a diffuse star shaped patters; the green arrows point towards the arms of the star shape from the center of the vortex core. (c) Atomic resolution topographic image of the CL imaged within the box shown in (b); the green arrows show the principal directions of the crystal lattice.

We believe that the reduced contrast in our images compared to refs. 22,23 is due to the presence of Co impurities which act as electronic scattering centers and smear the gap anisotropy through intra and inter-band scattering.

The six-fold symmetric vortex core structure explains why the ZFC VL gets oriented when we cycle through larger fields. As the magnetic field is increased the vortices come closer and feel the star shape of neighboring vortices. Since the star shape has specific orientation with respect to the CL, the VL also orients in a specific direction with respect to the CL. When the field is reduced, the vortices no longer feel the shape of neighboring vortices, but the lattice retains its orientation since there is no force to rotate it back. Since the 6-fold symmetry of the vortex core here is the same as the hexagonal Abrikosov lattice expected when the vortex cores are circular, we do not observe any field induced structural phase transition of the VL as observed in superconductors where the vortex core has four-fold symmetry[12].

## 5.6. Conclusion

In conclusion, we have used direct imaging of the crystal lattice and the VL using STM/S to show that the orientation of the VL in a conventional s-wave superconductor is strongly pinned to the crystal lattice. This orientational coupling influences both the equilibrium state at low fields and the order-disorder transition at high field. While at low fields, locally misoriented domains are observed in the ZFC state, these domains get oriented along the CL when the system is cycled through a larger field. In addition, the persistence of orientational order in the VL at high fields even after proliferation of topological defects, clearly suggests that this coupling can be energetically comparable to the random pinning potential and cannot be ignored in realistic models of the VL in weakly pinned Type II superconductors.

# Chapter 5

---

[23] I. Guillamon, H. Suderow, F. Guinea, S. Vieira, Phys. Rev. B **77**, 134505 (2008); I. Guillamon, H. Suderow, S. Vieira, P. Rodiere, *Physica* C **468,** 537 (2008).

# Chapter 6

# Conclusions and future directions

We have shown by performing real space imaging across the magnetic field driven peak effect that the vortex lattice in a 3-dimensional, weakly pinned Type-II superconductor disorders through two distinct topological phase transitions. In the first transition, the vortex solid having quasi-long range positional order and long range orientational order, namely the 'Bragg glass' phase goes into a state having short range positional order and quasi-long range orientational order, namely the 'Orientational glass' phase. This state across the second transition transforms into a state having short-range positional and orientational order, namely 'Vortex glass' phase which gradually evolves to isotropic 'Vortex liquid' phase with clear evidence of softening. The nature of these phase transformations were inferred upon by performing thermal hysteresis. Clear evidence of superheating and supercooling with different topological property of the metastable states from the actual ground states indicated the nature of both the phase transformations to be of first order.

The sequence of disordering of VL observed in our experiment is strikingly similar to the two-step BKTHNY transition observed in 2-D systems, where a hexatic fluid exists as an intermediate state between the solid and the isotropic liquid. However, since BKTHNY mechanism of two-step melting is not applicable in this system since it requires logarithmic interaction between vortices which is not realized in a 3-D VL, in our case the two-step disordering is essentially induced by the presence of quenched random disorder in the crystalline lattice, which provides random pinning sites for the vortices. In this alternative viewpoint, it was speculated that in the presence of weak pinning the transition can be driven by point disorder generating positional entropy rather than temperature. In this scenario topological defects proliferate in the VL through the local tilt of vortices caused by point disorder, creating an "entangled solid" of vortex lines. Here, in contrast to conventional thermal melting, the positional entropy generates instability in the ordered VL driving it into a disordered state, even when thermal excitation alone is not sufficient to induce a phase transition.

As the phase boundaries obtained from the bulk magnetisation measurements and the microscopic real space images matches extremely well. The magnetisation measurements performed on different samples with varying degree of disorder indicated for cleaner samples

Chapter 6

with increasingly less pinning, the phase space available for intermediate 'Orientational glass' phase is monotonically reducing leading us to speculate that in the limit of zero-pinning, the two phase transitions will merge into single first order phase transition.

In the theoretical treatment of disorder induced melting, only the role of dislocations were considered. It would be interesting to also theoretically investigate the role of disclinations in the VL, which has not been explored in detail so far.

We have also demonstrated the role of underlying crystalline lattice symmetry by showing that it aligns the VL along preferential directions which is the lattice vectors of the CL itself. Its reason being the anisotropic shape of the vortex core arising due to the anisotropy in the Ferni surface which reflects the hexagonal CL symmetry. In the less disordered samples, this orientational coupling is shown to provide robust orientational order even in the disclination dominated phase. In this way, we have shown that in addition to vortex-vortex and vortex-impurity interaction energies, we also need to consider the VL-CL interaction energy for constructing the complete vortex phase diagram. We hope that future theoretical studies will quantitatively explore the magnitude CL-VL coupling energy scale with respect to vortex–vortex and pinning energies and its effect on the vortex phase diagram of Type-II superconductors.

Theoretically, it would also be interesting to explore to what extent these concepts can be extended to other systems, such as colloids, charge-density waves and magnetic arrays where a random pinning potential is almost always inevitably present.